\documentclass{amsart}
\usepackage{epsfig}
\usepackage{hyperref}
\usepackage{verbatim,amssymb,amsfonts,amsthm,amsmath,amscd,ifthen}
\def\sgn{\qopname \relax o{sgn}}
{     \theoremstyle{plain}
        \newtheorem{theorem}{Theorem}[section]
         \newtheorem{lemma}[theorem]{Lemma} %[section]
          \newtheorem{corollary}[theorem]{Corollary} %[section]
          \newtheorem{proposition}[theorem]{Proposition} %[section]%}
}
{       \theoremstyle{remark}
        \newtheorem{remark}{Remark}[section]
}
\makeatletter
\@addtoreset{equation}{section}
\makeatother
\renewcommand{\thetheorem}{\arabic{section}.\arabic{theorem}}
\renewcommand{\theremark}{\arabic{section}.\arabic{remark}}
\renewcommand{\theequation}{\arabic{section}.\arabic{equation}}
\begin{document}
\def\spIon{Q_{1,1}} %{I_{19}}
\def\spIIon{Q_{1,2}} %{II_{19}}
\def\spIIIon{Q_{1,3}} %{III_{19}}
\def\spItt{Q_{2,1}} %{I_{23}}
\def\spIItt{Q_{2,2}} %{II_{23}}
\def\spIIItt{Q_{2,3}} %{III_{23}}
\def\spJtop{Q_{3,1}} %{J_{31}^{\prime}}
\def\spJtopp{Q_{3,2}} %{J_{31}^{\prime\prime}}
\def\spItf{Q_{4,1}} %{I_{35}}
\def\spIItf{Q_{4,2}} %{II_{35}}
\def\spIIItf{Q_{4,3}} %{III_{35}}
\def\spIte{Q_{5,1}} %{I_{38}}
\def\spIIte{Q_{5,2}} %{II_{38}}
\def\spIfz{Q_{6,1}} %{I_{40}}
\def\spIIfz{Q_{6,2}} %{II_{40}}
\def\spIIIfz{Q_{6,3}} %{III_{40}}
\def\spIst{Q_{7,1}} %{I_{62}}
\def\spIIst{Q_{7,2}} %{II_{62}}
\def\specIthree{Q_{8,1}} %{I_3 }
\def\specIIthree{Q_{8,2}} %{II_3 }
\def\specIIIthree{Q_{8,3}} %{III_3 }
\def\specIsix{Q_{9,1}} %{I_6 }
\def\specIIsix{Q_{9,2}} %{II_6 }
\def\specIIIsix{Q_{9,3}} %{III_6 }
\def\specIseven{Q_{10,1}} %{I_7 }
\def\specIIseven{Q_{10,2}} %{II_7 }
\def\specIeight{Q_{11,1}} %{I_8 }
\def\specIIeight{Q_{11,2}} %{II_8 }
\def\specIIIeight{Q_{11,3}} %{III_8 }
\def\too{{\dot{\to}}}
\def\trace{\mathop{\rm Trace}}
\def\N{\mathbb{N}}
\def\R{\mathbb{R}}
\def\E{\mathbb{E}}
\def\ls{\left[}
\def\ZP{{\mathbb{Z}}_{\geq0}}
\def\Z{{\mathbb{Z}}}
\def\C{{\mathbb{C}}}
\def\rs{\right]}\def\lp{\left(}\def\rp{\right)}
\def\diag{\qopname \relax o{diag}}
\def\vol{\qopname \relax o{vol}}
\def\Tr{\qopname \relax o{Tr}}
\def\tr{\qopname \relax o{tr}}
\def\dist{\qopname \relax o{dist}}
\def\dom{\qopname \relax o{Dom}}
\def\Prob{\qopname \relax o{Prob}}
\def\spec{\qopname \relax o{spec}}
\def\Ai{\qopname \relax o{Ai}}
\def\ra{\rightarrow}
\def\eps{\epsilon}
\def\sgn{\qopname \relax o{sgn}}
\def\const{{\rm const}\relax }
\def\sno{S_N^{(1)}}
\def\snop{S_N^{(1)\prime}}
\def\snf{S_N^{(4)}}
\def\snt{S_N^{(2)}}
\def\ph{\phi}
\def\corrf{\qopname \relax o{Corr}_N^{(4)}}
\def\corro{\qopname \relax o{Corr}_N^{(1)}}
\newcommand{\nc}{\newcommand}
\nc{\onetwo}[2]{\left(\begin{array}{cc}#1&#2\end{array}\right)}
\nc{\twotwo}[4]{\left(\begin{array}{cc}#1&#2\\&\\#3&#4\end{array}\right)}
\nc{\twoone}[2]{\left(\begin{array}{c}#1\\\\#2\end{array}\right)}
\nc{\mtt}[4]{\begin{pmatrix}#1&#2\\#3&#4\end{pmatrix}}
\newcommand{\bp}{\begin{proof}}
\newcommand{\ep}{\end{proof}}
\newcommand{\bt}{\begin{theorem}}
\newcommand{\et}{\end{theorem}}
\newcommand{\be}{\begin{equation}}
\newcommand{\ee}{\end{equation}}
\newcommand{\bq}{\begin{equation}}
\newcommand{\eq}{\end{equation}}
\newcommand{\ba}{\begin{aligned}}
\newcommand{\ea}{\end{aligned}}
\newcommand{\la}[1]{\label{#1}}
\nc{\fh}{\hat{f}}
\nc{\ft}{\tilde{f}}
\nc{\fth}{\hat{\tilde{f}}}
\nc{\pht}{\tilde{\phi}}
\nc{\pth}{\hat{\tilde{\phi}}}
\nc{\cd}{\cdots}
\nc{\ov}{\over}
\newcommand{\er}{\eqref}

\title[Universality for orthogonal and symplectic
ensembles]
{Universality in Random Matrix Theory for
orthogonal and symplectic ensembles} 
%A proof of the universality conjecture for orthogonal and symplectic
%ensembles of random matrices with polynomial potentials.\\Full version}

\author[Deift and Gioev]
{Percy Deift and Dimitri Gioev} 
\address{Deift: Department of Mathematics, Courant Institute of Mathematical
Sciences, New York University, New York, NY 10012}
\email{deift@cims.nyu.edu}
\address{Gioev:
Department of Mathematics,
University of Pennsylvania,
Philadelphia, PA 19104\\
and Department of Mathematics, Courant Institute of Mathematical
Sciences, New York University, New York, NY 10012}
\email{gioev@math.upenn.edu, gioev@cims.nyu.edu}
\begin{abstract}
We give a proof of universality in the bulk for orthogonal ($\beta=1$)
and symplectic ($\beta=4$) ensembles of random matrices in the scaling limit 
for a class of weights $w(x)=e^{-V(x)}$ where $V$ is a polynomial,
$V(x)=\kappa_{2m}x^{2m}+\cdots$, $\kappa_{2m}>0$.
For such weights the associated 
equilibrium measure is supported on 
a single interval.
The precise statement of our results is given in Theorem~\ref{thmuniv} below.
For a proof of universality in the bulk
for unitary ensembles ($\beta=2$), for the same class of weights,
see \cite{DKMVZ2}.

Our starting point is Widom's representation \cite{W}
of the orthogonal
and symplectic correlation kernels in terms of the kernel
arising in the unitary case ($\beta=2$) 
plus a correction term
which is constructed out of derivatives and integrals 
of orthonormal polynomials (OP's) $\{p_j\}_{j\geq0}$
with respect to the weight $w(x)$.
The calculations in \cite{W} in turn depend on the earlier work of
Tracy and Widom \cite{TW2}.
It turns out (see \cite{W} and also Theorems~\ref{thm1} and \ref{thm2} below)
that only the OP's in the range $j=N+O(1)$, $N\to\infty$,
contribute to the correction term.
In controlling this correction term, and hence proving
universality for $\beta=1$ and $4$,
the uniform Plancherel--Rotach type asymptotics for the OP's 
found in \cite{DKMVZ2} play an important role,
but there are significant new analytical difficulties
that must be overcome which are not present
in the case $\beta=2$.
We note that we do not use skew orthogonal polynomials. 
\end{abstract}
\maketitle
\section{Introduction}
\la{sec1}
In this paper we will be concerned with ensembles
of  matrices $\{M\}$ with probability distributions
\be
\la{eq1p1}
     \mathcal{P}_{N,\beta}(M)\,dM 
  = \frac1{\mathcal{Z}_{N,\beta}}\,e^{-\tr V_\beta(M)}\,dM,
\ee
for $\beta=1$, $2$ and $4$, the so-called Orthogonal, Unitary
and Symplectic ensembles, respectively (see \cite{M}).
For $\beta=1$, $2$, $4$, the ensemble consists of $N\times{}N$
real symmetric matrices, $N\times{}N$
Hermitian matrices, and $2N\times{}2N$
Hermitian self-dual matrices, respectively.
In general the potential $V_\beta(x)$
is a real-valued function growing sufficiently rapidly as $|x|\to\infty$,
but we will restrict our attention henceforth to $V_\beta$'s
which are polynomials,
\be
\la{eq1p2}
   V_\beta(x)=\kappa_{2m,\beta}x^{2m}+\cdots,\qquad\kappa_{2m,\beta}>0.
\ee
In \eqref{eq1p1}, $dM$ denotes Lebesgue measure on the 
algebraically independent 
entries of $M$, and $\mathcal{Z}_{N,\beta}$ is a normalization constant.
The above terminology for $\beta=1$, $2$ and $4$
reflects the fact that \eqref{eq1p1}
is invariant under conjugation of $M$, $M\mapsto{}UMU^{-1}$,
by orthogonal, unitary and unitary-symplectic
matrices $U$.
It follows from \eqref{eq1p1} that the distribution of the 
eigenvalues $x_1,\cdots,x_N$
of $M$ is given (see \cite{M}) by
\begin{equation}
\label{eq_PN}
     P_{N,\beta}(x_1,\cdots,x_N) =  \frac1{{Z}_{N,\beta}}
         \prod_{1\leq{}j<k\leq{}N}
        |x_j-x_k|^\beta\prod_{j=1}^N w_\beta(x_j)
\end{equation}
where again $Z_{N,\beta}$ is a normalization constant (partition function).
Here
\be
\la{eq1p3}
   w_\beta(x)=\begin{cases}
                  e^{-V_\beta(x)},&\beta=1,2\cr
                  e^{-2V_\beta(x)},&\beta=4.
               \end{cases}
\ee
(The factor $2$ in $w_{\beta=4}$ reflects the fact that the
eigenvalues of self-dual Hermitian matrices come in pairs.)
Let $\{p_j\}_{j\geq0}$ be the normalized orthogonal polynomials (OP's) 
on $\R$ with respect to the weight $w\equiv{}w_{\beta=2}$,
and define $\phi_j\equiv{}p_j{}w^{1/2}$.
Note that  $(\phi_j,\phi_k)=\delta_{jk}$
where $(\cdot,\cdot)$ denotes the standard inner product in $L^2(\R)$.

For the unitary matrix ensembles ($\beta=2$) 
an important role is played by the
kernel
\bq\label{K} 
      K_N(x,y)\equiv K_{N,2}(x,y)=\sum_{k=0}^{N-1}\ph_k(x)\,\ph_k(y).
\eq 
In particular the probability density \eqref{eq_PN},
the $l$-point correlation function $R_{N,l,2}$ 
and also the gap
probability $E_2(0;J)$ that a set $J$ contains no eigenvalues,
can all be expressed in terms of $K_N$, see e.g.~\cite{M}.
For example 
\be\la{eq1a}
   R_{N,l,2}(x_1,\cdots,x_l)=\det(K_N(x_j,x_k))_{1\leq{}j,k\leq{}l}.
\ee

The Universality Conjecture, in our situation, 
states that the limiting statistical behavior of the eigenvalues
$x_1,\cdots,x_N$ distributed according to the law \eqref{eq_PN},
in the appropriate scale as $N\ra\infty$,
should be independent of the weight $w_\beta$,
and should depend only on the invariance properties of $\mathcal{P}_{N,\beta}$,
$\beta=1$, $2$ or $4$, mentioned above. 
Universality has been considered extensively in the physics literature,
see e.g.~\cite{BZ,Be,HW,SV}.

The kernel $K_N(x,y)$ can also be expressed
via the Christoffel--Darboux formula
\begin{equation}
\label{eq_KN}
     K_N(x,y)= b_{N-1}\,\frac{\phi_{N}(x)\phi_{N-1}(y)
         -\phi_{N-1}(x)\phi_{N}(y)}{x-y},
\end{equation}
where $b_{N-1}$ is a coefficient 
in the three-term recurrence relation for OP's (see \eqref{eqthreeterm} below).
In view of the preceding remarks it follows that
in the case $\beta=2$,
the study of the large $N$ behavior of 
$P_{N,2}$, and in particular the proof of universality,
reduces to the 
asymptotic analysis of $b_{N-1}$ and the OP's $p_{N+j}$ with $j=0$ or $-1$.
By a fundamental observation
of Fokas, Its and Kitaev [FoIKi]
 the OP's solve 
a Riemann--Hilbert problem (RHP) of a type
that is amenable to the steepest descent method
introduced by Deift and Zhou in \cite{DZ} and further developed in \cite{DVZ}. 
In \cite{DKMVZ,DKMVZ2} the authors 
analyzed the asymptotics of OP's for very general classes of weights.
In particular they proved the Universality Conjecture 
in the case $\beta=2$ for weights
$w(x)=e^{-V(x)}$ where $V(x)$ is a polynomial as above,
and also for $w(x)=e^{-NV(x)}$ where $V(x)$ is real analytic and 
$V(x)/\log|x|\to+\infty$, as $|x|\to\infty$.
The (bulk) scaling limit $N\to\infty$
is described in terms of the so-called {\em sine kernel\ }
$K_\infty(x-y)$ where 
\be\la{eqsinek}
            K_\infty(t)\equiv\frac{\sin\pi t}{\pi t}.
\ee
For example \cite[Theorem~1.4]{DKMVZ2}, 
for $w(x)=e^{-V(x)}$, $V(x)$ polynomial,
and for any $l=2,3,\cdots$ and $r,y_1,\cdots,y_l$ in a compact
set, one has as $N\ra\infty$
\be\la{eq1b}
%\begin{aligned}
   \frac{1}{(K_N(0,0))^l}\,R_{N,l,2}\Big(r+\frac{y_1}{K_N(0,0)},
   \cdots,\,%&
r+\frac{y_l}{K_N(0,0)}\Big)%\\
%    &
\ra\det(K_\infty(y_j-y_k))_{1\leq j,k\leq l}.
%\end{aligned}
\ee
The scale $x=y/K_N(0,0)$ is chosen so that the expected number of
eigenvalues per unit $y$-interval is one: This scaling is standard
in Random Matrix Theory.
Indeed for any Borel set $B\subset\R$,
\be\la{eqs3p1}
  \int_{B} R_{N,l=1,2}(x)\,dx = \E\{\,\textrm{number of eigenvalues in $B$}\,\}.
\ee
Thus by \eqref{eq1a} $K_N(0,0)=R_{N,1,2}(0)$ gives the density
of the expected number of eigenvalues near zero.
In other words, in the appropriate scale,
the large $N$ behavior of the eigenvalues is {\em universal\ }(i.e.~independent of $V$).
Pioneering mathematical work on the Universality Conjecture
was done in \cite{PS}
and for the case of quartic two-interval potential $V(x)=N(x^4-tx^2)$,
$t>0$ (sufficiently) large, in \cite{BI}.
We note again that
 all these results apply only in the case $\beta=2$.

In the case $\beta=1$ and $4$ the situation is more complicated.
In place of \eqref{K} one must use
$2\times2$ matrix kernels (see e.g. \cite{M,TW2})
\bq\label{1K}
K_{N,1}(x,y)=\twotwo{S_{N,1}(x,y)}{(S_{N,1}D)(x,y)}
                   {(\eps S_{N,1})(x,y)-\frac12\sgn(x-y)}{S_{N,1}(y,x)},
   \quad N\textrm{ even},
\eq
and
\bq\label{4K} 
  K_{N,4}(x,y)=\frac12\twotwo
     {S_{N,4}(x,y)}{(S_{N,4}D)(x,y)}
    {(\eps S_{N,4})(x,y)}{S_{N,4}(y,x)}.
\eq
Here  $S_{N,\beta}(x,y)$, $\beta=1,4$, are certain 
scalar kernels (see \eqref{eq3pp1}, \eqref{eq3pp2} below),
$D$ denotes the differentiation operator, and $\eps$ is 
the operator with kernel $\eps(x,y)=\frac12\sgn(x-y)$\footnote{We use the 
standard notation $\sgn x=1,$ $0$, $-1$ for $x>0$, $x=0$, $x<0$,
respectively.}.
Such matrix kernels were first introduced by Dyson \cite{Dy70}
in the context of circular ensembles with a 
view to computing correlation functions.
Dyson's approach was extended to Hermitian ensembles,
first by Mehta \cite{Me71} for $V(x)=x^2$,
and then for more general weights by Mahoux and Mehta in \cite{MMe91}.
A more direct and unifying approach to the 
results of Dyson--Mahoux--Mehta was given 
by Tracy and Widom in \cite{TW2},
where formulae \eqref{eq3pp1}, \eqref{eq3pp2} below were derived.
We see that once the kernels
$S_{N,\beta}(x,y)$ are known, then so are the other kernels in $K_{N,\beta}$.
As in the case $\beta=2$, the kernels $K_{N,\beta}$
give rise to explicit formulae for $R_{N,l,\beta}$ and $E_\beta(0;J)$.
For example 
\be
\la{eqR1beta}
        R_{N,1,\beta}(x)\equiv R_{1,\beta}(x) = \frac12\tr K_{N,\beta}(x,x)
\ee
and
$$
        R_{N,2,\beta}(x,y) 
         = \frac14\,\big(\tr K_{N,\beta}(x,x)\big)\big(\tr K_{N,\beta}(y,y)\big) 
           - \frac12 \tr\big( K_{N,\beta}(x,y) K_{N,\beta}(y,x)\big), 
$$
and so on, see \cite{TW2}.
As indicated above, formula \eqref{1K} only applies to the case when $N$
is even. When $N$ is odd, there is a similar, but slightly more complicated,
formula (see \cite{AFNvM}). Throughout this paper, for $\beta=1$, 
we will restrict our attention 
{\em to the case when $N$ is even.\ }We expect that the methods in this paper 
also extend to the case $\beta=1$, $N$ odd, and we plan to consider
this situation in a later publication.
Of course, in situations where the asymptotics of \eqref{1K}
has been analyzed (e.g.{} $V(x)=x^2$) for all $N$ as $N\to\infty$,
the limiting behavior of $R_{N,l,\beta=1}$ 
is indeed seen to be independent of 
the parity of $N$ (see e.g.{} \cite{M,NW}).

Let $\{q_j(x)\}_{j\geq0}$ be any sequence of polynomials
of exact degree $j$, $q_j(x)=q_{j,j}x^j+\cdots$, $q_{j,j}\neq0$.
For $j=0,1,2,\cdots$, set 
\be
\la{eq3p1}
        \psi_{j,\beta}(x)=\begin{cases}
                 q_j(x) w_1(x),&\beta=1\cr
                 q_j(x) (w_4(x))^{1/2},&\beta=4.
                                     \end{cases}
\ee 
Let $M_{N,1}$ denote the $N\times{}N$ matrix with entries
\be
\la{eq3p2}
       (M_{N,1})_{jk} = (\psi_{j,1},\eps\psi_{k,1}),\qquad 0\leq j,k\leq N-1,
\ee
and let $M_{N,4}$ denote the $2N\times{}2N$ matrix with entries
\be
\la{eq3p3}
       (M_{N,4})_{jk} = (\psi_{j,4},D\psi_{k,4}),\qquad 0\leq j,k\leq 2N-1,
\ee
where again $(\cdot,\cdot)$ denotes the standard real inner product on $\R$.
The matrices $M_{N,1}$ and $M_{N,4}$ are invertible 
(see e.g.~\cite[(4.17), (4.20)]{AvM}).
Let $\mu_{N,1}$, $\mu_{N,4}$ denote the inverses of $M_{N,1}$, $M_{N,4}$
respectively.
With these notations we have \cite{TW2} the following formulae
for $S_{N,\beta}$ in \eqref{1K}, \eqref{4K}
\be
\la{eq3pp1}
      S_{N,1}(x,y) = -\sum_{j,k=0}^{N-1}
             \psi_{j,1}(x)\,(\mu_{N,1})_{jk}\,(\eps\psi_{k,1})(y)
\ee
\be
\la{eq3pp2}
      S_{N,4}(x,y) = \sum_{j,k=0}^{2N-1}
             \psi_{j,4}^\prime(x)\,(\mu_{N,4})_{jk}\,\psi_{k,4}(y).
\ee
An essential feature of the above formulae is that the polynomials $\{q_j\}$
are arbitrary
and we are free to choose them conveniently to facilitate the 
asymptotic analysis of \eqref{1K}, \eqref{4K} as $N\to\infty$.
In view of the unitary case $\beta=2$ described above, specific issues
that must be addressed by the choice of the $q_j$'s, are the following:

(a) We need an analog of the Christoffel--Darboux formula
to convert $S_{N,\beta}$ into a form similar to \eqref{eq_KN},
which depends {\em only\ } on polynomials of high order as $N\to\infty$.

(b) By (a), the large $N$ behavior of \eqref{1K}, \eqref{4K} 
becomes {\em purely\ }a question of the asymptotics of polynomials
of high order. Thus we need to choose polynomials whose asymptotic behavior can 
be analyzed.

(c) Finally the inverses of the matrices $M_{N,1}$ and $M_{N,4}$
must be controlled as $N\to\infty$.

Property (c) is new in the cases $\beta=1$ and $4$, and does not arise
in the unitary case $\beta=2$. Following \cite{W}, we will choose the $q_j$'s
to be orthogonal polynomials with respect to an appropriate weight. It turns out
that this choice addresses all three issues (a), (b), (c) simultaneously.

Previously the study of the 
orthogonal and symplectic ensembles was often
carried out via 
so-called {\it skew orthogonal polynomials (SOP's)\ }(see
\cite{M} for classical references
and, e.g., \cite{NW,AFNvM,AvM,FNH} for more
recent work). 
Skew orthogonal polynomials are characterized by the property that
(in our situation) $M_{N,\beta}$ has the form of a direct sum of scalar multiples
of the matrix $\mtt{0}{1}{-1}{0}$.
For many classical weights $w_\beta$, 
e.g., $w_\beta(x)=e^{-x^2}$,
the corresponding SOP's can be computed explicitly \cite{AFNvM,F}. 
The proof of universality, however, involves arbitrary
weights $w_\beta$ for which no such explicit formulae 
for the SOP's are available.
As noted above, the crucial ingredient in the proof of universality for the 
$\beta=2$ ensemble is the asymptotic analysis of the OP's
corresponding to an arbitrary weight $w_2$.
For $\beta=1,4$, the problem of 
computing the asymptotics of SOP's
raises challenging technical difficulties which have not yet been 
fully overcome (however see the work \cite{E}).

In order to state our main result we need more notation.
For any $m\in\N$ let $V(x)$ be a polynomial of degree $2m$
\be
\la{eq3star}
    V(x) = \kappa_{2m}x^{2m}+\cdots,\qquad \kappa_{2m}>0
\ee
and let $w(x)\equiv{}w_{\beta=2}(x)=e^{-V(x)}$ as before.
Let $p_j(x)$, $j\geq0$,
denote the OP's with respect to $w$,
and set $\phi_j(x)\equiv{}p_j(x)(w(x))^{1/2}$, $j\geq0$, as above. 
For $\beta=1,4$ set
\be
\la{eq3pppp1}
  V_\beta(x)\equiv \frac12{V(x)}
\ee
and let $N$ be even.
Then by \eqref{eq1p3}, $w_4=e^{-2V_4}=e^{-V}$ and
$w_1=e^{-V_1}=e^{-V/2}$. 
This ensures that for the choice $q_j=p_j$ in \eqref{eq3p1}
\be\la{eq5p1}
 \psi_{j,\beta=1}(x)=\psi_{j,\beta=4}(x) = \phi_j(x),
\ee
which enables us in turn to handle $S_{N,1}$ and
$S_{N/2,4}$ in \eqref{eq3pp1}, \eqref{eq3pp2}
simultaneously (see Remark~\ref{remSTAR} below). 
Henceforth and throughout the paper, $K_N$ 
denotes the Christoffel--Darboux kernel \eqref{K}, \eqref{eq_KN}
constructed out of these functions $\phi_j$.

The main result we state here (Theorem~\ref{thmuniv} below)
concerns universality for the kernels $K_{N,\beta}$ 
in the bulk scaling limit. 
Universality for other standard statistical quantities
also holds, for example for the $l$-point
cluster functions and for the gap probability,
 see Corollaries~\ref{cor1.1} and \ref{cor1.2} below.

As in the case $\beta=2$ above, we scale
so that the expected number of eigenvalues per unit interval
is one. Note that formula \eqref{eqs3p1}
also holds for $\beta=1$ and $4$ and so $R_{N,l=1,\beta}(0)$
gives the density of the expected number of (simple) eigenvalues near zero.
For $\beta=1,4$, in view of \eqref{eqR1beta} and \eqref{1K}, \eqref{4K}
\be
\la{eq14scaling}
\ba
        R_{N,1,1}(r) &= S_{N,1}(r,r)\\
        R_{N/2,1,4}(r) &= \frac12\,S_{N/2,4}(r,r).
\ea
\ee
It follows from \eqref{eqsc1}, \eqref{eqsc4} below
that as (even) $N\to\infty$ for $r$ in a compact set
\be\la{eqscsec1}
\ba
        S_{N,1}(r,r)&= K_N(0,0)(1+O(N^{-1/2}))\\
        S_{N/2,4}(r,r)&= K_N(0,0)(1+O(N^{-1/2}))
\ea
\ee
and hence
\be
\la{eqSTARSTAR}
\ba
        R_{N,1,1}(r) &= K_N(0,0)(1+O(N^{-1/2}))\quad\textrm{and}\\
        R_{N/2,1,4}(r) &= \frac12K_N(0,0)(1+O(N^{-1/2}))
\ea
\ee
give the scaling in Theorem~\ref{thmuniv} below.
\begin{remark}
The asymptotics in \eqref{eqscsec1}, \eqref{eqSTARSTAR}
and in many related situations (see e.g.{} Remark~\ref{remm1} below)
remains true for $r=o(N^{1/(2m)})$,
but we do not need this fact in what follows. 
\end{remark}
It turns out that the off-diagonal elements in $K_{N,\beta}$
scale differently as $N\to\infty$.
On the other hand, the statistics of the ensembles are clearly invariant
under the conjugation
$$
    K_{N,\beta}\mapsto K_{N,\beta}^{(\lambda)}\equiv
       \mtt{\lambda^{-1}}{0}{0}{\lambda}\cdot K_{N,\beta} \cdot
      \mtt{\lambda}{0}{0}{\lambda^{-1}}
     =\mtt{(K_{N,\beta})_{11}}{\lambda^{-2}(K_{N,\beta})_{12}}
               {\lambda^{2}(K_{N,\beta})_{21}}{(K_{N,\beta})_{22}}
$$
for any scalar $\lambda=\lambda_{N,\beta}$.
For example, this is obviously true for the cluster functions $T_{N,l,\beta}$
which have the form
\be\la{eqclust}
    T_{N,l,\beta}(y_1,\cdots,y_l)=\frac1{2l}\sum_{\sigma}
       \tr \Big( K_{N,\beta}(y_{\sigma_1},y_{\sigma_2})
                 K_{N,\beta}(y_{\sigma_2},y_{\sigma_3})\cdots 
            K_{N,\beta}(y_{\sigma_l},y_{\sigma_1})\Big)
\ee
where the sum is taken over all permutations of $\{1,\cdots,l\}$ (see
\cite[p.~816]{TW2}), etc.

Set 
$$
      \lambda_{N,1}=(R_{N,1,1}(r))^{1/2},\qquad
          \lambda_{N,4}=(R_{N/2,1,4}(r))^{1/2}.
$$
\begin{theorem}
\label{thmuniv}
Let $\beta=1$ or $4$.
For any $V(x)$ of degree $2m$ as in \eqref{eq3star}
define $V_\beta(x)$ and $w_\beta(x)$ as in \eqref{eq3pppp1}, \eqref{eq1p3}.
Then for $r$, $\xi$, $\eta$ in a compact set, as (even) $N\to\infty$
\be
\la{equniv1}
\ba
  \mathcal{E}_{N,1}&\equiv
     \frac{1}{\lambda_{N,1}^2} 
   K_{N,1}^{(\lambda_{N,1})}
   \bigg(r+\frac{\xi}{\lambda_{N,1}^2},
      r+  \frac{\eta}{\lambda_{N,1}^2}\bigg)
- K^{(1)}(\xi,\eta)\to0\\
  \mathcal{E}_{N,4}&\equiv
     \frac{1}{\lambda_{N,4}^2} 
   K_{N/2,4}^{(\lambda_{N,4})}
   \bigg(r+\frac{\xi}{\lambda_{N,4}^2},
      r+  \frac{\eta}{\lambda_{N,4}^2}\bigg)
  - K^{(4)}(\xi,\eta)\to0
\ea
\eq
where
\be
\la{M.19.2}
\ba
   K^{(1)}(\xi,\eta)&\equiv
     \twotwo{K_{\infty}(\xi-\eta)}
   {\frac{\partial}{\partial\xi}\big(K_\infty(\xi-\eta)\big)}
   {\int_0^{\xi-\eta} K_\infty(t)\,dt - \frac12\sgn(\xi-\eta)}
       {K_{\infty}(\eta-\xi)}
\\
     K^{(4)}(\xi,\eta)&\equiv
   \twotwo{K_{\infty}\big(2(\xi-\eta)\big)}
   {\frac{\partial}{\partial\xi}\big(K_\infty(2(\xi-\eta))\big)}
   {\int_0^{\xi-\eta} K_\infty(2t)\,dt}
       {K_{\infty}\big(2(\eta-\xi)\big)}. 
\ea
\ee
For $\beta=1,4$ and $r$, $\xi,\eta$ in a compact set, 
we have as $N\to\infty$
\be\la{eqs100}
       \mathcal{E}_{N,\beta} 
       =\twotwo{O\big(N^{-1/2}\big)}
             {O\big(N^{-{1/2}}\big)}
             {O(N^{-1/4})}
            {O\big(N^{-1/2}\big)}.
\ee
\end{theorem}
\begin{remark}
\la{remm1}
In the above theorem, we do not need to restrict $r$
to a compact set. The theorem remains true for $r=o(N^{1/(2m)})$.
\end{remark}
\begin{remark}
\la{remSTAR}
The choice in \eqref{eq3pppp1} 
implies that $V_1(x)=V_4(x)$.
Other authors make different choices.
For example, in the Gaussian case \cite{M},
Mehta considers $V_4(x)=2V_1(x)$.
Our choice insures that we can use the same 
auxiliary OP's in \eqref{eq5p1} for $\beta=1$ and $\beta=4$.
On the other hand, Mehta's choice ensures that the density of 
states for $P_{N,\beta}(x_1,\cdots,x_N)$ 
in \eqref{eq_PN} is the same 
for $\beta=1$ and $\beta=4$.
Nevertheless we obtain precisely the 
same answer \eqref{equniv1}
%\eqref{equniv1}, \eqref{equniv4} 
in Theorem~\ref{thmuniv}
as computed in \cite{M} for $\beta=1$ and $\beta=4$ in the Gaussian case.
Of course, this is due to the fact that in the end
we use the same scaling as in \cite{M}, viz., the expected number
of eigenvalues per unit interval is one.
\end{remark}

From the formula for the cluster functions
\eqref{eqclust} we immediately have the following result. % (no proof needed).
\begin{corollary}\la{cor1.1}
Let $\beta=1$ or $4$. Let $V$ be a polynomial of degree $2m$
and let $K^{(\beta)}$, $\beta=1,4$
 be as in Theorem~\ref{thmuniv}.
Then for $\beta=1$ and $l=2,3,\cdots$
we have, uniformly for $r$ and $\xi_1,\cdots,\xi_l$
in a compact set
\be\la{M.19.1}
\ba
   \lim_{N\to\infty} 
       \frac1{(\lambda_{N,1}^2)^l}\,
          &T_{N,l,1}\bigg(r+\frac{\xi_1}{\lambda_{N,1}^2},
                                                           \cdots, 
                  r+\frac{\xi_l}{\lambda_{N,1}^2}\bigg)\\
    &=\frac1{2l}\sum_{\sigma}
       \tr \Big( K^{(1)}(\xi_{\sigma_1},\xi_{\sigma_2})
                 K^{(1)}(\xi_{\sigma_2},\xi_{\sigma_3})\cdots 
            K^{(1)}(\xi_{\sigma_l},\xi_{\sigma_1})\Big).
\ea
\ee
For $\beta=4$, the same result is true provided we replace
$T_{N,l,1}\to{}T_{N/2,l,4}$, $\lambda_{N,1}\to\lambda_{N,4}$, and
$K^{(1)}\to{}K^{(4)}$.
\end{corollary}

Together with some additional estimates 
(see Section \ref{sectthree}),
Theorem~\ref{thmuniv} also yields the following universality
result for the gap probability.
Recall that for a $2\times2$ block operator 
$A=(A_{ij})_{i,j=1,2}$
with $A_{11},A_{22}$ in trace class and $A_{12},A_{21}$
Hilbert--Schmidt, the regularized $2$-determinant
(see e.g.{} \cite{Simon})
is defined by ${\det}_2(I+A)\equiv\det((I+A)e^{-A})\,e^{\tr(A_{11}+A_{22})}$.
\begin{corollary}\la{cor1.2}
Let $\beta=1$ or $4$. Let $V$ be a polynomial of degree $2m$
and let $K^{(\beta)}$, $\beta=1,4$
 be as in Theorem~\ref{thmuniv}.
Fix $\theta>0$ and $r$.
Then
\be\la{M.19.4}
\ba
   \lim_{N\to\infty} \Prob\bigg\{\textrm{no eigenvalues in }
    \bigg(
       r-\frac{\theta}{\lambda_{N,\beta}^2},
           r+\frac{\theta}{\lambda_{N,\beta}^2}
        \bigg) \bigg\}
   =\sqrt{{\det}^{(\beta)}(I-K_\theta^{(\beta)})}
\ea
\ee
where $K_\theta^{(\beta)}$ denotes the operator with kernel
$K^{(\beta)}(\xi,\eta)$ acting in $L^2(-\theta,\theta)$.
Here ${\det}^{(4)}(I-K_\theta^{(4)})$
 is the regular determinant of the trace class operator 
$K_\theta^{(4)}$, but  ${\det}^{(1)}(I-K_\theta^{(1)})$
 is the regularized $2$-determinant  ${\det}_2(I-K_\theta^{(1)})$ 
defined above.
\end{corollary}
\begin{remark}
The regularized $2$-determinant is needed for $\beta=1$
because the operator with kernel $\frac12\sgn(\xi-\eta)$
is Hilbert--Schmidt but not trace class on $L^2(-\theta,\theta)$
(see proof of Corollary~\ref{cor1.2} in Section~\ref{sectthree}).
\end{remark}

We complete this introduction with a description of Widom's result \cite{W}
which is basic for our approach in this paper.
Widom's method applies to general weights $w_\beta$
with the property that $w_\beta^\prime/w_\beta$ is a rational function.
This property certainly holds for our weights as 
in \eqref{eq1p3}, \eqref{eq1p2}.
Introduce  the semi-infinite matrices
\bq\label{DE}
  D_\infty\equiv{}((D\phi_j,\phi_k))_{j,k\geq0},\qquad
  \eps_\infty\equiv{}((\eps\phi_j,\phi_k))_{j,k\geq0}.
\eq 
It follows from \cite[Section~6]{TW1} that the matrix $D_\infty$
is banded with bandwidth $2n+1$ where 
\be
\la{eqn}
            n\equiv{}2m-1.
\ee
Thus $(D_\infty)_{jk}=0$ if $|j-k|>n$.
Next, let $N$ be greater than $n$, and 
introduce the following  $N$-dependent $n$-column vectors
\bq\label{phi}
\begin{aligned}
   \Phi_1(x)&\equiv{}(\phi_{N-n}(x),\cdots,\phi_{N-1}(x))^T\\
   \Phi_2(x)&\equiv{}(\phi_{N}(x),\cdots,\phi_{N+n-1}(x))^T\\
   \eps\Phi_1(x)&\equiv{}(\eps\phi_{N-n}(x),\cdots,\eps\phi_{N-1}(x))^T\\
   \eps\Phi_2(x)&\equiv{}(\eps\phi_{N}(x),\cdots,\eps\phi_{N+n-1}(x))^T
\ea
\eq
and the following $2n\times2n$ matrices 
consisting of four $n\times n$ blocks
\bq\label{B}
  B\equiv{}\twotwo{B_{11}}{B_{12}}
                      {B_{21}}{B_{22}} 
   =((\eps\phi_j,\phi_k))_{N-n\leq j,k\leq N+n-1}.
\eq 
and
\bq\label{A}
  A\equiv{}\twotwo{0}{A_{12}}
                      {A_{21}}{0} 
   =\twotwo{0}{D_{12}}
                      {-D_{21}}{0}
\eq 
where
$
  \twotwo{D_{11}}{D_{12}}
                      {D_{21}}{D_{22}} 
   \equiv{}((D\phi_j,\phi_k))_{N-n\leq j,k\leq N+n-1}.
$
Finally, set 
$$ 
   C=\twotwo{C_{11}}{C_{12}}
                      {C_{21}}{C_{22}} 
     \equiv{}\twotwo{I_n + (BA)_{11}}{(BA)_{12}}
                      {(BA)_{21}}{(BA)_{22}}.
$$
Note that 
$$ 
     C_{11}=I_n+B_{12}A_{21}=I_n-B_{12}D_{21}.
$$
The main result in \cite{W} is the following pair
of formulae for $S_{N,1}$ and $S_{N/2,4}$
\bq
\label{W1}
\ba
   S_{N,1}(x,y) = K_N(x,y)
     - (\Phi_1(x)^T,0^T)&\cdot 
        (AC(I_{2n}-BAC)^{-1})^T \\&\cdot(\eps\Phi_1(y)^T,\eps\Phi_2(y)^T)^T
\ea
\eq
and
\bq
\label{W4}
\ba
   S_{N/2,4}(x,y) = K_N(x,y) &+ \Phi_2(x)^T\cdot D_{21}\cdot\eps\Phi_1(y)\\
          &+\Phi_2(x)^T\cdot D_{21}C_{11}^{-1}B_{11}D_{12}\cdot\eps\Phi_2(y).
\ea
\eq
Observe that $S_{N,1}$ and $S_{N/2,4}$ are sums of the $\beta=2$
kernel $K_N(x,y)$ together with correction terms that depend
only on $\phi_{N+j}$ for $j\in\{-n,\cdots,n-1\}$.
Together with the Christoffel--Darboux formula \eqref{eq_KN}
we see immediately that the representations \eqref{W1}, \eqref{W4}
address issue (a) above. Moreover, the $\phi_{N+j}$'s are proportional
to orthonormal polynomials $p_{N+j}$, and hence can be evaluated
asymptotically as $N\to\infty$ by the methods in \cite{DKMVZ2}.
This addresses issue (b).
Finally we see that issue (c), the question of the control
of the inverses of the (large) matrices $M_{N,1}$, $M_{N,4}$,
reduces, via \eqref{W1}, \eqref{W4} to a question
of controlling the inverses of matrices
of {\em fixed\ }size $2n\times{}2n$ and $n\times{}n$, respectively.
It turns out that these fixed size matrices converge as $N\ra\infty$.
The proof that in both cases the limiting matrices are invertible
for any $V$ of type \eqref{eq3star}, 
is not obvious, and constitutes a significant part of the proof
of Theorem~\ref{thmuniv} in this paper.

In Section \ref{secttwo}, we state a variety of auxiliary results which we use
in the proof of Theorem~\ref{thmuniv}.
The results, some of which are of independent interest,
are of two types: 

(i) asymptotics of $(D\phi_{N+j},\phi_{N+k})$
and $(\eps\phi_{N+j},\phi_{N+k})$ for fixed $j,k$ as $N\to\infty$;

(ii) the equality of four specific determinants that arise in the analysis:
This reduces the proof of the invertibility of the limiting matrices mentioned
above to the proof of the invertibility of a single matrix $T_{m-1}$
(see \eqref{Tm1} below).

In Section \ref{sectthree}, we prove Theorem \ref{thmuniv}
(and hence Corollary \ref{cor1.1})
and Corollary \ref{cor1.2}
assuming the validity of the results in Section \ref{secttwo} (and \ref{sectfour}).
In Section \ref{sectfour}, we prove the results of type (i) above.
%(The proof of the $\eps$ case is not obvious at all and constitutes about $2/5$ of this paper.)
%
In Section \ref{sectfive}, we prove the equality of the determinants
in (ii).
Finally, in Section \ref{sectsix} we prove that for 
any $V(x)=\kappa_{2m}x^{2m}+\cdots$, the determinant of $T_{m-1}$
is nonzero.
In the course of the proof in Section \ref{sectsix},
we need to estimate certain explicit integrals:
A rigorous analysis of the error bounds in these estimates 
is given in an Appendix.
\begin{remark}
After the second version of this paper was 
posted on the arXiv, A.~Boutet de Monvel
informed us of the work of A.~Stojanovic (\cite{St1,St2,St3})
in which universality (in the bulk and also at the edge)
is proved for $\beta=1$ and $\beta=4$ in the even, quartic (two-interval) case
considered previously by Bleher and Its \cite{BI}
for $\beta=2$.
The method uses a variant of the formulae in \cite{W}
together with the asymptotics for OP's in \cite{BI}.

In \cite[Remark~2.4]{St2}, an interesting connection
is noted between the problem of controlling 
$\lim_{N\to\infty}\det{C_{11}}$ (in our notation)
and the problem of estimating partition functions $Z_{N,\beta}$
(see \eqref{eq_PN}) as $N\to\infty$.
Indeed from the formula $C_{11}=I_n-B_{12}D_{21}$ above
and \eqref{W} below, we see that $\det{C_{11}}=\det{\eps_N}\det{D_N}$,
where ${\eps_N},D_N$ are the leading $N\times{}N$ sections
of the matrices ${\eps_\infty},{D_\infty}$, respectively.
But $\det{\eps_N}$ and $\det{D_N}$ can be expressed in terms 
of partition functions (see \cite{AvM})
and this leads directly to a formula for $\det{C_{11}}$ as a ratio
of $Z_{N,\beta}$'s, $\beta=1,2,4$.
Thus one may try to use the methods of statistical mechanics
to control $\lim_{N\to\infty}\det{C_{11}}$.
The estimates in \cite{J}
show that the partition functions $Z_{N,\beta}$
have leading order
asymptotics of the form $e^{a_\beta{}N^2(1+o(1))}$
as $N\to\infty$, and moreover, their combined contributions
to $\det{C_{11}}$ cancel to this order.
Thus, in order to show that $\lim_{N\to\infty}\det{C_{11}}\neq0$,
one needs higher order asymptotics for the $Z_{N,\beta}$'s.
In one of our early attempts to prove universality,
for $\beta=1$ and $4$ we indeed tried to derive
such higher order asymptotics, but we were not successful.
The statistical mechanical 
approach to prove $\lim_{N\to\infty}\det{C_{11}}\neq0$,
remains an unresolved, and intriguing, possibility.
\end{remark}
\textbf{Notational remark:} Throughout this paper $c,C,C(m),c_1,c_2,\cdots$
refer to constants independent of $N,P$. 
More specifically, the symbols $c,C$ refer to generic constants,
whose precise value may change from one inequality
to another.
The symbol $c_N$ however
always refers to the $N$-dependent constant \eqref{cN} below.

{\bf Acknowledgments.} 
The authors would like to thank Harold Widom for alerting
one of us (P.D.) to his paper \cite{W},
and for suggesting that formulae in this paper,
together with the asymptotics for OP's in \cite{DKMVZ2},
might be used to prove universality for $\beta=1$ and $4$.
We would also like to thank
Alexei Borodin for suggesting a short proof 
of Lemma \ref{lembor}.
The work was supported in part by
an NSF grant DMS--0296084.
In 2002--03 the second author was supported in full 
by a postdoctoral scholarship from the Swedish Foundation for International
Cooperation in Research and Higher Education (STINT), Dnr.~PD2001--128.
The second author would like to thank the University of Pennsylvania
where he had a postdoctoral appointment in 2001--02 and 2003--04.
The second author would also like to thank
the Courant Institute,
New York University, where he has spent the year 2002--03 
and parts of 2003--04, for hospitality and financial support.
\section{Auxiliary results}
\la{secttwo}
In order to evaluate the asymptotic behavior
of the correction terms in \eqref{W1} and \eqref{W4},
one first needs to determine the asymptotics of 
$(D\phi_{N+j},\phi_{N+k})$ and $(\eps\phi_{N+j},\phi_{N+k})$
for fixed $j,k\in\{-n,\cdots,n-1\}$, as $N\ra\infty$. 
Such asymptotics are given by the next two theorems
which are proved in Section~\ref{sectfour} below.
Let $b_j$ be the coefficients in the three term 
recurrence relation \cite{Sz} satisfied by the OP's $p_j$
corresponding to the weight $w(x)=e^{-V(x)}$ (and hence also by the
functions $\phi_j=p_j w^{1/2}$)
\begin{equation}
\label{eqthreeterm}
   x p_j(x) = b_{j-1} p_{j-1}(x) + a_j p_j(x) + b_{j} p_{j+1}(x),
       \qquad j\geq0,
\end{equation}
($b_{-1}\equiv0$).
\begin{theorem}
\label{thm1}
Let $V(x)=\kappa_{2m}x^{2m}+\cdots$, $\kappa_{2m}>0$, 
and set $n\equiv{}2m-1$ as before. 
Then the matrix $D_\infty$ has $2n+1$ bands and
is asymptotic to the product of a diagonal matrix and a Toeplitz matrix. 
More precisely, for any fixed $j,k\in\Z$, as $N\to\infty$
\be
\la{Das}
\ba
     \null&(D\phi_{N+j},\phi_{N+k})\\
      &= m \kappa_{2m} b_{N+j}^n
         \cdot\begin{cases}
                      0, &j-k=0\hbox{\rm{ or }}|j-k|\geq n+1\cr
                       \sgn(j-k)\binom{n}{(n-|j-k|)/2}+o(1),&|j-k|=1,3,\cdots,n\cr
                     o(1), &|j-k| =2,4,\cdots,n-1.
          \end{cases}
\ea
\ee
%HERE $o(1)=O(N^{-1/(2m)})$.
\end{theorem}
To formulate the second theorem we need information
on the equilibrium measure $d\mu_N^{(\textrm{eq})}(x)$ 
(see e.g.~\cite{SaTo}) for OP's 
corresponding to the rescaled weight $e^{-NV_N(x)}$, 
$V_N=\frac1NV(c_Nx+d_N)$,
where $c_N$, $d_N$ are the so-called Mhaskar--Rakhmanov--Saff
(MRS) numbers (see \cite{MhSa,Ra}).
For $V(x)=\kappa_{2m}x^{2m}+\kappa_{2m-1}x^{2m-1}+\cdots$
as in \eqref{eq3star}, we have \cite{DKMVZ2}
to any order $q$ as $N\to\infty$
\be\la{cN}
   c_N = \bigg(\frac1{\kappa_{2m}}
       \frac{(2m)!!}{m(2m-1)!!}\bigg)^{1/(2m)}\,N^{1/(2m)}
             + \sum_{j=0}^q c_{(j)} N^{-j/(2m)} + O(N^{-(q+1)/(2m)})
\ee
and
\be\la{dN}
   d_N = -\frac{\kappa_{2m-1}}{2m\kappa_{2m}}
             + \sum_{j=1}^q d_{(j)} N^{-j/(2m)} + O(N^{-(q+1)/(2m)}).
\ee
As $N\ra\infty$, the equilibribum measure is absolutely continuous
with respect to Lebesgue measure, $d\mu_N^{(\textrm{eq})}(x)
=\psi_N^{(\textrm{eq})}(x)\,dx$, and is supported on the (single)
interval $[-1,1]$,
\be
\la{psiN}
      \psi_N^{(\textrm{eq})}(x)\equiv\psi_N(x) 
                     =\frac1{2\pi}|1-x^2|^{1/2}\chi_{[-1,1]}(x)
          \,h_N(x)
\ee
(see \cite{DKMVZ2}) where $h_N(x)$ is a real polynomial of degree $2m-2$
\be\la{eq25prime}
   h_N(x) = \sum_{k=0}^{2m-2}h_{N,k}x^k
\ee
and the coefficients $h_{N,k}$ can be expanded to any order in
powers of $N^{-1/(2m)}$ as above.
In particular, to any order $q=1,2,\cdots$,
as $N\ra\infty$, we find uniformly for $x$ in
compact sets 
\be\la{eq9pSTAR}
   h_N(x) = h(x) + \sum_{j=1}^{q} N^{-j/(2m)}\,h_{(j)}(x) 
         + O(N^{-(q+1)/(2m)})
\ee
where (see \cite[(2.7)]{DKMVZ2})
\be
\la{hz}
    h(x)= \sum_{k=0}^{m-1} \beta_k x^{2k},\qquad 
   \beta_k \equiv 2\frac{(2m)(2(m-1))\cdots(2(m-k))}
                             {(2m-1)(2(m-1)-1)\cdots(2(m-k)-1)}.
\ee
Also by \cite[p.~1501, Remark 3]{DKMVZ2}
\be\la{hvalues}
         h(1) = 4m,\qquad h(0)=\frac{4m}{2m-1}.
\ee
Note that $h(x)$ depends only on the degree $2m$
and is independent of the coefficients of $V$.
Note also that $h(x)\geq h(0)>0$ for all $x\in\R$.
\begin{theorem}
\label{thm2}
Let $V(x)=\kappa_{2m}x^{2m}+\cdots$, $\kappa_{2m}>0$
as before. 
Then, for any fixed $j,k\in\Z$, as $N\to\infty$
\be
\la{Eas}
\ba  \null&(\eps\phi_{N+j},\phi_{N+k})\\
      &= c_{N+j} (N+j)^{-1}
         \cdot\begin{cases}
                    \frac{(-1)^{N+j}}{2m} - I(j-k)  + o(1),
                                   &j-k\hbox{\rm{ odd}}\cr
                           o(1), &j-k\hbox{\rm{ even}},j-k\neq0\cr
                           0, &j-k=0
          \end{cases}
\ea
\ee
where for $q=\pm1,\pm3,\cdots$ we define
\be
\la{I}
   I(q)\equiv{} \frac2\pi\sin\frac{q\pi}2\int_{-1}^1
         \frac{\cos(q\arcsin x)}{h(x)}
         \frac{dx}{1-x^2}.
\ee
\end{theorem}
In the proof of \eqref{Eas} we compute a double integral
over all the different asymptotic regions in $\R$
for the OP's $\{p_j\}$ 
obtained in \cite{DKMVZ2} (see Subsection~\ref{ssec42} et seq.{} below).

We see from the above, that in a fixed size
neighborhood of the diagonal the matrices $D_{\infty}$ and $\eps_\infty$ 
do not depend asymptotically on lower order terms in $V(x)$.
Next note that the coefficient in \eqref{eqthreeterm} 
satisfies \cite[(2.11)]{DKMVZ2}
\be\la{bN} 
        b_N=\bigg(\frac12+o(1)\bigg)\,c_N,
       \qquad a_N=d_N + o(1) = O(1),\qquad N\ra\infty.
\ee
Combining our two theorems 
we can find an explicit expression for $C_{11}$,
and also for all other matrices involved in the $\beta=1$ and $4$ 
correction terms in \eqref{W1}, \eqref{W4}. In particular,
 as (even) $N\ra\infty$
\be
\la{BAformula}
\ba  
   B_{12} &= %\frac{2(m!)^2}{(2m)!}
 \frac{c_N}{N}\,
\left[\frac1{2m}
\begin{pmatrix}
&-1 &0&-1&\cdots&-1 \\
&  0 &1&0&\cdots&0\\ 
&\cdots\\ 
&-1 &0&-1&\cdots&-1
\end{pmatrix} 
\right.\\
&\qquad\qquad+ 
\left.\begin{pmatrix}
&I(n) &0&I(n+2)&\cdots&I(2n-1) \\
&  0 &I(n)&0&\cdots&0\\ 
&\cdots\\ 
&I(1) &0&I(3)&\cdots&I(n) 
\end{pmatrix} + o(1)\right]\\
  A_{21} &= -D_{21} =
-\frac{m\kappa_{2m}}{2^{2m-1}}\,c_N^{2m-1}\,
\left[\begin{pmatrix}
&\binom{n}{0} &0&\binom{n}{1}&\cdots&\binom{n}{(n-1)/2} \\
&  0 &1&0&\cdots&0\\ 
&\cdots\\ 
&0 &0&0&\cdots&1 
\end{pmatrix} +o(1)\right].
\ea
\ee
The second and the third matrices are Toeplitz.

By \eqref{cN} the asymptotic orders of $B_{12}$, $D_{21}$
are exactly opposite. The same is true for all 
the blocks $B_{ij}$ and $D_{ij}$, $i,j=1,2$,
and so whenever we have a product of elements
from $B$ and $A$ in \eqref{W1}, \eqref{W4},
the product is asymptotically constant. 
Also
\be\la{s5eq80}
\ba
   \lim_{N\to\infty} \frac{c_N}{N}
\frac{m\kappa_{2m}}{2^{2m-1}}\,c_N^{2m-1}
&=  \frac{(2m)!!}{2^{2m-1}(2m-1)!!}\\
      &= \frac{2^m m!}
     {2^{2m-1}1\cdot3\cdot\ldots\cdot(2m-1)}
          \,\frac{1\cdot2\cdot\ldots\cdot m}
         {1\cdot2\cdot\ldots\cdot m}
    = \frac{2(m!)^2}{(2m)!}
\ea
\ee
so that in $BA$ the coefficient $\kappa_{2m}$
cancels out, and as $N\to\infty$, $BA$ converges to a
constant matrix that depends only on $m$.

Recall that in each of the cases $\beta=1$ and $4$,
in order to compute the correction terms in \eqref{W1}, \eqref{W4}
we need to invert a certain matrix. 
Now we know that the $\beta=4$
matrix to be inverted, $I_n+B_{12}A_{21}$, converges as $N\to\infty$.
In the $\beta=1$ case the matrix to be inverted is $I_{2n}+BAC$,
which also converges as $N\to\infty$.
It is not a priori clear whether
the two limiting matrices are related. 
However our structure Theorems
\ref{thm1} and \ref{thm2}, and also Theorem~\ref{thm6} below,
imply the following.
Let $R\equiv{}R_n$ denote the $n\times n$
matrix with all zero entries apart from
ones on the anti-diagonal (thus $R_{i,j}=1$
if $j=n-i+1$, $1\leq i\leq n$, and $R_{i,j}=0$ otherwise).
\begin{theorem}
\label{thm4}
As $N\to\infty$, $N$ even, 
we have $(BA)_{22}=-R(BA)_{11}R+o(1)$.
Also, quite remarkably,
\bq
\la{BAC}
 BAC=\twotwo{0}{0}{(BA)_{21}+o(1)}{(BA)_{22}+o(1)},
\eq
so that in particular $BAC$ and $BA$ asymptotically
have the same two lower blocks.
\end{theorem}
It is interesting that the $11$ and the $21$ entries of $BAC$
are identically zero, not just $o(1)$.

Now we see that inverting $I_{2n}+BAC$ boils down
to inverting 
$$
     I_n-(-R(BA)_{11}R)+o(1) = R(I_n+B_{12}A_{21})R + o(1)
$$
so that the determinants of the limiting matrices
in the $\beta=1$ and $4$ cases that are to be inverted
are the same. Thus
 in both cases we have to check that
$\det(I_n+B_{12}A_{21})=\const+o(1)$, $\const\neq0$.

Next, we observe that the zero pattern in
 $B_{12}$ and $A_{21}$ in \eqref{BAformula} implies
that
$$ 
        \det(I+B_{12}A_{21}) 
                 = (\det T_m^{\prime}+o(1))\cdot 
                      (\det T_{m-1}+o(1))
$$
as $N\ra\infty$, $N$ even, where 
\bq
\la{Tm}
\ba
T_m^{\prime}\equiv I_{m}-&\frac{2(m!)^2}{(2m)!}\left[\frac1{2m}
\begin{pmatrix}
& -1 &-1&\cdots&-1 \\
& -1 &-1&\cdots&-1\\ 
&\cdots\\ 
&-1&-1&\cdots&-1
\end{pmatrix} 
\right.\\
&\qquad\qquad+ 
\left.\begin{pmatrix}
&I(n) &I(n+2)&\cdots&I(2n-1) \\
&I(n-2) &I(n)&\cdots&I(2n-3)\\ 
&\cdots\\ 
&I(1) &I(3)&\cdots&I(n) 
\end{pmatrix} \right]\\
  &\times
\begin{pmatrix}
&\binom{n}{0} &\binom{n}{1}&\cdots&\binom{n}{(n-1)/2} \\
& 0 &\binom{n}{0}&\cdots&\binom{n}{(n-3)/2} &\\ 
&\cdots\\ 
&0 &0&\cdots&1 
\end{pmatrix}
\ea
\ee
and
\bq
\la{Tm1}
\ba
T_{m-1}\equiv I_{m-1}-\frac{2(m!)^2}{(2m)!}&\left[\frac1{2m}
\begin{pmatrix}
& 1 &1&\cdots&1 \\
& 1 &1&\cdots&1\\ 
&\cdots\\ 
&1&1&\cdots&1
\end{pmatrix} 
\right.\\
&\qquad\qquad+ 
\left.\begin{pmatrix}
&I(n) &I(n+2)&\cdots&I(2n-3) \\
&I(n-2) &I(n)&\cdots&I(2n-5)\\ 
&\cdots\\ 
&I(3) &I(5)&\cdots&I(n) 
\end{pmatrix} \right]\\
  &\times
\begin{pmatrix}
&\binom{n}{0} &\binom{n}{1}&\cdots&\binom{n}{(n-3)/2} \\
& 0 &\binom{n}{0}&\cdots&\binom{n}{(n-5)/2} &\\ 
&\cdots\\ 
&0 &0&\cdots&1 
\end{pmatrix}.
\ea
\ee
Quite remarkably, we also have
\begin{theorem}
\la{thm5}
For any $m\in\N$, $m\geq2$,
$$
         \det T_m^{\prime} = \det T_{m-1}.
$$
\end{theorem}
Thus 
it suffices to show that the {\em single\ }determinant, $\det T_{m-1}$,
is nonzero (see Theorem~\ref{thm7} below).
The proofs of Theorem~\ref{thm4} and \ref{thm5}
rely on the following identities.
Let $D_N$, $\eps_N$ denote the $N\times{}N$
matrices formed by the first $N$ rows and columns
of $D_\infty$, $\eps_\infty$, respectively.
Recall that by \eqref{eqn} $D_\infty$ has bandwidth $2n+1$.
Let $I_\infty$ denote the semi-infinite identity matrix.
\begin{theorem}
\la{thm6}
(i) The semi-infinite matrices satisfy 
$$ 
      D_\infty\eps_\infty = \eps_\infty D_\infty = I_\infty
$$
(the products are well-defined since $D_\infty$ is banded);

(ii) the sections for $N>n$ satisfy
\be
\la{W}
       \eps_N D_N = \twotwo{I_{N-n}}{-\eps_{N-n,n}^*D_{21}}
             {0}{I_n-B_{12}D_{21}}
\ee
where $\eps_{N-n,n}^*$ is the $(N-n)\times n$ matrix formed
by the rows $0,\cdots,N-n-1$ and columns $N,\cdots,N+n-1$
of the matrix $\eps_\infty$.
\end{theorem}
\begin{remark}
Assuming $N$ is even and denoting $W_N\equiv{}\eps_N D_N$
we see from \eqref{W} that 
\be
\la{W2}
      \eps_N^{-1}=D_N W_N^{-1},
      \qquad D_N^{-1}=W_N^{-1} \eps_N.
\ee
Thus the question of the invertibility of both $\eps_N$
and $D_N$, reduces to checking that 
$$
    \det(I_n-B_{12}D_{21})=\det(I+B_{12}A_{21})\neq0.
$$
On the other hand, one can show
 that the corrections in \eqref{W1}, \eqref{W4} can
be rewritten in terms of the lower-right $n\times n$
corner of the matrices $\eps_N^{-1}$ and $D_N^{-1}$,
 and so \eqref{W} and \eqref{W2}
explain why the same determinant should be checked
to be nonzero for both the $\beta=1$ and $4$ cases.
Also using the fact that 
$$
       W_N^{-1}= \twotwo{I_{N-n}}{*}
             {0}{(I_n-B_{12}D_{21})^{-1}}
$$
along with \eqref{W2}, the skew symmetricity of $\eps_N$,
and \eqref{W}, we can show that $\eps_{N}^{-1}$
coincides with $D_N$ everywhere apart from
the above mentioned lower-right $n\times n$ corner.
\end{remark}
Next we state our 
main technical result that ensures the invertibility
of the matrices in \eqref{W1}, \eqref{W4} in both
the $\beta=1$ and $4$ cases.
\begin{theorem}
\la{thm7}
For any $m\in\N$, $m\geq2$,
$$
                  \det T_{m-1}\neq0.
$$ 
\end{theorem}
\begin{remark}
From numerical computations for $m=2,\cdots,14$,
it seems that $\det{}T_{m-1}$ approaches the value $\frac1{\sqrt{2}}$
as $m$ grows,
but we have not been able to use our methods to establish
such a result.
\end{remark}
\begin{remark}
\la{rem2star}
The basic idea in proving Theorem~\ref{thm7}
is to show that for an appropriate norm $\|\cdot\|$,
$\|T_{m-1}-I_{m-1}\|<1$. 
Rather than analyzing the quantities $I(q)$ in \eqref{I}, \eqref{Tm1}
directly, we consider 
\be
\la{Itilde1}
   \tilde{I}(q)\equiv{}m I(q)-\frac12 =
  \frac2\pi\sin\frac{q\pi}2\int_{-\pi/2}^{\pi/2}
         \cos(q\theta)\,y_m(\theta)\,d\theta,\qquad q=3,5,\cdots 4m-5
\ee
where
\be
\la{ym}
   y_m(\theta)\equiv{}\frac{m}{\cos\theta}
         \bigg(\frac{1}{h(\sin\theta)} 
        - \frac1{4m} 
         - \frac{\cos^2\theta}{2}\bigg)
\ee
(cf.{} \eqref{Itilde} et seq.{} below).
It turns out that $h(x)$ is (essentially) a $\space_2F_1$
hypergeometric function (see \eqref{hzz})
and satisfies a {\em first\ }order differential equation (see \eqref{eqhz}).
This in turn implies that $y_m$ satisfies a Riccati
equation 
\be\la{eqR}
    y_m^\prime = \frac4{\sin\theta}\bigg(y_m+\frac{2m+1}{4}\cos\theta\bigg)
            \bigg(y_m+\frac{1}{2\cos\theta}\bigg),
\ee
where we see that the right-hand side conveniently
splits into a product with explicit factors.
Equation \eqref{eqR} plays a key role in proving that the magnitude of
$y_m$ is $O(\sqrt{m})$, a crucial fact in the proof of Theorem \ref{thm7}.
In estimating the magnitude of the quantities $\tilde{I}(q)$,
it is clear from \eqref{Itilde1} that we are in a double-scaling situation:
if $q$ is large compared to $m$, then $\tilde{I}(q)$ is small
by virtue of the usual decay of Fourier coefficients of real
analytic functions.
On the other hand if $q$ is small with respect to $m$,
the asymptotic behavior of $y_m$ dominates.
Both regimes must be analyzed separately.
These two effects are comparable when $q\sim\sqrt{m}$.
\end{remark}
\section{Proofs of Theorem \ref{thmuniv} and Corollary \ref{cor1.2}}
\la{sectthree}
We only consider the case $\beta=1$. The proofs 
%of the statements in Theorem \ref{thmuniv} 
for $\beta=4$ are similar and are left to 
the reader.

We need
the following result which is based
on the Plancherel--Rotach type asymptotics for OP's
in \cite{DKMVZ2} (see Section \ref{sectfour} for the proof).
\begin{theorem}
\label{thmthree}
Let $V(x)=\kappa_{2m}x^{2m}+\cdots$ as before.
Let $a=o(c_N)$ as $N\to\infty$.
Then,
%for any $\delta>0$ the following holds for $|x|\leq\delta$, 
as $N\to\infty$
\be\la{eqthm10}
     \|\phi_N\|_{L^\infty((-a,a))} = O(c_N^{-1/2}),\qquad
     \|\eps\phi_N\|_{L^\infty(\R)}  = O(c_N^{1/2}\cdot N^{-1/2}).
\ee
Also $\|\phi_N\|_{L^\infty(\R)} = O(c_N^{-1/2}N^{1/6})$.
\end{theorem}
\subsection{Proof of Theorem \ref{thmuniv}}
Set $q_N\equiv\lambda_{N,1}^2=R_{N,1,1}(r)$.
Observe from \eqref{eqSTARSTAR}
 and \eqref{M.5.2}, \eqref{eqsc1}
that $q_N$ satisfies \eqref{M.1.2} with $\alpha=1/2$.
Thus $\alpha^*$ in \eqref{M.12.2} is $1/4$.
In view of \eqref{1K}, \eqref{W1}, we have
\be\la{M.13.1}
\ba
  \null &\frac{1}{q_N} 
   K_{N,1}^{(\lambda_{N,1})}
   \bigg(r+\frac{\xi}{q_N},
      r+  \frac{\eta}{q_N}\bigg)\\
  &=\twotwo{\frac{1}{q_N} S_{N,1}
                            \big(r+\frac{\xi}{q_N},
                              r+  \frac{\eta}{q_N}\big)}
                    {\frac{1}{q_N^2} (S_{N,1}D)
                            \big(r+\frac{\xi}{q_N},
                              r+  \frac{\eta}{q_N}\big)}
                    {  (\eps S_{N,1})
                            \big(r+\frac{\xi}{q_N},
                              r+  \frac{\eta}{q_N}\big) -\frac12\sgn(\xi-\eta)}
                   {\frac{1}{q_N} S_{N,1}
                            \big(r+\frac{\eta}{q_N},
                              r+  \frac{\xi}{q_N}\big)}
\ea
\ee
where 
\be\la{M.13.2}
\ba
   \frac{1}{q_N} S_{N,1} \bigg(r+\frac{\xi}{q_N},
                              r+  \frac{\eta}{q_N}\bigg)
   = &\frac{1}{q_N} K_N \bigg(r+\frac{\xi}{q_N},
                              r+  \frac{\eta}{q_N}\bigg)\\
     &- \frac{1}{q_N} (\Phi_1(r+\xi/q_N)^T,0^T)\cdot 
        (AC(I_{2n}-BAC)^{-1})^T \\
       &\qquad\qquad\qquad\cdot(\eps\Phi_1(r+\eta/q_N)^T,
                  \eps\Phi_2(r+\eta/q_N)^T)^T
\ea
\ee
The convergence of the derivatives and integrals
of the Christoffel--Darboux kernel $\frac{1}{q_N} 
K_N \big(r+\frac{\xi}{q_N},
                              r+  \frac{\eta}{q_N}\big)$
with uniform error estimates for $\xi,\eta,r\in[-2L_0,2L_0]$,
for a fixed $L_0<\infty$,
is established in Subsection~\ref{subsec_for_sec_3},
see \eqref{M.4.5}, \eqref{M.6.2}, \eqref{M.6.3} and \eqref{M.13.0}
below.

We now consider the correction terms.
By the results of Section~\ref{secttwo}
(see, in particular, \eqref{BAformula} et seq.), 
 $BA$ and $C$ converge as $N\to\infty$.
Moreover $I_{2n}-\lim_{N\to\infty}BAC$ is invertible
and hence $(I_{2n}-BAC)^{-1}=O(1)$ as $N\to\infty$.
From \eqref{Das} and \eqref{cN},
 $A\sim  b_N^n\sim  c_N^n=O(N^{1-1/(2m)})$
and from \eqref{eqthm10}
\be\la{M.14.1}
   \max_{|\xi|,|r|\leq2L_0} |\Phi_1(r+\xi/q_N)| = O(c_N^{-1/2}),\qquad
           \|\eps\Phi_2\|_{L^\infty(\R)} = O(c_N^{1/2}N^{-1/2}).
\ee
Thus the correction term 
for the $11$ and $22$ entries in \eqref{M.13.1}
 is bounded by
\be\la{M.14.2}
     \const\cdot\frac1{N^{1-1/(2m)}}\cdot c_N^{-1/2}\cdot N^{1-1/(2m)}
          \cdot( c_N^{1/2}N^{-1/2}) 
    = O\bigg(\frac1{\sqrt{N}}\bigg)
\ee
as $N\to\infty$, uniformly for $\xi,\eta,r\in[-2L_0,2L_0]$.

The correction term in the $12$ entry of \eqref{M.13.1}
has the form
$$
\ba 
\frac{1}{q_N^2} (\Phi_1(r+\xi/q_N)^T,0^T)
       &\cdot 
           (AC(I_{2n}-BAC)^{-1})^T \\
       &\cdot(\Phi_1(r+\eta/q_N)^T,
                    \Phi_2(r+\eta/q_N)^T)^T
\ea
$$
and is bounded by
\be\la{M.14.3}
     \const\cdot\frac1{(N^{1-1/(2m)})^2}\cdot c_N^{-1/2} \cdot N^{1-1/(2m)}
          \cdot c_N^{-1/2} = O\bigg(\frac1{N}\bigg)
\ee
as $N\to\infty$, uniformly for $\xi,\eta,r\in[-2L_0,2L_0]$.

Finally, to analyze the $21$ entry of \eqref{M.13.1}
(see \eqref{W1}),
we use the following observation. By \eqref{eq3pp1}, $(\eps S_{N,1})(x,y)$
is skew symmetric. Thus
\be\la{M.tmp0}
   (\eps S_{N,1})(x,y) = (\eps S_{N,1})(x,y)-(\eps S_{N,1})(y,y)
                        =-\int_{x}^y S_{N,1}(t,y)\,dt.
\ee
After setting $x=r+\xi/q_N$, $y=r+\eta/q_N$ we see that the correction
term has the form
\be\la{M.tmp}
\ba 
   -\bigg(\Big(-\int_{r+\xi/q_N}^{r+\eta/q_N}\Phi_1(t)\,dt\Big)^T,0^T\bigg)
       &\cdot 
           (AC(I_{2n}-BAC)^{-1})^T \\
       &\cdot\big(\eps \Phi_1(r+\eta/q_N)^T,
                   \eps \Phi_2(r+\eta/q_N)^T\big)^T.
\ea
\ee
The integration is over a subset of the region $I_3$ (see \eqref{eqdoms} below).
Hence as in \eqref{M.14.2},
 but now using \eqref{eqI.18.1} below, 
\eqref{M.tmp} is bounded by
\be\la{M.14.4}
     \const
          \cdot c_N^{1/2}N^{-1}\cdot N^{1-1/(2m)}
             \cdot c_N^{1/2}N^{-1/2} = O(N^{-1/2})
\ee
as $N\to\infty$.
We compute finally the contribution of the Christoffel--Darboux to
\eqref{M.tmp0}. We have
\be\la{M.tmp1}
        -\int_{x}^{y}K_N(t,y)\,dt = (\eps K_N)(x,y) - (\eps K_N)(y,y)
\ee
where $x=r+\xi/q_N$, $y=r+\eta/q_N$. By \eqref{M.6.4}, \eqref{M.13.0} below
with $\alpha=1/2,\alpha^*=1/4$, \eqref{M.tmp1} equals as $N\to\infty$
$$
      \bigg(\int_0^{\xi-\eta}K_\infty(t)\,dt 
           - \int_0^{\eta-\eta}K_\infty(t)\,dt\bigg) + O(N^{-1/4})
          = \int_0^{\xi-\eta}K_\infty(t)\,dt + O(N^{-1/4})
$$
where the estimates are uniform for $r,\xi,\eta\in[-2L_0,2L_0]$.

Assembling the above estimates we see that we have proved
Theorem \ref{thmuniv} in the case $\beta=1$ with error term
\be\la{M.17.1}
       \mathcal{E}_{N,1} 
       =\twotwo{O\big(N^{-1/2}\big)}
             {O\big(N^{-1/2}\big)}
             {O(N^{-1/4})}
            {O\big(N^{-1/2}\big)}
\ee
as $N\to\infty$, uniformly for  $\xi,\eta,r\in[-2L_0,2L_0]$.
\subsection{Proof of Corollary \ref{cor1.2}}
In \cite{TW2} the authors prove that if $\chi_B$
denotes the characteristic function of a bounded Borel
set $B$ then for finite $N$ in the case $\beta=1$
$$
 \Prob\{ \textrm{ no eigenvalues in B } \} 
   = \sqrt{\det\bigg( (1-K_{N,1}\chi_B)\twotwo{1}{0}{-\eps\chi_B}{1}\bigg)}
$$
where again $\eps$ denotes the operator with 
kernel $\frac12\sgn(x-y)$.
Now the product of the operators above is of the type
identity plus trace class,
but the individual operators are not.
This is because $\eps$ is Hilbert--Schmidt but not trace class.
However, as indicated in Section~\ref{sec1},
 we can use regularized $2$-determinants.
One easily sees, moreover,
 that ${\det}_2\twotwo{1}{0}{-\eps\chi_B}{1}=1$.
Thus
$$
\ba
   \det\bigg( (1-&K_{N,1}\chi_B)\twotwo{1}{0}{-\eps\chi_B}{1}\bigg)
         = {\det}_2\bigg( (1-K_{N,1}\chi_B)\twotwo{1}{0}{-\eps\chi_B}{1}\bigg)\\
      &={\det}_2 (1-K_{N,1}\chi_B) 
          \,{\det}_2\twotwo{1}{0}{-\eps\chi_B}{1}
      ={\det}_2 (1-K_{N,1}\chi_B).
\ea
$$
For $\beta=4$ such regularization issues do not arise.
For $\beta=1$ (cf.~\eqref{M.19.4})
\be\la{M.21.1}
\ba
  \Prob\bigg\{\textrm{ no eigenvalues in }
    &\bigg(
       r-\frac{\theta}{\lambda_{N,1}^2},
           r+\frac{\theta}{\lambda_{N,1}^2}
        \bigg) \bigg\}\\
   &=\sqrt{{\det}^{(1)}\Big(1-K_{N,1}
          \chi_{\big(r-\frac{\theta}{\lambda_{N,1}^2},
                     r+\frac{\theta}{\lambda_{N,1}^2}\big)}\Big)}.
\ea
\ee
For $\beta=4$, the same formula is true provided we replace
$\lambda_{N,1}\to\lambda_{N,4}$, $\det^{(1)}\to{}\det^{(4)}$, 
$K_{N,1}\to{}K_{N/2,4}$ (cf.~\eqref{M.19.1}).
In order to prove Corollary \ref{cor1.2}
for $\beta=1$ it is clearly sufficient
to prove that the entries of $\frac1{q_N}
              K_{N,1}^{(\lambda_{N,1})}\big(r+\frac\xi{q_N},r+\frac\eta{q_N}\big)$,
$q_N\equiv\lambda_{N,1}^2$,
 apart from the term $\frac12\sgn(\xi-\eta)$,
converge to the corresponding entries in 
$K^{(1)}(\xi,\eta)$ (cf.~\eqref{M.13.1}, \eqref{M.13.2}, \eqref{M.19.2})
in trace norm for $L^2(-\theta,\theta)$.
The case $\beta=4$ is similar, mutatis mutandis, 
and is left to the reader.

We consider $r,\xi,\eta$ in the compact 
 set $[-2L_0,2L_0]$ with $L_0\equiv\theta$.
We have
$$
  \frac1{q_N}S_{N,1} \big(r+\frac\xi{q_N},
   r+\frac\eta{q_N}\big)
            = \frac1{q_N}K_N \big(r+\frac\xi{q_N},
   r+\frac\eta{q_N}\big)+\textrm{ correction term}.
$$
By \eqref{M.13.2}, the correction term corresponds to a finite rank
operator, and hence its trace norm is bounded by the $L^2(-\theta,\theta)$
norms of the vectors
$(\Phi_1(r+\xi/q_N)^T,0^T)
$,
$(\eps\Phi_1(r+\eta/{q_N})^T,
   \eps\Phi_2(r+\eta/{q_N})^T)^T
$
which in turn may be estimated by the bounds in \eqref{M.14.1}.
We conclude that the trace norm of the correction term decays
as $N^{-1/2}$ uniformly for $|r|\leq2\theta$.

The fact that $\frac1{q_N}K_N \big(r+\frac\xi{q_N},
   r+\frac\eta{q_N}\big)\to K_\infty(\xi-\eta)$ in trace norm
was proved in the case $V(x)=x^{2m}$ in \cite{D}.
We now give a proof for general $V(x)=\kappa_{2m}x^{2m}+\cdots$,
which in addition also applies to the other entries 
of $\frac1{q_N}K_{N,1}^{(\lambda_{N,1})}$.
Note first that it is sufficient to prove that 
  $\chi\frac1{q_N}K_N \big(r+\frac\cdot{q_N},
   r+\frac{\cdot\cdot}{q_N}\big)\chi$
converges to $\chi K_\infty(\cdot-\cdot\cdot)\chi$
in trace norm for any $C_0^\infty(-2\theta,2\theta)$
function $\chi$ with $\chi(\xi)=1$ for $\xi\in(-\theta,\theta)$.
Let $P\equiv iD\equiv i\frac{\partial}{\partial\xi}$
denote the self-adjoint operator of differentiation
with periodic boundary conditions on $[-2\theta,2\theta]$.
Then $D+I=\frac1i  P+I$ is an invertible operator with
eigenvalues $\gamma_k=\frac{\pi k}{i2\theta}+1$, $k\in\Z$.
In particular $\frac1{D+I}$ is Hilbert--Schmidt
on $L^2(-2\theta,2\theta)$. Clearly $\chi\frac1{q_N}K_N\chi$
maps $L^2(-2\theta,2\theta)$ into the domain of $P$,
$\dom P=\{f\in L^2(-2\theta,2\theta)\,:\,
   f^\prime\in L^2(-2\theta,2\theta),\,f(-2\theta)=f(2\theta)\}$
 and so
\be\la{M.23.1}
\ba
   \chi\frac1{q_N}K_N\chi 
   &=\frac1{D+I} (D+I)\chi\frac1{q_N}K_N\chi \\
   &=\frac1{D+I} \chi\frac1{q_N}K_N\chi
             + \frac1{D+I} \chi^\prime \frac1{q_N}K_N\chi
         +\frac1{D+I} \chi\Big(\frac\partial{\partial\xi} \frac1{q_N}K_N\Big)\chi
\ea
\ee
But by \eqref{M.4.5} below,
 $\frac1{q_N}K_N \big(r+\frac\xi{q_N},
   r+\frac\eta{q_N}\big)$
converges to $K_\infty(\xi-\eta)$
uniformly for $|r|,|\xi|,|\eta|\leq2\theta$ and hence 
$\chi\frac1{q_N}K_N\chi\to \chi K_\infty\chi$
in Hilbert--Schmidt norm. 
But then $\frac1{D+I} \chi \frac1{q_N}K_N\chi \to
 \frac1{D+I} \chi K_\infty\chi $
in trace norm. The same is clearly true for the second term in \eqref{M.23.1},
and also for the third term (take $j=1$, $k=0$ in \eqref{M.4.5} below).
This shows that $\frac1{q_N}K_N\to K_\infty$ in trace norm
and completes the proof that the $11$ entry of
 $\frac1{q_N}K_N \big(r+\frac\xi{q_N},
   r+\frac\eta{q_N}\big)\to K_\infty(\xi-\eta)$
converges in trace norm to $(K^{(1)})_{11}$, as desired.
Clearly the same is true for the $22$ entry.
Similar considerations using \eqref{M.6.2}
in place of \eqref{M.4.5} prove the corresponding result 
for the $12$ entry.
Finally, for the $21$ entry we observe that $Y_N(\xi,\eta)$
in \eqref{M.13.0} is a sum of two terms $A_N(\xi,\eta)+B_N(\eta)$.
The first term can be controlled in trace norm using the operator
$P=iD$ as above, together with the estimate \eqref{M.13.0}(a).
On the other hand $B_N(\eta)$ corresponds to a rank one operator in
$L^2(-\theta,\theta)$ %OR $2\theta$ ?? 
and hence its trace norm goes to zero
as $N\to\infty$ by \eqref{M.13.0}(b).
Finally, the correction term in the $21$ entry can be controlled as above.
This completes the proof of Corollary \ref{cor1.2}.
\section{Asymptotics of integrated OP's, 
proofs of Theorem~\ref{thm1} and Theorems~\ref{thm2}, \ref{thmthree}}
\la{sectfour}
\textbf{Notational remark:} Throughout this Section, the notation 
$N,P\too\infty$ means that $N,P\to\infty$, $N\geq P$, 
$N-P$ is fixed (and finite).
\subsection{Proof of Theorem~\ref{thm1}}
Let $V(x)=\kappa_{2m}x^{2m}+\cdots$, $\{p_j\}_{j\geq0}$,
$\{\phi_j\}_{j\geq0}$ be as before. 
The following observation is due to A.~Borodin.
\begin{lemma}
\la{lembor}
For any $j,k\geq0$
\be\la{eqlembor}
             (D\phi_j,\phi_k) = \frac12\sgn(j-k)\,(V^\prime\phi_j,\phi_k).
\ee
\end{lemma}
\begin{proof}
For $j\leq k$ we have 
$$
\begin{aligned} 
    (D\phi_j,\phi_k) 
         &= \int \big(p_j(x)e^{-V(x)/2}\big)^\prime p_k e^{-V(x)/2}\,dx\\
         &= \int p_j^\prime(x) p_k(x) e^{-V(x)}\,dx -\frac12(V^\prime\phi_j,\phi_k)
\end{aligned}
$$
and the last integral vanishes since $p_k$ is orthogonal to
all polynomials of degree $<k$, and we have assumed that $j\leq k$.
Observe that $(V^\prime\phi_k,\phi_k)=0$ as it should be.
The case $j>k$ follows by skew symmetry.
\end{proof}
\begin{remark}
Clearly Lemma \ref{lembor} is true for any differentiable $V(x)$
such that 
$$
     \int |x|^q\,(1+|V^\prime(x)|)\,e^{-V(x)}\,dx < \infty,\qquad q=0,1,2,\cdots.
$$
\end{remark}
As $V^\prime(x)$ is a polynomial of degree $n=2m-1$, it 
follows immediately from \eqref{eqlembor}
that $((D\phi_j,\phi_k))_{j,k\geq0}$ is a banded batrix,
$(D\phi_j,\phi_k)=0$ for $|j-k|>n$.

For the proof of Theorem \ref{thm1} we need to evaluate
the leading asymptotics of $(V^\prime\phi_{N+j},\phi_{N+k})$,
$j,k$ fixed, as $N\to\infty$.
This clearly reduces to computing the large $N$ asymptotics 
of $(x^q\phi_{N+j},\phi_{N+k})$, $j,k$ fixed, $0\leq{}q\leq{}n$.
In order to compute such asymptotics we will use the following relations
for the recurrence coefficients in \eqref{eqthreeterm}
\begin{equation}
\label{eq1thmxjcond}
   b_{N}/b_{N-1}\to1,\qquad\textrm{and}\qquad a_N =o(b_N),
       \qquad \textrm{as }N\to\infty.
\end{equation}
These relations
 certainly hold for our potential $V(x)=\kappa_{2m}x^{2m}+\cdots$,
see \eqref{bN}, \eqref{cN}, \eqref{dN}.
\begin{lemma}
\label{thmxjnotmonom}
For any fixed $q\in\N$, as $N\to\infty$
\be
\la{eq50}
\ba
x^q\phi_N(x) = &b_N^{q}\cdot\Bigg[
         \sum_{l=0}^q \binom{q}{l}(1+o(1))
                           \cdot\phi_{N-q+2l}(x)\\
            &+\sum_{l=1}^q o(1)
                           \cdot\phi_{N-q+(2l-1)}(x)
\Bigg]
\ea
\ee
where the notation $o(1)$ indicates terms which are independent 
of $x$ and uniform for $0\leq{}l\leq{}q$.
(The $o(1)$ terms above can be replaced with $O(N^{-1/(2m)})$.)
\end{lemma}
\begin{proof}
By \eqref{eq1thmxjcond} the three term recurrence relation \eqref{eqthreeterm}
takes the form as $N\to\infty$,
\be
\la{eq31}
\ba
x\phi_N(x) &= b_N\,
        \big[\phi_{N+1}(x)+
                (1+o(1))\,\phi_{N-1}(x)\\
      &\qquad+o(1)\, \phi_{N}(x)\big]
\ea
\ee
which coincides with \eqref{eq50} for $q=1$.
A simple induction on $q$ using \eqref{eq31} and Pascal's triangle,
now gives the result.
\end{proof}
We are now in a position to prove Theorem \ref{thm1}.
Write $V(x)=\sum_{q=0}^{2m}v_qx^q$, $v_{2m}=\kappa_{2m}$.
Then by Lemma~\ref{lembor}, Lemma \ref{thmxjnotmonom}
and the orthogonality of the $\phi_j$'s,
\be
\la{eq15p2}
\ba
  (D\phi_{N+j},\phi_{N+k}) 
        =&\frac12\sgn(j-k)\sum_{q=1}^{2m}
              qv_q(x^{q-1}\phi_{N+j},\phi_{N+k})\\
      =&\frac12\sgn(j-k)\sum_{q=1}^{2m}
              qv_q b_N^{q-1}\,\Bigg[
         \sum_{l=0}^{q-1} \binom{q-1}{l}(1+o(1))\,
                           \delta_{2l,k-j+q-1}\\
            &\qquad\qquad\qquad\qquad+\sum_{l=1}^{q-1} o(1)\,
                           \delta_{2l,k-j+q}
\Bigg],
\ea
\ee
as $N\to\infty$.
Observe that for $|j-k|\leq n$ and $q=2m=n+1$,
$k-j+q-1=k-j+n$
lies between $0$ and $2n$.
Hence if $|j-k|\leq n$ and $j-k$ is odd, we 
see that the square bracket in \eqref{eq15p2}
for $q=2m$ gives rise to a leading contribution 
$\binom{n}{(n+k-j)/2}\,(1+o(1))$. However if 
$|j-k|\leq n$ and $j-k$ is even, the contribution is $o(1)$.
But by \eqref{bN}, \eqref{cN}, $b_N\to\infty$, and hence
the terms corresponding to $q=1,2,\cdots,2m-1$ in \eqref{eq15p2}
contribute to lower order. This completes the proof of Theorem~\ref{thm1}.
\subsection{Integrating the Plancherel--Rotach type asymptotics for OP's
and the proof of Theorems \ref{thmthree} and~\ref{thm2}}
\la{ssec42}
\subsubsection{Auxiliary estimates on the integrated 
Plancherel--Rotach asymptotics}
\la{4.2.1}
For the convenience of the reader, we recall relevant results
from \cite{DKMVZ2}.
Assume that $V(x)=\kappa_{2m}x^{2m}+\cdots$,
$m\in\N$, $\kappa_{2m}>0$. 
Let $p_N$ be the $N$th
OP~on $\R$ with respect to the weight $e^{-V(x)}$ and set
$\phi_N(x)\equiv{}p_N(x)e^{-V(x)/2}$ as before.
 In \cite[Theorem~2.2]{DKMVZ2} an asymptotic expansion,
as $N\to\infty$, is derived for $\phi_N$ in the whole complex plane,
and the leading term in the expansion in each region (see below)
is computed explicitly.
Here we are only interested in real values of the argument.
Fix $\delta_0>0$ and sufficiently small (see \cite{DKMVZ2}).
For $0<\delta\leq\delta_0$ define
\be
\la{eqdoms}
\ba
      I_5&\equiv{} (-\infty,-1-\delta),\qquad I_4\equiv{}(-1-\delta,-1+\delta)\\
      I_3&\equiv{} (-1+\delta,1-\delta),\qquad I_2\equiv{} (1-\delta,1+\delta)\\
      I_1&\equiv{} (1+\delta,+\infty)
\ea
\ee
and let $\psi_N(x)$, $h_N(x)$, $h(x)$ 
be as in \eqref{psiN}, \eqref{eq9pSTAR} and \eqref{hz}, respectively.
In \cite{DKMVZ2}, each of the regions $I_2$ and $I_4$
is split further into two regions, $I_2=C_{1,\delta}\cup{}C_{2,\delta}$, 
$I_4=D_{1,\delta}\cup{}D_{2,\delta}$,
the estimates in $C_{2,\delta}$ and $D_{2,\delta}$ being finer than those
in $C_{1,\delta}$ and $D_{1,\delta}$, respectively.
For our purposes it is sufficient to conflate the regions and
use the estimates in $C_{1,\delta}$ and $D_{1,\delta}$ for all points
in $I_2$ and $I_4$, respectively. The apparent singularities
in the formulae below for $z=\pm1$ are of course removable
(see e.g.{} \eqref{eqI8.1} and \eqref{4.12} below).
In what follows, notation of the type $O(1/N)$ as $N\to\infty$, 
means that the estimate holds for
$x$ uniformly in the respective region \eqref{eqdoms} under consideration.
Also the estimates are uniform for $\delta$ in {\em compact\ }subsets
of $(0,\delta_0]$: in this connection see the important Remark~\ref{rem4.139p}
following \eqref{eqN.69.1} below.
Finally, the constants $c_1,c_2,\cdots$
below may depend on $\delta$,
but they are independent of $x$ and $N,P$.

We are ready to state \cite[Theorem~2.2]{DKMVZ2}. 
First, we consider the ``exponential'' regions $I_{1,5}$.
Uniformly for all $|x|>1+\delta$
$$
\begin{aligned}
    \phi_N(c_Nx+d_N)
=  &\frac1{\sqrt{4\pi c_N}}\, e^{-(N/2)\int_1^x|y^2-1|^{1/2}h_N(y)dy}\\
    &\times\Big( \Big|\frac{x-1}{x+1}\Big|^{1/4}
                  +\Big|\frac{x+1}{x-1}\Big|^{1/4} \Big)
     \,\Big(1+O\Big(\frac1N\Big)\Big),\qquad x>1+\delta
\end{aligned}
$$
and
$$
\begin{aligned}
    \phi_N(c_Nx+d_N)
=  &\frac{(-1)^N}{\sqrt{4\pi c_N}}\, 
        e^{-(N/2)\int_x^{-1}|y^2-1|^{1/2}h_N(y)dy}\\
    &\times\Big( \Big|\frac{x-1}{x+1}\Big|^{1/4}
                  +\Big|\frac{x+1}{x-1}\Big|^{1/4} \Big)
     \,\Big(1+O\Big(\frac1N\Big)\Big),\qquad x<-1-\delta.
\end{aligned}
$$
In particular, as 
\be\la{hestbelow}
      h_N(x)\geq h_{\textrm{min}}>0, \qquad x\in\R,\qquad N\geq N_1(V)
\ee
(\cite[Prop. 5.3]{DKMVZ2}), we have for $|x|\geq1+\delta$
\be
\la{eqI7.1}
 |\phi_N(c_Nx+d_N)|
   \leq  \frac{c_{1}}{\sqrt{4\pi c_N}}\, 
        e^{-Nc_{3}}\,e^{-(N/4)c_{2}\,(x^2-(1+\delta)^2)},  
\ee
where we have used the fact that
 $|y^2-1|^{1/2}\geq{}c_{2}|y|$ for $|y|\geq1+\delta$.

Next, consider the ``Airy'' region on the right, $I_{2}$.
Uniformly for all $1-\delta<x<1+\delta$
\be\la{eqI2reg}
\begin{aligned}
    \phi_N(c_Nx+d_N)
=  \frac1{\sqrt{c_N}}\bigg(
          &\Big|\frac{x+1}{x-1}\Big|^{1/4} |f_N(x)|^{1/4} \Ai(f_N(x))
                            \,\Big(1+O\Big(\frac1N\Big)\Big)\\
         &- \Big|\frac{x-1}{x+1}\Big|^{1/4} \frac1{|f_N(x)|^{1/4}} 
                    \Ai^\prime(f_N(x))
                            \,\Big(1+O\Big(\frac1N\Big)\Big)
       \bigg)
\end{aligned}
\ee
where
\be\la{eqI8.1}
        f_N(x) = \alpha_NN^{2/3}\,(x-1)\fh_N(x)
\ee
and in turn (see (the proof of) \cite[Proposition~7.3]{DKMVZ2})
\begin{enumerate}
\item
 $\fh_N(x)$ is real analytic on $(1-\delta,1+\delta)$,
and to any order $q=0,1,2,\cdots$
$$
   \fh_N(x) = \sum_{j=0}^q N^{-j/(2m)}\,\fh_{(j)}(x) + O(N^{-(q+1)/(2m)}) 
$$
uniformly for $x$ in the interval. Moreover, the functions $\fh_{(j)}(x)$
are also real analytic on $1-\delta< x< 1+\delta$ 
\item 
to any order $q=1,2,\cdots$ 
$$
         \alpha_N\equiv \big(h_N^2(1)/2\big)^{1/3}
            =2m^{2/3} +  \sum_{j=1}^q N^{-j/(2m)}\,\alpha_{(j)}
                   + O(N^{-(q+1)/(2m)}) 
$$
\item
 $f_N^\prime(x)=-\alpha_N N^{2/3}U_N(x)$,
where $U_N(x)=\fh_N(x) + (x-1)\fh_N^\prime(x)$ also has 
an expansion uniform in $x$ to any order $q=0,1,2,\cdots$ as above 
$$
  U_N(x) = \sum_{j=0}^q N^{-j/(2m)}\,U_{(j)}(x) + O(N^{-(q+1)/(2m)}).
$$
The terms $U_{(j)}(x)$ are real analytic on $1-\delta< x< 1+\delta$
\item
 $\max_{k=0,1,2}\max_{1-\delta\leq x\leq1+\delta}
  |d^k\fh_N(x)/dx^k|\leq M<\infty$ for $N\geq N_2(V)$ 
\item 
 $\fh_N(1)=1=U_N(1)$ and $\min_{1-\delta\leq x\leq1+\delta}\fh_N(x)\geq\frac12$
for $N\geq N_2(V)$.
Also $\fh_{(0)}(1)=1=U_{(0)}(1)$.
\end{enumerate}
Similarly, in the left ``Airy'' region $I_{4}$
we have uniformly for all $-1-\delta<x<-1+\delta$
\be\la{eqI.8.0}
\begin{aligned}
    \phi_N(c_Nx+d_N)
=  \frac{(-1)^N}{\sqrt{c_N}}\bigg(
          &\Big|\frac{x-1}{x+1}\Big|^{1/4} |\ft_N(x)|^{1/4} \Ai(-\ft_N(x))
                            \,\Big(1+O\Big(\frac1N\Big)\Big)\\
         &- \Big|\frac{x+1}{x-1}\Big|^{1/4} \frac1{|\ft_N(x)|^{1/4}} 
                    \Ai^\prime(-\ft_N(x))
                            \,\Big(1+O\Big(\frac1N\Big)\Big)
       \bigg)
\end{aligned}
\ee
where
\be
\la{4.12}
        \ft_N(x) = \tilde{\alpha}_N
          N^{2/3}\,(x+1)\fth_N(x)
\ee
and in turn
\begin{enumerate}
\item $\fth_N(x)$ is real analytic on $(-1-\delta,-1+\delta)$,
and to any order $q=0,1,2,\cdots$
$$
   \fth_N(x) = \sum_{j=0}^q N^{-j/(2m)}\,\fth_{(j)}(x) + O(N^{-(q+1)/(2m)}) 
$$
uniformly for $x$ in the interval. Moreover, the functions $\fth_{(j)}(x)$
are also real analytic on $-1-\delta< x< -1+\delta$ 
\item to any order $q=1,2,\cdots$ 
$$
         \tilde{\alpha}_N\equiv \big(h_N^2(-1)/2\big)^{1/3}
            =2m^{2/3} +  \sum_{j=1}^q N^{-j/(2m)}\,\tilde{\alpha}_{(j)}
                   + O(N^{-(q+1)/(2m)}) 
$$
\item
 $\ft_N^\prime(x)=-\tilde{\alpha}_N N^{2/3}\tilde{U}_N(x)$,
where $\tilde{U}_N(x)=\fth_N(x) + (x+1)\fth_N^\prime(x)$ also has 
an expansion uniform in $x$ to any order $q=0,1,2,\cdots$ 
as above
$$
  \tilde{U}_N(x) = \sum_{j=0}^q N^{-j/(2m)}\,
          \tilde{U}_{(j)}(x) + O(N^{-(q+1)/(2m)}).
$$
The terms $\tilde{U}_{(j)}(x)$ are real analytic on $1-\delta< x< 1+\delta$
\item
 $\max_{k=0,1,2}\max_{-1-\delta\leq x\leq-1+\delta}
  |d^k\fth_N(x)/dx^k|\leq M<\infty$ for $N\geq N_2(V)$ 
\item
 $\fth_N(-1)=1=\tilde{U}_N(-1)$ and $\min_{-1-\delta\leq x\leq-1+\delta}\fth_N(x)\geq\frac12$
for $N\geq N_2(V)$.
Also $\fth_{(0)}(-1)=1=\tilde{U}_{(0)}(-1)$.
\end{enumerate}

Finally, in the middle region $I_3$
we have uniformly for all $-1+\delta<x<1-\delta$
\be\la{eqI.9.1}
\begin{aligned}
    \phi_N(c_Nx+d_N)
&=  \sqrt{\frac2{\pi c_N}}\frac1{|1-x^2|^{1/4}}\\
    &\times \bigg(
          \cos\Big(\frac{N}2\int_1^x|1-y^2|^{1/2}h_N(y)dy
                       + \frac12\arcsin x\Big)
                            \,\Big(1+O\Big(\frac1N\Big)\Big)\\
         &\qquad\qquad+ \sin\Big(\frac{N}2\int_1^x|1-y^2|^{1/2}h_N(y)dy 
                             - \frac12\arcsin x\Big)
                 \,O\Big(\frac1N\Big)
       \bigg).
\end{aligned}
\ee

In the proof of Theorem~\ref{thm2} (and Theorem~\ref{thmthree})
 we need the following two results that hold
as $N\ra\infty$, 
\be\la{eqI.19.2}
\ba
    \int_{-\infty}^{+\infty}\phi_N(y)\,dy
     &\equiv c_N \int_{-\infty}^{+\infty}\phi_N(c_Ny+d_N)\,dy\\
      &=c_N^{1/2}\,N^{-1/2}\,(2m)^{-1/2}\,(1+(-1)^N+O(N^{-1/2})
               +O(N^{-1/(2m)}))
\ea
\ee
and uniformly for $x\in\R$
\be\la{eqI.19.1}
    \bigg| c_N \int_{-\infty}^x \phi_N(c_Ny+d_N)\,dy\bigg|
      = O\big(c_N^{1/2}\,N^{-1/2}\big)
\ee
or, equivalently, for any interval $K\subset\R$
$$
       \bigg| c_N \int_K \phi_N(c_Ny+d_N)\,dy\bigg|
      = O\big(c_N^{1/2}\,N^{-1/2}\big).
$$
\textbf{Notational remark:}
 In \eqref{eqI.19.2} above the reader
may wonder why we write $O(N^{-1/2})+O(N^{-1/(2m)})$
rather than just $O(N^{-1/(2m)})$.
 The reason is that the $O(N^{-1/(2m)})$
term arises simply from the evaluation
of the constants $\alpha_N,\tilde{\alpha}_N$
arising in the asymptotics using
\eqref{eqI8.1}(2), \eqref{4.12}(2). The term $O(N^{-1/2})$,
however, constitutes the detailed estimate of errors involved
in evaluating the integral over all the different
asymptotic regions \eqref{eqdoms}.
In order to separate out these two very different sources of error
we have adopted the convention of writing error estimates
in the form $O(N^{-1/2})+O(N^{-1/(2m)})$ as above.
We use this convention in all the estimates that follow in this 
Section.
In the proof of Theorem \ref{thm2},
 error estimates of the type $O(N^{-1/(2m)})$ also arise
from the approximation of $h_N(x)$ in \eqref{eq9pSTAR}
 (see e.g.{} \eqref{p.45.1} below).

Note that \eqref{eqI.19.2}, \eqref{eqI.19.1} 
are direct consequences of the following 
more detailed statement, which is also used in the proof of Theorem~\ref{thm2}.
\begin{proposition}
\la{propnew1}
The following holds uniformly for $x$ in the respective regions as $N\to\infty$.
In $I_1$:
\be
\la{eqI.10.1}
\ba
  \bigg|c_N\int_{1+\delta}^{x}\phi_N(c_Ny+d_N)\,dy\bigg|
   &\leq c_N\int_{1+\delta}^{x}\big|\phi_N(c_Ny+d_N)\big|\,dy\\
  &=  O\big( (c_N/N)^{1/2} 
        \,e^{-Nc_{3}}\big),
               \qquad x\geq 1+\delta;
\ea
\ee
in $I_2$:
\be\la{eqI.17.1}
\ba
 c_N\int_{1-\delta}^{1+\delta}\phi_N(c_Ny+d_N)\,dy 
           &= c_N^{1/2}N^{-1/2}(2m)^{-1/2}(1+O(N^{-1/2})+O(N^{-1/(2m)}))\\
 \bigg|c_N\int_{1-\delta}^{x}\phi_N(c_Ny+d_N)\,dy\bigg| 
           &\leq O\bigg(\frac{c_N^{1/2}}{N^{1/2}}\bigg),
                    \qquad 1-\delta\leq x\leq 1+\delta;
\ea
\ee
in $I_3$:
\be\la{eqI.18.1}
\bigg|c_N\int_{-1+\delta}^{x}\phi_N(c_Ny+d_N)\,dy\bigg| 
           \leq O\bigg(\frac{c_N^{1/2}}{N}\bigg),
                     \qquad -1+\delta\leq x\leq 1-\delta;
\ee
in $I_4$:
\be
\la{eqI.16.1}
\ba
 c_N\int_{-1-\delta}^{-1+\delta}\phi_N(c_Ny+d_N)\,dy 
           &= c_N^{1/2}N^{-1/2}(2m)^{-1/2}(-1)^N(1+O(N^{-1/2})+O(N^{-1/(2m)}))\\
 \bigg|c_N\int_{-1-\delta}^{x}\phi_N(c_Ny+d_N)\,dy\bigg| 
           &\leq O\bigg(\frac{c_N^{1/2}}{N^{1/2}}\bigg),
                     \qquad -1-\delta\leq x\leq -1+\delta;
\ea
\ee
in $I_5$:
\be
\la{eqI.10.1prime}
\ba
\bigg|c_N\int_{-\infty}^{x}\phi_N(c_Ny+d_N)\,dy\bigg|
 &\leq c_N\int_{-\infty}^{x}\big|\phi_N(c_Ny+d_N)\big|\,dy\\
  &=   O\Big((c_N/N)^{1/2}\,e^{-Nc_{3}}\Big),\qquad
               x\leq -1-\delta.
\ea
\ee
\end{proposition}
\begin{proof}
We refer to \eqref{eqI7.1} 
and note that 
$$
     \int_{1+\delta}^\infty
                      e^{-(N/4)c_{2}\,(y^2-(1+\delta)^2)}\,dy
        \leq O(N^{-1/2})
$$
which proves \eqref{eqI.10.1}. The 
proof of \eqref{eqI.10.1prime} is similar.

We now prove \eqref{eqI.18.1}. 
Uniformly for $-1+\delta\leq x\leq1-\delta$,
by \eqref{eqI.9.1},
$$
\begin{aligned}
  c_N&\int_{-1-\delta}^{x}\phi_N(c_Ny+d_N)\,dy\\
&=  c_N^{1/2}\sqrt{\frac2{\pi}}\int_{-1+\delta}^x \bigg[
          \cos\Big(\frac{N}2\int_1^y|1-t^2|^{1/2}h_N(t)dt
                       + \frac12\arcsin y\Big)
                            \,\Big(1+O\Big(\frac1N\Big)\Big)\\
         &\qquad\qquad+ \sin\Big(\frac{N}2\int_1^y|1-t^2|^{1/2}h_N(t)dt 
                             - \frac12\arcsin y\Big)
                 \,O\Big(\frac1N\Big)
       \bigg]\,\frac{dy}{|1-y^2|^{1/4}}\\
  &=c_N^{1/2}\sqrt{\frac2{\pi}}\int_{-1+\delta}^x
            \cos\Big(\frac{N}2\int_1^y|1-t^2|^{1/2}h_N(t)dt
                       + \frac12\arcsin y\Big)\,\frac{dy}{|1-y^2|^{1/4}}\\
          &\qquad\qquad+ O\Big(c_N^{1/2}N^{-1}\Big).
\end{aligned}
$$
Integrating by parts in the last integral we find
\be\la{eqI.18.0}
\begin{aligned}
  c_N^{1/2}\int_{-1+\delta}^x
            &\cos\Big(\frac{N}2\int_1^y|1-t^2|^{1/2}h_N(t)dt
                       + \frac12\arcsin y\Big)\,\frac{dy}{|1-y^2|^{1/4}}\\
          &=c_N^{1/2}\frac{\sin\Big(\frac{N}2\int_1^y|1-t^2|^{1/2}h_N(t)dt
                          + \frac12\arcsin y\Big)}
                     {\frac{N}2|1-y^2|^{3/4}h_N(y)
                          + \frac12|1-y^2|^{-1/4}}\bigg|_{y=-1+\delta}^{x}\\
     &\qquad-c_N^{1/2}\int_{-1+\delta}^x
            \sin\Big(\frac{N}2\int_1^y|1-t^2|^{1/2}h_N(t)dt
                       + \frac12\arcsin y\Big)\\
          &\qquad\qquad\times\frac{d}{dy}\bigg( 
          \frac1{ {\frac{N}2|1-y^2|^{3/4}h_N(y)
                          + \frac12|1-y^2|^{-1/4}}}\bigg)\,dy\\
       &=O(c_N^{1/2}N^{-1})
\end{aligned}
\ee
uniformly for $x\in{}I_3$,
which proves \eqref{eqI.18.1}. 
(We have used \eqref{hestbelow}
and also the uniform boundedness of $h_N^\prime(x)$
on $I_3$, as $N\to\infty$.)

Now we prove \eqref{eqI.16.1}. 
Uniformly for $-1-\delta\leq x\leq-1+\delta$,
by \eqref{eqI.8.0},
$$
\begin{aligned}
  c_N&\int_{-1-\delta}^{x}\phi_N(c_Ny+d_N)\,dy\\
&=  (-1)^Nc_N^{1/2} \int_{-1-\delta}^x 
   \bigg[
          \Big|\frac{y-1}{y+1}\Big|^{1/4} |\ft_N(y)|^{1/4} \Ai(-\ft_N(y))
                            \,\Big(1+O\Big(\frac1N\Big)\Big)\\
         &\qquad\qquad- \Big|\frac{y+1}{y-1}\Big|^{1/4} \frac1{|\ft_N(y)|^{1/4}} 
                    \Ai^\prime(-\ft_N(y))
                            \,\Big(1+O\Big(\frac1N\Big)\Big)
       \bigg]\,dy.
\end{aligned}
$$
Now
$$
\begin{aligned}
 \bigg|c_N^{1/2} &\int_{-1-\delta}^x 
           \Big|\frac{y-1}{y+1}\Big|^{1/4} |\ft_N(y)|^{1/4} \Ai(-\ft_N(y))
                            \,O\Big(\frac1N\Big)\,dy\bigg|\\
    &\leq C_1 c_N^{1/2} O\Big(\frac1N\Big)\,\int_{-1-\delta}^{-1+\delta}
           \Big|\frac{y-1}{y+1}\Big|^{1/4}\,dy = O(c_N^{1/2}N^{-1})
\end{aligned}
$$
where (cf.~\cite{Stegun})
$$
        C_1\equiv \sup_{y\in\R}|y|^{1/4}|\Ai(y)|<\infty.
$$
Next, in view of \eqref{4.12} and the properties of $\fth_N$,
$$
\begin{aligned}
 \bigg|c_N^{1/2} &\int_{-1-\delta}^x 
           \Big|\frac{y+1}{y-1}\Big|^{1/4} 
                     \frac{\Ai^\prime(-\ft_N(y))}{|\ft_N(y)|^{1/4}}
                 \, O\Big(\frac1N\Big)\,dy\bigg|\\
    &\leq C_2 c_N^{1/2} O\Big(\frac1N\Big)\,
        \int_{-1-\delta}^{-1+\delta}
           \Big|\frac{y+1}{y-1}\Big|^{1/4}
         \frac{1+|\ft_N(y)|^{1/4}}{|\ft_N(y)|^{1/4}}\,dy\\ 
    &= C_2 c_N^{1/2} O\Big(\frac1N\Big)\,
        \int_{-1-\delta}^{-1+\delta}
           \Big|\frac{y+1}{y-1}\Big|^{1/4}
         \frac{1+|\tilde{\alpha}_N N^{2/3}(y+1)\fth_N(y)|^{1/4}}
                     {|\tilde{\alpha}_N N^{2/3}(y+1)\fth_N(y)|^{1/4}}\,dy\\ 
    &\leq \const\cdot c_N^{1/2} N^{-1}
        \int_{-1-\delta}^{-1+\delta}
          \Big( N^{-1/6} + |{y+1}|^{1/4}\Big)\,dy
   = O(c_N^{1/2}N^{-1}),
\end{aligned}
$$
where (cf.~\cite{Stegun})
\be\la{eqC2}
        C_2\equiv \sup_{y\in\R}(1+|y|^{1/4})^{-1}|\Ai^\prime(y)|<\infty.
\ee
and we have used (see \cite[(8.72)]{DKMVZ2} and \eqref{4.12}(2))
\be\la{eqalpha}
    \tilde{\alpha}_N\equiv \big(h_N^2(-1)/2\big)^{1/3} = 2m^{2/3}+O(N^{-1/(2m)}).
\ee

We consider next
\be
\la{eqI.11.2}
   c_N^{1/2} \int_{-1-\delta}^x 
           \Big|\frac{y+1}{y-1}\Big|^{1/4} 
                     \frac{\Ai^\prime(-\ft_N(y))}{|\ft_N(y)|^{1/4}}
                 \,dy.
\ee
Make the change of variables (cf.~\eqref{4.12}, \eqref{eqalpha})
\be\la{eqI.11.3}
   u(y) \equiv -\ft_N(y) = -\tilde{\alpha}_N N^{2/3}(y+1)\fth_N(y)
\ee
which implies (cf.{} \eqref{4.12}(3))
\be\la{eqI.12.0}
\ba
   \frac{du}{dy} &= -\tilde{\alpha}_N N^{2/3}\tilde{U}_N(y)\\
   \tilde{U}_N(y) &\equiv \fth_N(y) + (y+1)\fth_N^\prime(y).
\ea
\ee
Note also that
\be\la{eqI.12.1}
   y(u) + 1 = \frac{-u}{\tilde{\alpha}_N N^{2/3}\fth_N(y(u))}.
\ee
Now set
\be\la{eqBPBN}
             B_N(y)\equiv \frac1{|y-1|^{1/4}|\fth_N(y)|^{1/4}\,\tilde{U}_N(y)}
\ee
where the denominator is bounded away from $0$
by \eqref{4.12} for $\delta>0$ small enough.
Then \eqref{eqI.11.2} becomes
\be\la{eqI.13.0}
\ba
   -&\frac{c_N^{1/2}}{\tilde{\alpha}_N^{5/4}N^{5/6}}
                 \int_{u(-1-\delta)}^{u(x)} \Ai^\prime(u)\,B_N(y(u))\,du\\
   &=-\frac{c_N^{1/2}}{\tilde{\alpha}_N^{5/4}N^{5/6}}
                 \bigg[ \Ai(u)\,B_N(y(u))\bigg|_{u(-1-\delta)}^{u(x)}
                 +\frac{1}{\tilde{\alpha}_N N^{2/3}}
           \int_{u(-1-\delta)}^{u(x)} \Ai(u)\,
                  \frac{B^\prime(y(u))}{\tilde{U}_N(y(u))}\,du\bigg]\\
   &= O(c_N^{1/2}N^{-5/6})
\ea
\ee
since the boundary term and 
the last integrand are uniformly bounded 
for $x\in I_4$
and the length of the interval of integration
is $O(N^{2/3})$.

Finally we consider the integral of $\Ai$. 
Set
\be\la{eqE}
          L_N(y)\equiv \frac{|y-1|^{1/4}|\fth_N(y)|^{1/4}}{\tilde{U}_N(y)}
\ee
and for future reference note that (see \eqref{4.12}(5))
\be\la{eq25star}
  L_N(-1) = 2^{1/4}.
\ee
Making the same change of 
variables \eqref{eqI.11.3}, \eqref{eqI.12.0}, \eqref{eqI.12.1} we find
\be
\la{eqI.13.1}
\ba
   c_N^{1/2}(-1)^N &\int_{-1-\delta}^x 
           \Big|\frac{y+1}{y-1}\Big|^{1/4} \,|\ft_N(y)|^{1/4}
                     \Ai(-\ft_N(y))                 \,dy\\
   &=\frac{c_N^{1/2}(-1)^N}{\tilde{\alpha}_N^{3/4}N^{1/2}}
                 \int_{u(x)}^{u(-1-\delta)} \Ai(u)\,L_N(y(u))\,du.
\ea
\ee

In what follows we use the standard estimates
(see \cite{Stegun})
\be\la{eqI.14.1}
\ba
  |\Ai(t)|&\leq Ce^{-(2/3)t^{3/2}},\qquad 
      |\Ai^\prime(t)|\leq  C(1+t^{1/4})e^{-(2/3)t^{3/2}},
                    \qquad t\geq0,\\
  |\Ai(t)|&\leq    C(1+|t|)^{-1/4}, \qquad 
         |\Ai^\prime(t)|\leq    C(1+|t|)^{1/4}, \qquad t<0
\ea
\ee
and
\be\la{eqI.14.2}
   \Big|\int_{-\infty}^t \Ai(s)\,ds\Big|
             \leq C (1+|t|)^{-3/4},\qquad t\leq0.
\ee
Note that by \eqref{eqI.11.3}, $u(y)>0$ for $y<-1$ and $u(y)<0$ for $y>-1$.
Consider first \eqref{eqI.13.1} for $x<-1$:
\be\la{eqI.14.3}
\ba
      \bigg|\frac{c_N^{1/2}}{\tilde{\alpha}_N^{3/4}N^{1/2}}
                 &(-1)^N\int_{u(x)}^{u(-1-\delta)} \Ai(u)\,L_N(y(u))\,du\bigg|\\
         &\leq \frac{c_N^{1/2}}{\tilde{\alpha}_N^{3/4}N^{1/2}}\const
                 \int_{0}^{\infty} |\Ai(u)|\,du = O(c_N^{1/2}N^{-1/2}).
\ea
\ee
Next we consider $x=-1$.
Note that $u(-1)=0$ and $\lim_{N\to\infty}u(-1-\delta)=+\infty$.
Also for any fixed $u\in[0,u(-1-\delta)]$,
$\lim_{N\to\infty}{L_N(y(u))}=L_N(-1)=2^{1/4}$
(see \eqref{eqI.12.1}, \eqref{eqE}, \eqref{eq25star},
 and also \eqref{eqalpha}).
We have
$$
\ba
 \int_0^{u(-1-\delta)} &\Ai(u)\,L_N(y(u))\,du
     = L_N(-1) \int_0^{u(-1-\delta)} \Ai(u)\,du\\
              &+\int_0^{u(-1-\delta)} \Ai(u)\,\big(L_N(y(u))-L_N(-1)\big)\,du
     \equiv Q_{0,1} + Q_{0,2}.
\ea
$$
Now by \eqref{eqI.12.1}
$$
\ba
   |Q_{0,2}|\leq &\int_0^{u(-1-\delta)} | \Ai(u)|
         \,\Big(\max_{-1-\delta\leq t\leq-1}|L_N^\prime(t)|\Big)
            \, |y(u)+1| \,du\\
      &\leq c \int_0^{u(-1-\delta)} | \Ai(u)|
         \,\frac{u}{|\tilde{\alpha}_NN^{2/3}\fth(y(u))|}\,du\\
    &\leq cN^{-2/3} \int_0^{\infty} | u\Ai(u)|\,du = O(N^{-2/3}).
\ea
$$
Also
$
  Q_{0,1} =  2^{1/4}\big(\int_0^{\infty}
               - \int_{u(-1-\delta)}^\infty \big) \Ai(u)\,du$.
Taking \eqref{eqI.11.3}, \eqref{eqI.14.1}, \eqref{4.12}(2)
 into account, we conclude that 
\be
\la{eqI.14.4}
\ba
      \frac{c_N^{1/2}(-1)^N}{\tilde{\alpha}_N^{3/4}N^{1/2}}
                 &\int_{0}^{u(-1-\delta)} \Ai(u)\,L_N(y(u))\,du\\
         &= \frac{c_N^{1/2}}{N^{1/2}}
              \frac{2^{1/4}}{(2m^{2/3})^{3/4}}
                 (-1)^N\Big(\int_{0}^{\infty} \Ai(u)\,du
            + O(N^{-2/3})+O(N^{-1/(2m)})+O(e^{-cN})\Big).
\ea
\ee

Now consider $-1<x<-1+\delta$. We have $u(x)<0$. 
Then \eqref{eqI.13.1} becomes
\be\la{eqI.15.0}
\ba
    \frac{c_N^{1/2}(-1)^N}{\tilde{\alpha}_N^{3/4}N^{1/2}}
                \bigg( \int_{0}^{u(-1-\delta)}+\int_{u(x)}^{0}\bigg)
                    \Ai(u)\,L_N(y(u))\,du
\ea
\ee
where the first integral was evaluated in \eqref{eqI.14.4},
and for the second we write
\be\la{eqI.15.1}
\ba
   \int_{u(x)}^0 \Ai(u)\,L_N(y(u))\,du 
   = &\Big(\int_{-\infty}^u \Ai(t)dt\Big)\,L_N(y(u))\bigg|_{u(x)}^0\\ 
         &-\int_{u(x)}^0 \Big(\int_{-\infty}^u \Ai(t)dt\Big)\,
               \frac{L_N^\prime(y(u))}{-\tilde{\alpha}_N N^{2/3} \tilde{U}_N(y(u))}\,du\\
   = &2^{1/4}\Big(\int_{-\infty}^0 \Ai(t)dt\Big)
 - \Big(\int_{-\infty}^{u(x)} \Ai(t)dt\Big)\,L_N(y(u(x)))\\
         &+O\bigg( \int_{u(x)}^0 \frac1{(1+|u|)^{3/4}}\,\frac{du}{N^{2/3}}\bigg)
\ea
\ee
by \eqref{eqI.14.2} and the uniform boundness of $L_N^\prime$.
From this we draw two conclusions.
 First, uniformly for $-1\leq x\leq -1+\delta$
\be\la{eqI.15.1prime}
   \int_{u(x)}^0 \Ai(u)\,L_N(y(u))\,du 
        = O(1), \qquad N\ra\infty,
\ee
and, second, again using \eqref{eqI.14.2},
 for any fixed $-1<x\leq-1+\delta$,
and in particular for $x=-1+\delta$,
\be\la{eqI.15.1bis}
\ba
   \int_{u(x)}^0 \Ai(u)\,L_N(y(u))\,du 
   = 2^{1/4}\Big(\int_{-\infty}^0 \Ai(t)dt\Big)
   +O(N^{-1/2})
\ea
\ee
since
\be\la{equinfty}
   u(x) = -\tilde{\alpha}_N N^{2/3} (x+1) \fth_N(x) \leq -C(x)\, N^{2/3}\to-\infty.
\ee
Hence we conclude that the uniform estimate in \eqref{eqI.16.1} holds.
Also recalling \eqref{eqI.14.4}, \eqref{eqalpha} we find
$$
\ba
    \frac{c_N^{1/2}(-1)^N}{\tilde{\alpha}_N^{3/4}N^{1/2}}
                2^{1/4}&\Big( \int_{0}^{+\infty}\Ai(t)\,dt 
            + \int_{-\infty}^{0}\Ai(t)\,dt + O(N^{-1/2}) \Big)\\
    &=\frac{c_N^{1/2}}{N^{1/2}}\,(2m)^{-1/2}\,
    \Big( (-1)^N +O(N^{-1/2}) +O(N^{-1/(2m)})\Big),
  \qquad N\to\infty,
\ea
$$
since $\int_{-\infty}^\infty\Ai(t)\,dt=1$ (see \cite{Stegun}),
 which proves the 
asymptotic formula in \eqref{eqI.16.1}.

The proof of \eqref{eqI.17.1} is similar.
\end{proof}
\subsubsection{Proof of Theorem \ref{thmthree}}
First we prove the second statement.
By the definition of $\eps$
\be\la{s5eqeps}
\ba
   \eps\phi_N(x) &= \frac12\Big[\int_{-\infty}^x\phi_N(y)\,dy
                       -\int_{x}^{+\infty}\phi_N(y)\,dy\Big]\\
     &= \frac12 \int_{-\infty}^{+\infty}\phi_N(y)\,dy
                       -\int_{x}^{+\infty}\phi_N(y)\,dy.
\ea
\ee
Making the change of variables $y\to{}c_Ny+d_N$, we see that we have
 to estimate 
\be\la{s5eq1}
       \frac12 c_N \int_{-\infty}^{+\infty}\phi_N(c_Ny+d_N)\,dy
\ee
and 
\be\la{s5eq1prim}
        c_N \int_{(x-d_N)/c_N}^{+\infty}\phi_N(c_Ny+d_N)\,dy
\ee
The second statement now follows 
from \eqref{eqI.19.2}, \eqref{eqI.19.1} above.

Now set $y=c_Nx+d_N$, i.e.{} $x=(y-d_N)/c_N$. 
Note that if $y\in[-a,a]$ where $a=o(c_N)$, then
 $x=o(1)$, and hence $\phi_N(y)=\phi_N(c_Nx+d_N)$
is given asymptotically by \eqref{eqI.9.1}. 
Thus $\|\phi_N(y)\|_{L^\infty([-a,a])}=O(c_N^{-1/2})$ 
which proves the first statement. 

Finally to prove the third statement (which is not used in this paper,
but is of independent interest)
we again introduce for $y$ the rescaled variable $x$,
$y=c_Nx+d_N$, and note that by \eqref{eqI7.1},
$$
      \sup_{|x|\geq1+\delta}|\phi_N(c_Nx+d_N)|
               =O(c_N^{-1/2}\,e^{-Nc_{3}}),
$$
and by \eqref{eqI.9.1}, 
$$
      \sup_{|x|\leq1-\delta}|\phi_N(c_Nx+d_N)| =O(c_N^{-1/2}).
$$
Thus $\|\phi_N\|_{L^\infty(\R)}$ is determined by the Airy regions.
Let us consider $-1-\delta<x<-1+\delta$ (the neighborhood of $x=1$ is treated
in the same way).
We refer to \eqref{eqI.8.0} and note that by \eqref{eqI.11.3}
 the $\Ai$ term is of order
\be\la{eqnotn1}
        \big| c_N^{-1/2} |x+1|^{-1/4} 
                        |\tilde{\alpha}_N N^{2/3}(x+1)\fth_N(x)|^{1/4} 
                       \,\Ai(u(x))(1+O(N^{-1})\big|
          = O(c_N^{-1/2}N^{1/6})
\ee
since $\Ai$ is bounded (and is not of a smaller order 
since e.g.~$\Ai(u(-1))=\Ai(0)\neq0$). 
Also by \eqref{eqC2} for the $\Ai^\prime$ term
\be\la{eqnotn2}
\ba
        \bigg| c_N^{-1/2} &|x+1|^{1/4} 
                        \frac{\Ai^\prime(-\ft_N(x))}{|\ft_N(x)|^{1/4}} 
                            \,(1+O(N^{-1})\bigg|\\
          &\leq C_2 c_N^{-1/2} |x+1|^{1/4} 
               \frac{(1+|\ft_N(x)|^{1/4})}{|\ft_N(x)|^{1/4}} \,(1+O(N^{-1})
              = O(c_N^{-1/2})
\ea
\ee
by \eqref{eqI.11.3}.
\begin{remark} 
Note that in \cite{Sz} the 
estimate $\|\phi_N\|_{L^\infty(\R)}=O(N^{-1/12})$ 
is proved for the case of Hermite polynomials,
$V(x)=x^2$. 
Of course in view of \eqref{cN} for $m=1$ this is consistent with
our estimate $O(c_N^{-1/2}N^{1/6})$. 
Also from our estimate and \eqref{cN} it follows that, 
for all $m\geq2$, $\|\phi_N\|_{L^\infty(\R)}$ blows up as $N^{1/6-1/(4m)}$,
as $N\to\infty$.
\end{remark}
\subsubsection{Proof of Theorem \ref{thm2}}
\la{sssecpfthm2}
From \eqref{Eas} we anticipate the leading order terms to be of order
$\frac{c_N}{N}\sim\frac{(c_Nc_P)^{1/2}}{(NP)^{1/2}}$.
Using \eqref{s5eqeps} we find
\be\la{s5eq29}
\ba
   (\phi_N,\eps\phi_P) = &\frac12 
            \Big(\int_{-\infty}^{+\infty}\phi_N(x)\,dx\Big)
            \Big(\int_{-\infty}^{+\infty}\phi_P(y)\,dy\Big)\\
     &- \int_{-\infty}^{+\infty}\phi_N(x)\,dx
                       \int_{x}^{+\infty}\phi_P(y)\,dy.
\ea
\ee
By \eqref{eqI.19.2}, as $N,P\too\infty$,
 the first term in \eqref{s5eq29} equals
\be\la{s5eq30}
\ba
     \frac12 \frac{(c_Nc_P)^{1/2}}{(NP)^{1/2}}
       \,\frac1{2m}\,&(1+(-1)^N+O(N^{-1/2})+O(N^{-1/(2m)}))\\
        &\times(1+(-1)^P+O(P^{-1/2})+O(P^{-1/(2m)})).
\ea
\ee
The estimation of the second term in \eqref{s5eq29} involves
an extended region-by-region calculation, and concludes eventually
at equation \eqref{eqN.69.1} below.

The following result will be used repeatedly throughout this Subsection.
\begin{lemma}
\la{lemprec2} 
For $V(x)=\kappa_{2m}x^{2m}+\cdots$, $\kappa_{2m}>0$
as before, as $N,P\too\infty$
\be\la{eqN.34.1}
        \frac{c_N}{c_P} - 1 = \frac1{2m}\frac{N-P}{P} + O(P^{-1-1/(2m)})
\ee
and 
\be\la{eqN.35.1}    
          \frac{d_N-d_P}{c_P} = O(P^{-1-1/m}).
\ee
\end{lemma}
\begin{proof}
Note first that for any fixed $\alpha>0$, 
as $N,P\too\infty$
\be\la{eqNPalpha}
\ba
        \frac1{N^{\alpha}}- \frac1{P^{\alpha}}
      =        \frac1{(P+(N-P))^{\alpha}}- \frac1{P^{\alpha}}
%\\ &= \frac1{P^\alpha}\bigg[\bigg(1+\frac{N-P}P\bigg)^{-\alpha} - 1\bigg]\\&
     = &\frac1{P^\alpha}\bigg[-\alpha\frac{N-P}P 
                 + O\bigg(\frac1{P^2}\bigg)\bigg]\\
         &=O(P^{-1-\alpha}).
\ea
\ee 
This together with \eqref{cN} for $q=2m$ implies
$$
\ba
    c_N - c_P =& c_{(-1)}\big[N^{1/(2m)}-P^{(1/2m)}\big]
              + \sum_{k=1}^{2m}\big[c_{(k)}\cdot 
                     (N^{-k/(2m)}-P^{-k/(2m)})\big] + O(P^{-1-1/(2m)})\\
             &= c_{(-1)}\big[N^{1/(2m)}-P^{(1/2m)}\big]
              + \sum_{k=1}^{2m}\big[c_{(k)}\cdot 
                     O(N^{-1-k/(2m)})\big] + O(P^{-1-1/(2m)}).
\ea
$$
Therefore by \eqref{eqNPalpha}
$$
\ba
    \frac{c_N-c_P}{c_P} 
            =  &\frac{c_{(-1)}\big[N^{1/(2m)}-P^{(1/2m)}\big] 
               + O(P^{-1-1/(2m)})}{c_{(-1)}P^{1/(2m)}(1+O(P^{-1/(2m)})} \\
             &= \big[ ((N/P)^{1/(2m)}-1) + O(P^{-1-1/m})\big]\,
                          (1+O(P^{-1/(2m)})) \\
             &= \Big[ \frac1{2m}\frac{N-P}{P} +O(P^{-2}) 
                                     + O(P^{-1-1/m})\Big]\,
                          (1+O(P^{-1/(2m)})) \\
             &= \frac1{2m}\frac{N-P}{P} +O(P^{-1-1/(2m)}).
\ea
$$
The second statement follows similarly from \eqref{eqNPalpha}
and \eqref{dN} with $q=2m$.
\end{proof}
We begin the estimation of the second term in \eqref{s5eq29} 
by making the
change of variables $x\to{}c_Nx+d_N$, $y\to{}c_Py+d_N$
in \eqref{s5eq29}
to find
\be\la{s5eq31}
\ba
   \int_{-\infty}^{+\infty}\phi_N(x)\,dx
                       \int_{x}^{+\infty}\phi_P(y)\,dy
         &=   c_Nc_P\int_{-\infty}^{+\infty}\phi_N(c_Nx+d_N)\,dx
                       \int_{X(x)}^{+\infty}   \phi_P(c_Py+d_P)\,dy\\
         &=   c_Nc_P \sum_{j=1}^5
          \int_{I_j} \phi_N(c_Nx+d_N)\,dx
                       \int_{X(x)}^{+\infty}   \phi_P(c_Py+d_P)\,dy
\ea
\ee
where
\be\la{eqN.1.1}
      X(x)\equiv x\frac{c_N}{c_P}+\frac{d_N-d_P}{c_P}.
\ee
Note that by Lemma \ref{lemprec2}
$$ 
          \frac{c_N}{c_P} = 1+ \frac{N-P}{2m}\frac1P+O(P^{-1-1/(2m)}) > 1,
                   \qquad  \frac{d_N-d_P}{c_P}=O(P^{-1-1/m})
$$
and so for any fixed $x_0>0$ (e.g.{} $x_0=1/2$) and $N,\,P$ large enough
\be\la{eqXx}
          X(x)>x\quad\textrm{for }x\geq x_0,
 \qquad \qquad         X(x)<x\quad\textrm{for }x\leq- x_0.
\ee
Now \eqref{eqI.10.1}, \eqref{eqI.10.1prime}, \eqref{eqI.19.1} imply
\be\la{eqN.2.1}
\ba
   c_Nc_P &\bigg|
          \Big(\int_{I_1}+\int_{I_5}\Big) \phi_N(c_Nx+d_N)\,dx
                       \int_{X(x)}^{+\infty}   \phi_P(c_Py+d_P)\,dy \bigg|\\
   &\leq c_Nc_P \bigg(
          \Big(\int_{I_1}+\int_{I_5}\Big) \big|\phi_N(c_Nx+d_N)\big|\,dx\bigg)
                       \cdot\sup_{x\in\R}\bigg|\int_{x}^{+\infty}   \phi_P(c_Py+d_P)\,dy\bigg|\\
   &=O\big( (c_Nc_P)^{1/2}(NP)^{-1/2}\,e^{-Nc_{3}}\big).
\ea
\ee
Hence in \eqref{s5eq31} we are left with three types of integrals
\be\la{eqN.2.2}
\ba
  c_Nc_P \bigg(&\int_{I_4}+\int_{I_3}+\int_{I_2}\bigg)
           \phi_N(c_Nx+d_N)\,dx
                       \int_{X(x)}^{+\infty}   \phi_P(c_Py+d_P)\,dy\\
          &\equiv J_4+J_3+J_2.
\ea
\ee
By \eqref{eqXx}, uniformly for $x\in{}I_4$, we have
 $X(x)<x$ for $N$, $P$ sufficiently large. 
So
\be\la{eqN.3.2}
\ba
  J_4 = c_Nc_P  &\int_{I_4}
           \phi_N(c_Nx+d_N)\,dx
                     \bigg(\int_{X(x)}^{-1+\delta}
                  +\int_{I_3}+\int_{I_2}+\int_{I_1}\bigg)  \phi_P(c_Py+d_P)\,dy\\
          &\equiv J_{44}+J_{43}+J_{42}+J_{41}.
\ea
\ee
Now $J_{43}$, $J_{42}$, $J_{41}$ are products of \eqref{eqI.16.1} with
 \eqref{eqI.18.1}, \eqref{eqI.17.1}, \eqref{eqI.10.1} respectively,
 and hence 
\be\la{eqN.3.3}
    J_{42} = \frac{(-1)^N}{2m} \frac{(c_Nc_P)^{1/2}}{(NP)^{1/2}}
       \,(1+O(N^{-1/2})+O(N^{-1/(2m)}))
\ee
and
\be\la{eqN.3.4}
    J_{43} + J_{41}  = 
  \frac{(c_Nc_P)^{1/2}}{(NP)^{1/2}}
       \,O\big( P^{-1/2} +
        \,e^{-Pc_{3}}\big)  =   
   \frac{(c_Nc_P)^{1/2}}{(NP)^{1/2}}
       \,O\big( P^{-1/2} \big).
\ee
Next define $x_{N,P}\in{}I_4$ such that 
\be\la{eqN.3.1}
    X(x_{N,P}) \equiv  x_{N,P}\frac{c_N}{c_P}+\frac{d_N-d_P}{c_P} 
                        = -1-\delta.
\ee
By Lemma \ref{lemprec2}, we see for $N,P\too\infty$
\be\la{eq4.55prime}
   x_{N,P} = -(1+\delta)\Big( 1-\frac1{2m}\frac{N-P}{P}\Big)
               + O(P^{-1-1/(2m)})
\ee
and so $x_{N,P}$ indeed lies in $I_4=(-1-\delta,-1+\delta)$.
With this notation
\be\la{eqN.4.1}
\ba
  J_{44} =  \bigg(&\int_{-1-\delta}^{x_{N,P}}
               +\int_{x_{N,P}}^{-1+\delta}
           c_N\phi_N(c_Nx+d_N)\,dx
                   \Big(  \int_{X(x)}^{-1+\delta} c_P \phi_P(c_Py+d_P)\,dy\Big)\,dx\\
          &\equiv J_{44}^{\prime}+J_{44}^{\prime\prime}.
\ea
\ee
Observe that in $J_{44}^{\prime\prime}$, $-1-\delta<X(x)<-1+\delta$
and so all the points $y$ in the inner integral lie in $I_4$.

We need the bounds 
\be\la{eqN.4.abs}
           \sup_{\min(|x+1|,|x-1|)\geq\delta} \big|\phi_N(c_Nx+d_N)\big|
                    \leq C\,c_N^{-1/2}
\ee
and
\be\la{eqN.4.2}
           \sup_{{|x\pm1|\leq\delta}} \big|\phi_N(c_Nx+d_N)\big|
                    \leq \frac{C}{c_N^{1/2}}
                            \frac{N^{1/6}}{(1+|x\pm1|N^{2/3})^{1/4}}
\ee
which were almost proved in \eqref{eqnotn1}, \eqref{eqnotn2}.
Indeed, the two inequalities preceding \eqref{eqnotn1}
imply \eqref{eqN.4.abs}, and as
$$
        |\Ai(f_N(x))|\leq C\cdot(1+|f_N(x)|)^{-1/4},\qquad 
           |\Ai^\prime(f_N(x))|\leq C\cdot(1+|f_N(x)|)^{1/4},
$$
we have for $|x-1|\leq\delta$, by \eqref{eqI8.1},
\be\la{eqN.4.Ai}
\ba
       \frac{|f_N(x)|^{1/4}|\Ai(f_N(x))|}{|x-1|^{1/4}} 
            &\leq \frac{C|f_N(x)|^{1/4}}{|x-1|^{1/4}(1+|f_N(x)|)^{1/4}}
             \leq \frac{C(N^{2/3})^{1/4}} {(1+|x-1|N^{2/3})^{1/4}}\\
    \frac{|x-1|^{1/4}|\Ai^\prime(f_N(x))|}{|f_N(x)|^{1/4}} 
           &\leq \frac{C(1+|x-1|N^{2/3})^{1/4}}{N^{1/6}}
             \leq \const.
\ea
\ee
Inserting \eqref{eqN.4.Ai} and the analogous inequalities 
for $|x+1|\leq\delta$ into \eqref{eqI2reg}, we obtain \eqref{eqN.4.2}.

Now by \eqref{eqI.19.1}, \eqref{eqN.4.2}
$$%\be\la{eqN.2.1}
\ba
   \big| J_{44}^{\prime} \big|
        \leq c_N &\int_{-1-\delta}^{x_{N,P}}
        \frac{C\, N^{1/6}}{c_N^{1/2}}
                \frac{C\, c_P^{1/2}}{P^{1/2}}\,dx\\
   &= \const\cdot \frac{(c_Nc_P)^{1/2}}{P^{1/2}}\,N^{1/6}
            \,\big(x_{N,P}+(1+\delta)\big).
\ea
$$%\ee
But by \eqref{eq4.55prime}
$$
    x_{N,P}+1+\delta = \frac{d_P-d_N}{c_P}
    + x_{N,P}\,\frac{c_P-c_N}{c_P} = O(P^{-1}).
$$
Thus
\be\la{eqN.5.1}
    \big| J_{44}^{\prime} \big|
       \leq \const\cdot \frac{(c_Nc_P)^{1/2}}{P^{1/2}}\,N^{1/6}P^{-1}
                =  \frac{(c_Nc_P)^{1/2}}{(NP)^{1/2}}\,O\big(N^{-1/3}\big).
\ee
Next
\be\la{eqN.6.0}
\ba
    J_{44}^{\prime\prime} =
  (-1)^{N}c_N^{1/2} \int_{x_{N,P}}^{-1+\delta}
       \bigg(
          &\Big|\frac{x-1}{x+1}\Big|^{1/4} |\ft_N(x)|^{1/4} \Ai(-\ft_N(x))
                            \,\Big(1+O\Big(\frac1N\Big)\Big)\\
         &\quad- \Big|\frac{x+1}{x-1}\Big|^{1/4} \frac1{|\ft_N(x)|^{1/4}} 
                    \Ai^\prime(-\ft_N(x))
                            \,\Big(1+O\Big(\frac1N\Big)\Big)
       \bigg)\,dx\\
         &\times\int_{X(x)}^{-1+\delta} c_P\phi_P(c_py+d_P)\,dy
\ea
\ee
We consider first the two $O(N^{-1})$ terms.
As 
\be\la{eqN.6.1}
   \big| |\ft_N(x)|^{1/4} \Ai(-\ft_N(x))\big|\leq C^\prime
\ee
we have
\be\la{eq.31.1}
\ba
   c_N^{1/2} \int_{x_{N,P}}^{-1+\delta}
          &\Big|\frac{x-1}{x+1}\Big|^{1/4} |\ft_N(x)|^{1/4} |\Ai(-\ft_N(x))|
                       \,O\Big(\frac1N\Big)\,dx\\
         &\quad\times\sup_{x\in{\R}}\bigg|\int_{X(x)}^{-1+\delta} 
                     c_P\phi_P(c_py+d_P)\,dy\bigg|\\
           &\leq \const\,C^\prime\frac{(c_Nc_P)^{1/2}}{P^{1/2}} O(N^{-1})
            \int_{-1-\delta}^{-1+\delta} \Big|\frac{x-1}{x+1}\Big|^{1/4}\,dx = 
                \frac{(c_Nc_P)^{1/2}}{(NP)^{1/2}}O(N^{-1/2}).
\ea
\ee
Also by the analog of the
 second inequality in \eqref{eqN.4.Ai} for $|x+1|\leq\delta$,
\be\la{eqN.7.00}
            \frac{|x+1|^{1/4}}{|\ft_N(x)|^{1/4}}
                   \big|\Ai^\prime(-\ft_N(x))\big|\leq C^\prime
\ee
and hence
\be\la{eq.32.1}
\ba
   c_N^{1/2} \int_{x_{N,P}}^{-1+\delta}
          &\Big|\frac{x+1}{x-1}\Big|^{1/4} \frac1{|\ft_N(x)|^{1/4}} |\Ai^\prime(-\ft_N(x))|
                       \,O\Big(\frac1N\Big)\,dx\\
         &\quad\times\sup_{x\in{\R}}\bigg|\int_{X(x)}^{-1+\delta} 
                     c_P\phi_P(c_py+d_P)\,dy\bigg|\\
           &\leq \const\,C^\prime \frac{(c_Nc_P)^{1/2}}{P^{1/2}} O(N^{-1})
            \int_{-1-\delta}^{-1+\delta} \frac1{|x+1|^{1/4}}\,dx = 
                \frac{(c_Nc_P)^{1/2}}{(NP)^{1/2}}O(N^{-1/2}).
\ea
\ee
Now we substitute \eqref{eqI.8.0}
for $\phi_P(c_P y+d_P)$ in \eqref{eqN.6.0}.
Using an obvious schematic notation we 
note that, in view of \eqref{eq.31.1}, \eqref{eq.32.1},
we have shown
\be\la{eqex}
\ba
  J_{44}^{\prime\prime}
 =    &\int_{x_{N,P}}^{-1+\delta}\big( \Ai + \Ai^\prime\big) \,dx 
        \int_{X(x)}^{-1+\delta}
               \big(\Ai + \Ai^\prime\big)\,dy \\
 &+\int_{x_{N,P}}^{-1+\delta}\big( \Ai + \Ai^\prime\big) \,dx 
        \int_{X(x)}^{-1+\delta}
               \big(\Ai\cdot O(P^{-1}) + \Ai^\prime\cdot O(P^{-1})\big)\,dy \\
          &+\frac{(c_Nc_P)^{1/2}}{(NP)^{1/2}}O(N^{-1/2}),\qquad
     N,P\too\infty.
\ea
\ee
To estimate the second integral in \eqref{eqex}
% two $O(P^{-1})$ terms in the integral
we interchange the order of integration.
Set $y_0\equiv X(-1+\delta)\in (-1-\delta,-1+\delta)$
and note that $X(x)$ is a 1-1 function from
$[x_{N,P},-1+\delta]$ onto $[-1-\delta,y_0]$.
We conclude that the second integral in \eqref{eqex} takes the form
\be\la{eqN.16.extra}
\ba
  (-1)^{N+P}(c_Nc_P)^{1/2} 
           \int_{-1-\delta}^{y_0}
       \bigg(
          &\Big|\frac{y-1}{y+1}\Big|^{1/4} |\ft_P(y)|^{1/4} \Ai(-\ft_P(y))
                            \,O\Big(\frac1P\Big)\\
         &\quad- \Big|\frac{y+1}{y-1}\Big|^{1/4} \frac1{|\ft_P(y)|^{1/4}} 
                    \Ai^\prime(-\ft_P(y))
                            \,O\Big(\frac1P\Big)
       \bigg)\,dy\\
   \times \int_{x_{N,P}}^{X^{-1}(y)}
       \bigg(
          &\Big|\frac{x-1}{x+1}\Big|^{1/4} |\ft_N(x)|^{1/4} \Ai(-\ft_N(x))
                            \\
         &\quad- \Big|\frac{x+1}{x-1}\Big|^{1/4} \frac1{|\ft_N(x)|^{1/4}} 
                    \Ai^\prime(-\ft_N(x))
                            \bigg)\,dx\\
  +(-1)^{N+P}(c_Nc_P)^{1/2} 
           \int_{y_0}^{-1+\delta}
       \bigg(
          &\Big|\frac{y-1}{y+1}\Big|^{1/4} |\ft_P(y)|^{1/4} \Ai(-\ft_P(y))
                            \,O\Big(\frac1P\Big)\\
         &\quad- \Big|\frac{y+1}{y-1}\Big|^{1/4} \frac1{|\ft_P(y)|^{1/4}} 
                    \Ai^\prime(-\ft_P(y))
                            \,O\Big(\frac1P\Big)
       \bigg)\,dy\\
   \times \int_{x_{N,P}}^{-1+\delta}
       \bigg(
          &\Big|\frac{x-1}{x+1}\Big|^{1/4} |\ft_N(x)|^{1/4} \Ai(-\ft_N(x))
                            \\
         &\quad- \Big|\frac{x+1}{x-1}\Big|^{1/4} \frac1{|\ft_N(x)|^{1/4}} 
                    \Ai^\prime(-\ft_N(x))
                            \bigg)\,dx.
\ea
\ee
Note that it follows from \eqref{eqI.13.0}, 
\eqref{eqI.14.3}, \eqref{eqI.14.4}, \eqref{eqI.15.0}, 
 \eqref{eqI.15.1prime} that,
uniformly for all intervals $K\subset[-1-\delta,-1+\delta]$,
$$
\ba
   \bigg| \int_K
           \Big|\frac{y+1}{y-1}\Big|^{1/4} 
                     \frac{\Ai^\prime(-\ft_N(y))}{|\ft_N(y)|^{1/4}}
                 \,dy\bigg| &\leq C\,N^{-5/6}\\
   \bigg| \int_K
           \Big|\frac{y-1}{y+1}\Big|^{1/4} 
                     \,|\ft_N(y)|^{1/4}\,\Ai(-\ft_N(y))
                 \,dy\bigg| &\leq C\,N^{-1/2}.
\ea
$$
This together with \eqref{eqN.6.1}, \eqref{eqN.7.00}
implies that \eqref{eqN.16.extra} is of order
\be\la{eq.32.2}
          (c_Nc_P)^{1/2} P^{-1} N^{-1/2} \int_{-1-\delta}^{-1+\delta} \frac{dx}{|x+1|^{1/4}}
       =   \frac{(c_Nc_P)^{1/2}}{(NP)^{1/2}}\, O(N^{-1/2}).
\ee
Thus (cf.{} \eqref{eqex}) we have shown (again schematically)
\be
\la{eqN.17.1}
\ba
  J_{44}^{\prime\prime}
 =    &\int_{x_{N,P}}^{-1+\delta}\big( \Ai + \Ai^\prime\big) \,dx 
        \int_{X(x)}^{-1+\delta}
               \big(\Ai + \Ai^\prime\big)\,dy \\
 &+\frac{(c_Nc_P)^{1/2}}{(NP)^{1/2}}O(N^{-1/2}),\qquad
     N,P\too\infty.
\ea
\ee

Now we  consider the terms of the form $\Ai^\prime\times\Ai^\prime$. 
After changing variables
\be\la{eqchv}
         v=-\ft_P(y)=-\tilde{\alpha}_P P^{2/3} (y+1)\fth_P(y), 
        \qquad  u=-\ft_N(x)
\ee
and recalling \eqref{eqI.11.3}, \eqref{eqI.12.0}, \eqref{eqI.12.1} 
and \eqref{eqBPBN},
we find that the $\Ai^\prime\times\Ai^\prime$ term equals
 $(-1)^{N+P}(c_Nc_P)^{1/2}$ times
\be\la{eqN.19.1}
\ba
             \int_{x_{N,P}}^{-1+\delta}
  &\Big|\frac{x+1}{x-1}\Big|^{1/4} \frac1{|\ft_N(x)|^{1/4}} 
                    \Ai^\prime(-\ft_N(x))\,dx\\
   &\qquad\qquad\qquad\times\int_{X(x)}^{-1+\delta}
        \Big|\frac{y+1}{y-1}\Big|^{1/4} \frac1{|\ft_P(y)|^{1/4}} 
                    \Ai^\prime(-\ft_P(y))
   \,dy\\   
       &=  \frac{1}{(\tilde{\alpha}_N\tilde{\alpha}_P)^{5/4}}
           \frac{1}{(NP)^{5/6}} \int_{u(-1+\delta)}^{u(x_{N,P})}
                                                      B_N(x(u))\, \Ai^\prime(u)\,du
 \\  &\qquad\qquad\qquad\times          
\int_{v(-1+\delta)}^{v(X(x(u)))}
                                     B_P(y(v))\, \Ai^\prime(v) \,dv\\
     &=  \frac{1}{(\tilde{\alpha}_N\tilde{\alpha}_P)^{5/4}}
           \frac{1}{(NP)^{5/6}} \int_{u(-1+\delta)}^{u(x_{N,P})}
   B_N(x(u))\, \Ai^\prime(u)\,du\\
   &\qquad\qquad\qquad\times  \bigg(
         B_P(y(v(X(x(u)))))    \Ai(v(X(x(u)))) \\
            &\qquad\qquad\qquad\qquad-    B_P(y(v(-1+\delta)))    \Ai(v(-1+\delta))\\
               &\qquad\qquad\qquad\qquad+\frac1{\tilde{\alpha}_P P^{2/3}} 
                   \int_{v(-1+\delta)}^{v(X(x(u)))}
        \frac{B_P^\prime(y(v))\, \Ai(v)}{\tilde{U}_P(y(v))} \,dv
 \bigg)\\
  &\equiv \spIon+\spIIon+\spIIIon.
\ea
\ee
Note from \eqref{eq4.55prime} that $x_{N,P}+1=-\delta+o(1)$
as $N,P\too\infty$, and hence by \eqref{eqI.11.3}
\be\la{eq4.70prime}
 u(x_{N,P})\to+\infty,\qquad u(x_{N,P})=O(N^{2/3}).
\ee
Now
$$
\ba
   \spIIon =  -  &\frac{1}{(\tilde{\alpha}_N\tilde{\alpha}_P)^{5/4}}
           \frac{1}{(NP)^{5/6}}
         B_P(y(v(-1+\delta)))    \Ai(v(-1+\delta))\\
   &\qquad\qquad\qquad\times  \bigg(
         B_N(x(u))    \Ai(u)\bigg|_{u(-1+\delta)}^{u(x_{N,P})}\\
               &\qquad\qquad\qquad+\frac1{\tilde{\alpha}_N N^{2/3}} 
                   \int_{u(-1+\delta)}^{u(x_{N,P})}
        \frac{B_N^\prime(x(u))\, \Ai(u)}{\tilde{U}_N(x(u))} \,du
 \bigg)\\
  &\leq\frac{\const}{(NP)^{5/6}}
\ea
$$
since the boundary term and the integrand are 
uniformly bounded and also the length of the interval of integration 
is $O(N^{2/3})$. 
Hence 
\be\la{eqN.20.1}
       |\spIIon|\leq\frac\const{(NP)^{1/2}}\,O(N^{-2/3}).
\ee
 
In $\spIIIon$ we integrate $\Ai^\prime(u)$ 
and differentiate the other factors
to obtain
\be\la{eq19.10}
\ba
    \frac{1}{(\tilde{\alpha}_N\tilde{\alpha}_P)^{5/4}}
           &\frac{1}{(NP)^{5/6}}\frac{1}{\tilde{\alpha}_P P^{2/3}}
   \bigg[   \Big(    B_N(x(u))\, \Ai(u)
                           \int_{v(-1+\delta)}^{v(X(x(u)))}
        \frac{B_P^\prime(y(v))\, \Ai(v)}{\tilde{U}_P(y(v))} 
              \,dv\Big)     \bigg|_{u(-1+\delta)}^{u(x_{N,P})}\\
    &- \int_{u(-1+\delta)}^{u(x_{N,P})}
                   \Ai(u)\,\frac{ B_N^\prime(x(u))}{-\tilde{\alpha}_N N^{2/3} \tilde{U}_N(x(u))}\,du
%\\   &\qquad\qquad\qquad\times  
                \int_{v(-1+\delta)}^{v(X(x(u)))}
                              \frac{B_P^\prime(y(v))\, \Ai(v)}{\tilde{U}_P(y(v))} \,dv\\
    &- \int_{u(-1+\delta)}^{u(x_{N,P})}
          \Ai(u)\,B_N(x(u))\,\frac{ B_P^\prime(X(x(u))) \Ai(v(X(x(u))))}{\tilde{U}_P(X(x(u)))}\\
   &\qquad\qquad\qquad\times  
                    \frac{d}{du}\big[v(X(x(u)))\big]\,du
 \bigg] \equiv \spIIIon^\prime+
                     \spIIIon^{\prime\prime}+\spIIIon^{\prime\prime\prime}
\ea
\ee
where we have used 
\be\la{eqid}
              y\circ v = \textrm{id}.
\ee
Set 
\be\la{eqbeta}
         \beta(u)\equiv v(X(x(u))).
\ee
\begin{lemma}
\la{propbeta}
Uniformly for $u(-1+\delta)<u<u(-1-\delta)$,
we have as $N,P\too\infty$
\be\la{eqN.21.pp.1}
    \beta(u) = O(P^{-1/3}) + (1+O(P^{-1}))\,u
\ee
and
\be\la{eqN.21.p.1}
    \frac{d}{du}\beta(u) = 1+O\Big(\frac1P\Big).
\ee
\end{lemma}
\begin{proof} 
We have  in view of \eqref{eqI.12.0}, \eqref{eqN.1.1}
$$
\ba
   \frac{d}{du}\beta(u) &= v^\prime(X(x(u)))\,X^{\prime}(x(u))\,x^\prime(u) \\
              &=\frac{P^{2/3}}{N^{2/3}}
                     \frac{\tilde{\alpha}_P}{\tilde{\alpha}_N}\,
                     \frac{c_N}{c_P}\,
                       \frac{\tilde{U}_P(X(x(u)))}{\tilde{U}_N(x(u))}\\
              &=\frac{P^{2/3}}{N^{2/3}}
                     \frac{\tilde{\alpha}_P}{\tilde{\alpha}_N}
                     \frac{c_N}{c_P}
                     \Bigg( \frac{\tilde{U}_P(x(u))}{\tilde{U}_N(x(u))} + 
                        \frac{\tilde{U}_P(X(x(u)))- \tilde{U}_P(x(u))}{\tilde{U}_N(x(u))}\bigg).
\ea
$$
By Lemma \ref{lemprec2}, $\frac{c_N}{c_P}-1=O(P^{-1})$,
and also
$X(x(u))-x(u)=(\frac{c_N}{c_P}-1)x(u)+(d_N-d_P)/c_P=O(P^{-1})
$
since $x(u)\in[-1-\delta,-1+\delta]$ is uniformly bounded.
Also by \eqref{4.12}
$\max_{u\in[u(-1+\delta),u(-1-\delta)]}\tilde{U}_P^\prime(u)\leq\const$.
In addition, $\tilde{\alpha}_N$ also has a complete expansion in powers
of $N^{-1/(2m)}$ (see \eqref{4.12}),
and therefore by the same argument as in the proof of Lemma \ref{lemprec2},
$\frac{\tilde{\alpha}_P}{\tilde{\alpha}_N}=1+O(P^{-1})$, as $N,P\too\infty$. 
Finally as $\tilde{U}_N(x)$ has a complete expansion (see \eqref{4.12})
in powers of $N^{-1/(2m)}$, uniformly for $-1-\delta\leq x\leq-1+\delta$,
we again find $\frac{\tilde{U}_P(x(u))}{\tilde{U}_N(x(u))}=1+O(P^{-1})$.
Thus
$$
\ba
   \frac{d}{du}\beta(u) &=
         \big(1+O(P^{-1})\big)\,\big(1+O(P^{-1})\big)\,
          \big(1+O(P^{-1})\big)\,\big(1+O(P^{-1})\big)\\
     &=1+O(P^{-1})
\ea
$$
uniformly for $u(-1+\delta)\leq u\leq u(-1-\delta)$
 which proves \eqref{eqN.21.p.1}.

Finally by \eqref{eqN.21.p.1}
$$
\ba
       \beta(u) = \beta(0) + \int_0^u \beta^\prime(t)\,dt 
              = \beta(0) + u\cdot\big(1+O(P^{-1})\big)
\ea
$$
which together with
$$
\ba
       \beta(0) &= v(X(-1)) = v\big(-c_N/c_P + (d_N-d_P)/c_P\big)
               = v\big(-1 + O(P^{-1})\big)\\
              &= -\tilde{\alpha}_P P^{2/3} \big(1 - 1 + O(P^{-1})\big)
                        \,\fth_P\big(-1 + O(P^{-1})\big) = O(P^{-1/3})
\ea
$$
proves \eqref{eqN.21.pp.1}.
\end{proof}
Now we use the uniform boundedness of the functions $\Ai$,
$B_N$, $B_N^\prime$, $\frac1{\tilde{U}_N}$, and also \eqref{eqN.21.p.1},
to conclude the following. 
Note that the second estimate in \eqref{eqI.14.1} 
holds on the whole of $\R$ and that the lengths of the intervals of integration
in \eqref{eq19.10} are of order $O(N^{2/3})$.
 Hence the integral
in $\spIIIon^\prime$ (and also the whole boundary term)
is bounded by $O((N^{2/3})^{3/4})=O(N^{1/2})$.
The term $\spIIIon^{\prime\prime}$
is estimated in a similar way.
Finally, in $\spIIIon^{\prime\prime\prime}$ we just estimate the integrand
by a constant (note \eqref{eqN.21.p.1}). 
We then obtain
\be\la{eqN.21.2}
\ba
      |\spIIIon^\prime| &\leq 
               \frac{\const}{P^{7/3}}\,N^{1/2} = \frac1{(NP)^{1/2}}\,O(P^{-5/6})\\
      |\spIIIon^{\prime\prime}| &\leq 
               \frac{\const}{P^{7/3}}\frac1{N^{2/3}}
                \,(N^{2/3})^{3/4}\,(N^{2/3})^{3/4} 
           = \frac1{(NP)^{1/2}}\,O(P^{-1})\\
      |\spIIIon^{\prime\prime\prime}| &\leq 
               \frac{\const}{P^{7/3}}\,N^{2/3} = \frac1{(NP)^{1/2}}\,O(P^{-2/3}).
\ea
\ee
Finally we consider the term $\spIon$
in \eqref{eqN.19.1}. Using \eqref{eqid}, \eqref{eqbeta}
we rewrite $\spIon$ as
$$
\ba
    \frac{1}{(\tilde{\alpha}_N\tilde{\alpha}_P)^{5/4}}
           \frac{1}{(NP)^{5/6}} \bigg(
        &\int_{u(-1+\delta)}^{-1}
               +\int_{-1}^{u(x_{N,P})}\bigg)
                                                      B_N(x(u))\, \Ai^\prime(u)
 \\  &\qquad\qquad\qquad\times          
    B_P(X(x(u)))    \Ai(\beta(u))\,du\\
\equiv \spIon^\prime + \spIon^{\prime\prime}.
\ea
$$
Now in $\spIon^{\prime\prime}$, 
$u(x_{N,P})\to+\infty$, and we use the estimate for $\Ai^\prime$
in  \eqref{eqI.14.1} (which clearly also holds on $[-1,+\infty)$) 
together with the boundedness of the other factors to find
\be\la{eq4.77prime}
      |\spIon^{\prime\prime}|\leq \frac{\const}{(NP)^{5/6}} 
         = \frac{1}{(NP)^{1/2}} O(N^{-2/3}). 
\ee
Note that by \eqref{eqN.21.pp.1}, for $P$ and $N$ large enough,
\be\la{eqbeta1}
          C^{-1} \leq \frac{u}{\beta(u)} \leq C
                       \qquad\textrm{and}\qquad \beta(u)\leq-1/2
\ee
uniformly for $u(-1+\delta)<u<-1$.
Hence using \eqref{eqI.14.1},
and recalling 
that $B_N$, $B_N^\prime$ are uniformly bounded, 
and also using the properties of $\fth_N$ after \eqref{4.12} (cf. \eqref{equinfty}),
we find
\be\la{eqN.22.1}
    |\spIon^\prime| \leq \frac{\const}{(NP)^{5/6}}
              \int_{-\delta N^{2/3} \fth_N(-1+\delta)}^{-1} 
                 \bigg| \frac{u}{\beta(u)}\bigg|^{1/4}\,du
        \leq \const \frac\delta{(NP)^{1/2}}. % + \frac1{(NP)^{1/2}}\,O(P^{-2/3}).
\ee
We will see below
that it is important that we can choose $\delta$ to be arbitrarily small.
Collecting the estimates 
\eqref{eq4.77prime}, \eqref{eqN.22.1}, 
\eqref{eqN.20.1}, \eqref{eqN.21.2} in \eqref{eqN.19.1}
and recalling  \eqref{eqN.17.1} we conclude (schematically)
 \be
\la{eqN.17.1NEW}
\ba
  J_{44}^{\prime\prime}
 =    &\int_{x_{N,P}}^{-1+\delta} \Ai \,dx 
        \int_{X(x)}^{-1+\delta}
               \Ai \,dy \\
   &+ \int_{x_{N,P}}^{-1+\delta} \Ai \,dx 
        \int_{X(x)}^{-1+\delta}
                \Ai^\prime \,dy + \int_{x_{N,P}}^{-1+\delta} \Ai^\prime \,dx 
        \int_{X(x)}^{-1+\delta}
                \Ai \,dy \\
 &+\frac{(c_Nc_P)^{1/2}}{(NP)^{1/2}}O(\delta + N^{-1/2}),\qquad
     N,P\too\infty.
\ea
\ee

Now we  consider the term of the form $\Ai\times\Ai^\prime$
 in \eqref{eqN.17.1NEW}.
After changing variables as in \eqref{eqchv} and 
using \eqref{eqid}, we find that the $\Ai\times\Ai^\prime$ term 
equals $(-1)^{N+P}(c_Nc_P)^{1/2}$ times 
\be\la{eqN.23.1}
\ba
             \int_{x_{N,P}}^{-1+\delta}
  &\Big|\frac{x-1}{x+1}\Big|^{1/4}\,|\ft_N(x)|^{1/4}\, 
                    \Ai(-\ft_N(x))\,dx\\
   &\qquad\qquad\qquad\times\int_{X(x)}^{-1+\delta}
        \Big|\frac{y+1}{y-1}\Big|^{1/4} \frac1{|\ft_P(y)|^{1/4}} 
                    \Ai^\prime(-\ft_P(y))
   \,dy\\   
       &=  \frac{1}{\tilde{\alpha}_N^{3/4}\tilde{\alpha}_P^{5/4}}
           \frac{1}{N^{1/2}P^{5/6}} 
                       \int_{u(-1+\delta)}^{u(x_{N,P})}
                                  L_N(x(u))\, \Ai(u)\,du
 \\  &\qquad\qquad\qquad\times          
\int_{v(-1+\delta)}^{v(X(x(u)))}
                                     B_P(y(v))\, \Ai^\prime(v) \,dv\\
     &=    \frac{1}{\tilde{\alpha}_N^{3/4}\tilde{\alpha}_P^{5/4}}
           \frac{1}{N^{1/2}P^{5/6}} 
                         \int_{u(-1+\delta)}^{u(x_{N,P})}
                                   L_N(x(u))\, \Ai(u)\,du\\
   &\qquad\qquad\qquad\times  \bigg(
         B_P(X(x(u)))    \Ai(v(X(x(u)))) \\
            &\qquad\qquad\qquad\qquad-    B_P(-1+\delta)    \Ai(v(-1+\delta))\\
               &\qquad\qquad\qquad\qquad+\frac1{\tilde{\alpha}_P P^{2/3}} 
                   \int_{v(-1+\delta)}^{v(X(x(u)))}
        \frac{B_P^\prime(y(v))\, \Ai(v)}{\tilde{U}_P(y(v))} \,dv
 \bigg)\\
  &\equiv \spItt+\spIItt+\spIIItt.
\ea
\ee
In $\spIIItt$ we split both integrals into the parts
with positive and negative arguments.
Using \eqref{eqI.14.1} and \eqref{4.12} we find 
\be\la{eqN.24.1}
\ba
  |\spIIItt| \leq
       &\frac{\const}{N^{1/2}P^{5/6}}\frac1{P^{2/3}} 
                         \Big( \const + \int_{u(-1+\delta)}^0
               \frac{du}{(1+|u|)^{1/4}} \Big)\\
   &\qquad\qquad\qquad\times  
              \Big(    \const + \int_{v(-1+\delta)}^0
              \frac{dv}{(1+|v|)^{1/4}} \Big) \\
  & \leq     \frac{\const}{N^{1/2}P^{3/2}}
             \big(C+ (\delta N^{2/3})^{3/4}\big)
                 \big(C+ (\delta P^{2/3})^{3/4}\big)
\\ & 
   \leq \frac{1}{(NP)^{1/2}} \,O\big(\delta^{3/2} + P^{-1}\big).
\ea
\ee
Next by \eqref{eqid}
\be\la{eqN.24.2}
\ba
  |\spIItt| \leq
       &\frac{C}{N^{1/2}P^{5/6}}
                       \bigg(\int_{u(-1+\delta)}^{u(x_{N,P})}
                                   L_N(x(u)\, \Ai(u)\,du\bigg)
%\\   &\qquad\qquad\qquad\times 
         \,     B_P(-1+\delta)    \Ai(v(-1+\delta))\\
    & \leq \frac{C}{(NP)^{2/3}}
                       \bigg(\const + \int_{u(-1+\delta)}^{-1}
                                   L_N(x(u)\, \Ai(u)\,du\bigg)\,O(1)\\
   &=  \frac{1}{(NP)^{2/3}}\,O(1)
                       \bigg[\const + 
              \Big(\int_{-\infty}^u \Ai(t)dt\Big)\,
                       L_N(x(u))\bigg|_{u(-1+\delta)}^{-1}\\
                  &\qquad\qquad\qquad+\frac1{\tilde{\alpha}_N N^{2/3}}
                           \int_{u(-1+\delta)}^{-1}
                       \Big(\int_{-\infty}^u \Ai(t)dt\Big)\,        
                           \frac{ L_N^\prime(x(u))}{\tilde{U}_N(x(u))}\,du  \bigg]\\
    & \leq \frac{1}{(NP)^{2/3}}\,O(1)\,\const 
          = \frac{1}{(NP)^{1/2}}\,O(P^{-1/3})
\ea
\ee
since the length of the interval of integration is $O(N^{2/3})$
(again by \eqref{4.12}) 
and since $\int_{-\infty}^u\Ai(t)dt$ and all other functions
are uniformly bounded.

Finally splitting the interval as in \eqref{eqN.24.1},
and using \eqref{4.12}, \eqref{eqbeta1} and \eqref{eqI.14.1}, we find
\be\la{eqN.25.1}
\ba
  |\spItt| \leq
       &\frac{\const}{N^{1/2}P^{5/6}}
                         \Big( \const + \int_{-\delta N^{2/3}\fth_N(-1+\delta)}^{-1}
                                     |\Ai(u)|\,|\Ai(\beta(u))|\,du  \Big)\\
  & \leq \frac{\const}{(NP)^{1/2}}\frac1{P^{1/3}}
                         \Big( \const + \int_1^{\delta N^{2/3}\fth_N(-1+\delta)}
                                   \frac{du}{u^{1/2}}  \Big)\\
   &\leq  \frac{\const}{(NP)^{1/2}}\frac1{P^{1/3}}
                         \big( \const + (\delta N^{2/3})^{1/2}\big)\\
   & \leq  \frac{C}{(NP)^{1/2}} \,O\big(\delta^{1/2} + N^{-1/3}\big).
\ea
\ee
Note that the integral of the form $\Ai^\prime\times\Ai$ in \eqref{eqN.17.1NEW}
can be estimated similarly after changing
the order of integration as in \eqref{eqex} above:
the estimate is then the same as for the integral 
of the form $\Ai\times\Ai^\prime$.
Thus collecting the estimates 
\eqref{eqN.25.1}, \eqref{eqN.24.2}, 
\eqref{eqN.24.1} in \eqref{eqN.23.1},
and recalling  \eqref{eqN.17.1NEW}, 
we conclude (schematically)
 \be
\la{eqN.17.1NN}
\ba
  J_{44}^{\prime\prime}
 =    &\int_{x_{N,P}}^{-1+\delta} \Ai \,dx 
        \int_{X(x)}^{-1+\delta}
               \Ai \,dy \\
 &+\frac{(c_Nc_P)^{1/2}}{(NP)^{1/2}}
             O(\delta^{1/2}+\delta+\delta^{3/2} + P^{-1/3}),\qquad
     N,P\too\infty.
\ea
\ee
Note for future reference that we will eventually take a limit
$\delta\to0$, and hence we leave only the term
$O(\delta^{1/2})$ in the above
estimate (and in similar ones below).

It remains to analyze the 
$\Ai\times\Ai$ integral in \eqref{eqN.17.1NN}.
As we will see, this is the only term that contributes
to leading order in $J_{44}$.
Making the same change of variables that led to \eqref{eqN.23.1},
and recalling \eqref{eqE}, \eqref{eqbeta}, we write the
 $\Ai\times\Ai$ integral as $(-1)^{N+P}\sqrt{c_Nc_P}$ times
\be
\la{eqN.31.1}
\ba
       \frac{1}{(\tilde{\alpha}_N\tilde{\alpha}_P)^{3/4}}
           \frac{1}{(NP)^{1/2}}
                       &\bigg( \int_{u(-1+\delta)}^{0}
              +\int_0^{u(x_{N,P})}\bigg)
                                  L_N(x(u))\, \Ai(u)\,du
 \\  &\qquad\qquad\qquad\times          
\int_{v(-1+\delta)}^{\beta(u)}
                                     L_P(y(v))\, \Ai(v) \,dv\\
   &\equiv \frac{1}{(\tilde{\alpha}_N\tilde{\alpha}_P)^{3/4}}
           \frac{1}{(NP)^{1/2}}
       \Big( \spJtop + \spJtopp\Big).
\ea
\ee
Note that in $\spJtopp$, $u\geq0$ 
(recall $\lim_{N,P\too\infty}u(x_{N,P})=+\infty$).
It follows from \eqref{eqN.21.pp.1} that there is a number $u_0>0$,
$u_0=O(P^{-1/3})$ % ($u_0$ depends also on $N$), 
such that $\beta(u)\geq0$ for $u\geq u_0$. 
We write
\be\la{eqN.35.2}
\ba
    \spJtopp 
  = & \int_0^{u_0}           L_N(x(u))\, \Ai(u)\,du
                            \int_{v(-1+\delta)}^{\beta(u)}
                                     L_P(y(v))\, \Ai(v) \,dv\\
              &+\int_{u_0}^{u(x_{N,P})}
                                  L_N(x(u))\, \Ai(u)\,du
 \\  &\qquad\qquad\qquad\times\bigg(          
                              \int_{v(-1+\delta)}^0 +\int_{0}^{\beta(u)}\bigg)
                                     L_P(y(v))\, \Ai(v) \,dv\\
  &\equiv   \spItf + \spIItf + \spIIItf.
\ea
\ee
Now
\be\la{eq.I35}
\ba
   |\spItf| =   \bigg| &
               \int_0^{u_0} L_N(x(u)) \Ai(u)\,du\,
           \bigg[
         L_P(y(v))\Big(\int_{-\infty}^v \Ai(t)dt\Big)
                   \bigg|_{v(-1+\delta)}^{\beta(u)}\\
               &\qquad\qquad\qquad
         +\frac1{\tilde{\alpha}_P P^{2/3}} 
                   \int_{v(-1+\delta)}^{\beta(u)}
        \frac{L_P^\prime(y(v))}{\tilde{U}_P(y(v))} 
                  \Big(\int_{-\infty}^v \Ai(t)dt\Big)\,dv
 \bigg]\bigg|\\
  &\leq\frac{\const}{P^{1/3}}
\ea
\ee
since $u_0=O(P^{-1/3})$,
 all the integrands in the $du$ and $dv$ integrals are 
uniformly bounded, 
and the length of the inner interval of integration 
is $O(P^{2/3})$ (recall \eqref{eqN.21.pp.1}). 

To estimate $\spIIItf$ recall first from \eqref{eqI.12.1}
that $y(v)-(-1)=-\frac{v}{\tilde{\alpha}_P P^{2/3}\tilde{U}_P(y(v))}$.
Also by \eqref{4.12}, $|L_P^{\prime}|$ is uniformly bounded.
Hence
\be\la{eqN.31.2}
            \big|L_P(y(v))-L_P(-1)\big|
                  \leq \const \cdot |y(v)-(-1)| \leq \const \cdot|v|\cdot P^{-2/3}.
\ee
Thus, since $\beta(u)\geq0$, using \eqref{eqI.12.0},
 \eqref{eqI.14.1} and the fact that 
$\fth_N(-1)=1$, we obtain
$$
\ba
           \int_{0}^{\beta(u)}
                                     L_P(y(v))\, \Ai(v) \,dv
           =&L_P(-1) \int_{0}^{\beta(u)}
                                  \Ai(v) \,dv
             + O\bigg( \int_{0}^{\beta(u)} \frac{v\,e^{-(2/3)v^{3/2}}}{P^{2/3}}
                             \,dv\bigg)\\
  =&2^{1/4} \int_{0}^{\beta(u)}
                                  \Ai(v) \,dv
             + O( P^{-2/3}),
 \ea
$$
and so
\be\la{eqN.36.1}
\ba
    \spIIItf
  = & \int_{u_0}^{u(x_{N,P})}
                                  L_N(x(u))\, \Ai(u)
                \bigg(     2^{1/4} \int_{0}^{\beta(u)}
                                  \Ai(v) \,dv
             + O( P^{-2/3})   \bigg)\,du\\
  = & 2^{1/4}\int_{u_0}^{u(x_{N,P})}
                                  L_N(x(u))\, \Ai(u)\,
                             \Big(\int_{0}^{\beta(u)}
                                  \Ai(v) \,dv\Big)\,du
              + O( P^{-2/3})
\ea
\ee
again by \eqref{eqI.14.1}.
%NEW2 
Next
\be\la{eq.NEW2}
\ba
  \int_{u_0}^{u(x_{N,P})}
                                  &L_N(x(u))\, \Ai(u)\,du
           \int_{0}^{\beta(u)} \Ai(v) \,dv\\
  &=L_N(-1)\int_{u_0}^{u(x_{N,P})}
                                  \Ai(u)\,du
           \int_{0}^{\beta(u)} \Ai(v) \,dv\\
     &+\int_{u_0}^{u(x_{N,P})}
                                  \big(L_N(x(u))-L_N(-1)\big)\, \Ai(u)\,du
           \int_{0}^{\beta(u)}
                 \Ai(v) \,dv \equiv \spIte+\spIIte.
\ea
\ee
Now by the uniform boundedness of $L_N^\prime$ and \eqref{eqI.12.1},
and since $\int_0^{\infty}|v^j \Ai(v)|dv<\infty$, $j=0,1$,
$$
    |\spIIte| \leq c \int_{u_0}^{u(x_{N,P})}
      |x(u)+1|\,|\Ai(u)|\,du 
   \leq c_1 N^{-2/3} \int_{u_0}^{u(x_{N,P})} |u\Ai(u)|\,du = O(N^{-2/3}).
$$
Next
$$
\ba
  \spIte  &= 2^{1/4} \int_{u_0}^{u(x_{N,P})}
                                  \Ai(u)\,du
         \bigg(  \int_{0}^{u}
         + \int_{u}^{\beta(u)} \bigg) \Ai(v) \,dv
       \equiv \spIte^{\prime}+\spIte^{\prime\prime}.
\ea
$$
We have by \eqref{eqN.21.pp.1}
$$
   \Big|\int_{u}^{\beta(u)} \Ai(v) \,dv\Big|
    \leq c |\beta(u)-u| = O(P^{-1/3} + P^{-1}|u|)
$$
and hence
$$
    | \spIte^{\prime\prime}| \leq
        2^{1/4}\int_0^\infty |\Ai(u)| \, O(P^{-1/3} + P^{-1}|u|)\,du 
     = O(P^{-1/3}).
$$
Also 
$$
\ba
  \spIte^{\prime}
     = &2^{1/4}\bigg( \int_0^\infty - \int_0^{u_0}
             - \int_{u(x_{N,P})}^\infty\bigg)
                                  \Ai(u)\,du
          \int_{0}^{u}  \Ai(v) \,dv\\
       &= 2^{1/4}  \int_0^\infty 
                                  \Ai(u)\,du
          \int_{0}^{u}  \Ai(v) \,dv 
        + O(P^{-1/3}) + O(e^{-cN})
\ea
$$
where we have used the fact that $u_0=O(P^{-1/3})$,
the uniform boundedness of $\int_0^{u}\Ai(v)dv$ for $u\geq0$,
and the (super)exponential decay of $\Ai(v)$ for $v\geq0$.
Thus
\be\la{eqdomconv}
       \spIIItf = 2^{1/2}\int_0^{+\infty} \Ai(u)\,du
                      \int_0^u \Ai(v)\,dv + O(P^{-1/3}),
\ee
$N,P\too\infty$.
%ADD APPROPRIATE REFERENCES 
%END NEW2

To estimate $\spIItf$ we first write
\be\la{eqN.32.0}
\ba
              \int_{v(-1+\delta)}^0 
                                     &L_P(y(v))\, \Ai(v) \,dv\\
           &=L_P(-1) \int_{-\infty}^{0}
                                  \Ai(t) \,dt-L_P(y(v(-1+\delta))) 
               \int_{-\infty}^{v(-1+\delta)}
                                  \Ai(t) \,dt\\
             &\qquad+ \frac1{\tilde{\alpha}_P P^{2/3}}
                \int_{v(-1+\delta)}^0 \Big(\int_{-\infty}^{v}
                                  \Ai(t) \,dt\Big)\,
                  \frac{L_P^\prime(y(v))}{\tilde{U}_P(y(v))}\,dv.
 \ea
\ee
But by \eqref{eqI.14.2},
as $v(-1+\delta)=-\tilde{\alpha}_PP^{2/3}\delta \fth_P(-1+\delta)\to-\infty$,
$$
\ba
  \bigg| L_P(y(v(-1+\delta))) 
               \int_{-\infty}^{v(-1+\delta)}
                                  \Ai(t) \,dt \bigg|
      &\leq \const\,|\tilde{\alpha}_PP^{2/3}\delta \fth_P(-1+\delta)|^{-3/4}\\
        &\leq \const\,      \frac1{\delta^{3/4}P^{1/2}}
\ea
$$
and also 
\be\la{eqN.32.1}
\ba
 \bigg|\frac1{\tilde{\alpha}_P P^{2/3}}
                \int_{v(-1+\delta)}^0 
             \Big(&\int_{-\infty}^{v}
                                  \Ai(t) \,dt\Big)\,
                  \frac{L_P^\prime(y(v))}{\tilde{U}_P(y(v))}\,dv\bigg|\\
      &\leq \frac\const{P^{2/3}}\int_{-\tilde{\alpha}_PP^{2/3}\delta \fth_P(-1+\delta)}^0
              \frac{dv}{(1+|v|)^{3/4}}\\ 
      &\leq \const\,\frac{\delta^{1/4}}{P^{1/2}}.
\ea
\ee
Thus
$$
\ba
    \spIItf &= \int_{u_0}^{u(x_{N,P})}
                                  L_N(x(u))\, \Ai(u)\,du
 \\  &\qquad\qquad\times\bigg[         L_P(-1) \int_{-\infty}^{0}
                                  \Ai(v) dv
      +  O\Big(\frac{\delta^{1/4}}{P^{1/2}}
                +\frac1{\delta^{3/4}P^{1/2}}\Big)\bigg]\\
  &= 2^{1/4}\int_{u_0}^{u(x_{N,P})}
                                  L_N(x(u))\, \Ai(u)\,du
        \int_{-\infty}^{0}
                                  \Ai(v) dv
      +  O\Big( \frac1{\delta^{3/4}P^{1/2}}\Big)
\ea
$$
where we have again used the estimate in \eqref{eqI.14.1}
as in \eqref{eqN.36.1}.
Now as in the analysis of \eqref{eq.NEW2} we find
$$
\ba
      \int_{u_0}^{u(x_{N,P})} 
 &L_N(x(u))\, \Ai(u)\,du\\
   &= L_N(-1) \int_{u_0}^{u(x_{N,P})} \Ai(u)\,du
        +  \int_{u_0}^{u(x_{N,P})} \big( L_N(x(u))-L_N(-1)\big)\, \Ai(u)\,du\\
   &= 2^{1/4}\int_0^{\infty} \Ai(u)\,du + O(P^{-1/3}+e^{-cN}) + O(N^{-2/3}).
\ea
$$ 
We conclude that
\be\la{eq.II35}
     \spIItf = 2^{1/2}\int_{0}^{+\infty}
               \Ai(u)\,du
                        \int_{-\infty}^{0}
                                  \Ai(v) dv
      +  O(P^{-1/3}) + O\Big( \frac1{\delta^{3/4}P^{1/2}}\Big).
\ee
Thus as $N,P\too\infty$, the second integral in \eqref{eqN.31.1} behaves
as  %$(-1)^{N+P}\sqrt{c_Nc_P}$ times
\be\la{eqN.38.1}
\ba
   \spJtopp 
 =  \int_{0}^{+\infty}
               \Ai(u)\,du
                        \int_{-\infty}^{u}
                                  \Ai(v) dv
       +  O(P^{-1/3}) + O( \delta^{-3/4}P^{-1/2}).
\ea
\ee
%Note that the first factor becomes simply $\frac1{2m}$.

The first integral in \eqref{eqN.31.1} is given by (recall \eqref{eqid})
\be\la{eqN.39.1}
\ba
   \spJtop = &%\frac1{(\tilde{\alpha}_N\tilde{\alpha}_P)^{3/4}} \frac1{(NP)^{1/2}}
               \int_{u(-1+\delta)}^{0} L_N(x(u)) \Ai(u)\,du
         \int_{v(-1+\delta)}^{\beta(u)}
         L_P(y(v))\, \Ai(v)\,dv\\
  &=    L_N(x(u)) \Big(\int_{-\infty}^u \Ai(t)dt\Big)
         \,\Big(     \int_{v(-1+\delta)}^{\beta(u)}
                            L_P(y(v)) \Ai(v)\,dv \Big)
                   \bigg|_{u(-1+\delta)}^{0}\\
   &\qquad+ \frac1{\tilde{\alpha}_NN^{2/3}}
        \int_{u(-1+\delta)}^{0} 
            \Big(\int_{-\infty}^u \Ai(t)dt\Big) 
                  \frac{L_N^\prime(x(u))}{\tilde{U}_N(x(u))}
                              \,du
\\ &\qquad\qquad\qquad\qquad\qquad\qquad\times
             \Big(     \int_{v(-1+\delta)}^{\beta(u)}
                            L_P(y(v)) \Ai(v)\,dv \Big)\\
   &\qquad-        \int_{u(-1+\delta)}^{0} 
            \Big(\int_{-\infty}^u \Ai(t)dt\Big) 
                 \,L_N(x(u)) \,   L_P(X(x(u)))               
                     \, \Ai(\beta(u)) \, \beta^\prime(u)\,du
    \\
         &\equiv
  %   \frac1{(\tilde{\alpha}_N\tilde{\alpha}_P)^{3/4}} \frac1{(NP)^{1/2}}
     \spIfz+\spIIfz+\spIIIfz.
\ea
\ee
Observe that $u(-1+\delta)\leq u\leq0$, and so
$\beta(u)\leq\const<\infty$ by \eqref{eqN.21.pp.1},
\be\la{eqN.40.1}
\ba
   \bigg| \int_{v(-1+\delta)}^{\beta(u)}
         &L_P(y(v))\, \Ai(v)\,dv\bigg| 
   \leq\bigg|         L_P(y(v))
           \Big(\int_{-\infty}^v \Ai(t)dt\Big)
                   \bigg|_{v(-1+\delta)}^{\beta(u)}\,\bigg|\\
   &\qquad+ \frac1{\tilde{\alpha}_PP^{2/3}}
      \bigg|  \int_{v(-1+\delta)}^{\beta(u)}
            \Big(\int_{-\infty}^v \Ai(t)dt\Big) 
                  \frac{L_P^\prime(y(v))}{\tilde{U}_P(y(v))}
                              \,dv\bigg|\\
   &\leq \const + O\big( (N^{2/3})^{1/4} P^{-2/3})
      \leq\const
\ea
\ee
where we have used the bound on
the integrand \eqref{eqI.14.2}.
Thus again using \eqref{eqI.14.2},
\be\la{eqN.40.2}
   |\spIIfz| \leq \frac\const{N^{2/3}}
        \int_{v(-1+\delta)}^0 \frac{C}{(1+|u|)^{3/4}}\,du = O(N^{-1/2}).
\ee

Note next that in
$$
\ba
    \spIfz = &L_N(x(0))  \Big(\int_{-\infty}^0 \Ai(t)dt\Big)
              \cdot\int_{v(-1+\delta)}^{\beta(0)}
                              L_P(y(v))\, \Ai(v)\,dv\\
           &-L_N(x(u(-1+\delta)))  
                      \Big(\int_{-\infty}^{u(-1+\delta)} \Ai(t)dt\Big)
              \cdot\int_{v(-1+\delta)}^{\beta(u(-1+\delta))}
                              L_P(y(v))\, \Ai(v)\,dv
\ea
$$
the second term is $O(P^{-1/2})$
 by \eqref{eqI.14.2}, \eqref{equinfty}, \eqref{eqN.40.1}. 
As in \eqref{eqN.32.0},
 the first term can be written as
$$
\ba
    L_N(x(0))  &\Big(\int_{-\infty}^0 \Ai(t)dt\Big)
              \cdot L_P(y(\beta(0)))\bigg[ \Big(\int_{-\infty}^{\beta(0)}
                               \Ai(v)\,dv\Big) + O(\delta^{-3/4}P^{-1/2})\bigg]\\
      &= 2^{1/2}
               \Big(\int_{-\infty}^0 \Ai(t)dt\Big)
                \Big(\int_{-\infty}^{0}
                               \Ai(v)\,dv\Big) 
           + O(\delta^{-3/4}P^{-1/2}+P^{-1/3})
\ea
$$
in view of \eqref{eqI.12.1} and \eqref{eqN.21.pp.1}.
Thus
\be\la{eqN.41.1}
   \spIfz= 2^{1/2}
               \bigg(\int_{-\infty}^0 \Ai(t)dt\bigg)^2
           + O(\delta^{-3/4}P^{-1/2}+P^{-1/3}).
\ee

Finally,
\be\la{eqN.40.aux}
\ba
   \spIIIfz = &- \int_{u(-1+\delta)}^{0} 
            \Big(\int_{-\infty}^u \Ai(t)dt\Big) 
                 \,L_N(x(u)) \,   L_P(X(x(u)))               
                     \, \Ai(\beta(u)) \, \beta^\prime(u)\,du\\
     &=-\int_{u(-1+\delta)}^{0} 
            \Big(\int_{-\infty}^u \Ai(t)dt\Big) 
                 \,L_N(x(u)) \,   L_P(X(x(u)))               
                     \, \Ai(\beta(u)) \,du + O(P^{-1/3})
\ea
\ee
by \eqref{eqN.21.p.1}, the uniform boundedness
of the remaining factors in the integrand, and since the length of the
interval of integration is $O(N^{2/3})$.
Denote
$$
     A(u)\equiv \int_{u(-1+\delta)}^u
               \Ai(\beta(s)) \bigg(\int_{-\infty}^s \Ai(t)dt \bigg)\,ds.
$$
Since
$$
       \frac{d}{du} A(u)
       = \bigg(\int_{-\infty}^u \Ai(t)dt \bigg)\,\Ai(\beta(u))
$$
the remaining integral in \eqref{eqN.40.aux} after integration by
parts becomes
\be\la{eqN.40.aux2}
\ba
   -A(u) \,&L_N(x(u)) \,   L_P(X(x(u)))\bigg|_{u(-1+\delta)}^0 \\              
    &-\frac1{\tilde{\alpha}_NN^{2/3}}\int_{u(-1+\delta)}^{0} 
            A(u)\,\frac{L_N^\prime(x(u))}{\tilde{U}_N(x(u))}
                             \,   L_P(X(x(u)))\,du \\               
    &-\frac1{\tilde{\alpha}_NN^{2/3}}\int_{u(-1+\delta)}^{0} 
            A(u)\,L_N(x(u))\frac{ L_P^\prime(X(x(u)))\,X^\prime(x(u))}
 {\tilde{U}_P(x(u))}
                             \,  du.            
\ea
\ee
We need the following result.
\begin{lemma}\la{lemA}
As $N,P\too\infty$, we have uniformly
for $u(-1+\delta)\leq u\leq0$
\be\la{eqN.49.1}
\ba
   A(u) = &\int_{-\infty}^u \Ai(s)\Big(\int_{-\infty}^s \Ai(t)dt\Big)\,ds
%\\            &
      + O(\delta^{1/2})
                  + O(\delta^{-3/2}N^{-1}).
%      +O(N^{-1/3}).
\ea
\ee
\end{lemma}
\begin{proof}
For $u(-1+\delta) \leq u\leq-1$ we recall \eqref{eqchv}
\be\la{eqe10pf}
         u(-1+\delta) = -\tilde{\alpha}_N \delta \fth_N(-1+\delta) 
        = -c \delta N^{2/3}(1+o(1)),\qquad c>0.
\ee
For $u\leq0$
\be\la{eqe21pf}
\ba
  \bigg| \int_{u(-1+\delta)}^u &\Ai(\beta(s))\Big(\int_{-\infty}^s \Ai(t)dt\Big)\,ds
            -\int_{-\infty}^u \Ai(s)\Big(\int_{-\infty}^s \Ai(t)dt\Big)\,ds\bigg| \\
      &\leq   
   \bigg| \int_{-\infty}^{u(-1+\delta)} \Ai(s)
             \Big(\int_{-\infty}^s \Ai(t)dt\Big)\,ds\bigg|\\
    &\qquad+\bigg| \int_{u(-1+\delta)}^u \big[\Ai(\beta(s))-\Ai(s)\big]\,
                      \Big(\int_{-\infty}^s \Ai(t)dt\Big)\,ds \bigg|. 
\ea
\ee
Note that 
\be\la{eqAA}
         \int_{-\infty}^u \Ai(s)\Big(\int_{-\infty}^s \Ai(t)dt\Big)\,ds
              = \frac12\bigg(\int_{-\infty}^u \Ai(s)\,ds\bigg)^2
\ee
and so, by \eqref{eqI.14.2}, \eqref{eqe10pf},
 the first term in the RHS in \eqref{eqe21pf} equals
$$
         \frac12\Big(\int_{-\infty}^{u(-1+\delta)} \Ai(s)\,ds\Big)^2 
            = O((\delta N^{2/3})^{-3/2}) = O(\delta^{-3/2}N^{-1}).
$$

In view of \eqref{eqC2}, \eqref{eqN.21.pp.1}, 
we bound the second term in \eqref{eqe21pf} by
$$
\ba
     \int_{u(-1+\delta)}^u 
           &\big|\Ai(\beta(s))-\Ai(s)\big|\cdot
                      \Big|\int_{-\infty}^s \Ai(t)dt\Big|\,ds\\
     &\leq \int_{u(-1+\delta)}^0 
           \Big(\sup_{t\in[\beta(s),s]}|\Ai^\prime(t)|\Big)
                 \cdot|\beta(s)-(s)|\cdot
                      \Big|\int_{-\infty}^s \Ai(t)dt\Big|\,ds\\
     &\leq \const\int_{u(-1+\delta)}^1 
          (1+ |s|)^{1/4} \,\Big[ O(P^{-1/3}) + |s|O(P^{-1})\Big]\,
           \frac{ds}{1+|s|^{3/4}}\\
     &\leq  O(P^{-1/3}(\delta N^{2/3})^{1/2})  
              + O(P^{-1}(\delta N^{2/3})^{3/2})  
                = O( \delta^{1/2})
\ea
$$
 uniformly for $u(-1+\delta)\leq u\leq0$.
\end{proof}
From \eqref{eqAA},
Lemma \ref{lemA}, \eqref{eqN.40.aux2},
and the fact that $A(u(-1+\delta))=0$,
we conclude that 
\be\la{eqN.50.1}
   \spIIIfz = -\frac{2^{1/2}}2 \bigg(\int_{-\infty}^0 \Ai(s)\,ds\Big)^2
            +  O(\delta^{1/2}) +O(\delta^{-3/2}N^{-1}) 
\ee
again by the uniform boundedness of
the terms in the integrands, the integrability 
of $\big(\int_{-\infty}^u\Ai(s)\,ds\big)^2\sim|u|^{-3/2}$, $u\to-\infty$,
and the fact that $u(-1+\delta)=O(N^{2/3})$.
By \eqref{eqN.50.1}, \eqref{eqN.40.2}, \eqref{eqN.41.1}, \eqref{eqAA}
 we see that \eqref{eqN.39.1} is given by
\be\la{eqN.51.1}
\ba
  \spJtop =    
 %\frac1{2^{3/2}m}\frac{2^{1/2}}{(NP)^{1/2}} \bigg[   
  &\int_{-\infty}^0 \Ai(s)\Big(\int_{-\infty}^s \Ai(t)dt\Big)\,ds\\
            &+  O(\delta^{1/2}) 
            + O(\delta^{-3/4}N^{-1/2}) + O(\delta^{-3/2}N^{-1})
       + O(P^{-1/3}). % ??? \bigg].
\ea
\ee
Thus by \eqref{eqN.31.1}, \eqref{eqN.51.1}, \eqref{eqN.38.1},
the $\Ai\times\Ai$ integral
in \eqref{eqN.17.1NN} equals $(-1)^{N+P}
\frac{(c_Nc_P)^{1/2}}{(\tilde{\alpha}_N\tilde{\alpha}_P)^{3/4}(NP)^{1/2}}$
 times
\be\la{eqN.51.2}
\ba
     {}&\Big(\int_{-\infty}^0+\int_0^{+\infty}\Big) 
                    \Ai(s)\Big(\int_{-\infty}^s \Ai(t)dt\Big)\,ds
                                 + O(P^{-1/3})+ O\bigg(\delta^{1/2}
                      + \frac1{\delta^{3/4}P^{1/2}}+ \frac1{\delta^{3/2}P}
                  \bigg)
         \\
   &= %\frac1{2m}
    \int_{-\infty}^{+\infty} \Ai(s)\Big(\int_{-\infty}^s \Ai(t)dt\Big)\,ds
                   + O(P^{-1/3})+ O\bigg(\delta^{1/2}
                      + \frac1{\delta^{3/4}P^{1/2}}+ \frac1{\delta^{3/2}P}\bigg)
          \\
   &= %\frac1{2m}
    \frac12\Big(\int_{-\infty}^{+\infty} \Ai(s)\,dx\Big)^2
          + O(P^{-1/3})+ O\bigg(\delta^{1/2}
                      + \frac1{\delta^{3/4}P^{1/2}}+ \frac1{\delta^{3/2}P}\bigg)
         \\
   &= \frac12 %{4m}
     \bigg[ 1+ O(P^{-1/3}) + O\bigg(\delta^{1/2}
                      + \frac1{\delta^{3/4}P^{1/2}}+ \frac1{\delta^{3/2}P}\bigg)
          \bigg]
\ea
\ee
by \eqref{eqAA}, and 
since $\int_{-\infty}^{+\infty}\Ai(t)dt=1$ (see \cite{Stegun}).
Thus from \eqref{4.12}(2), \eqref{eqN.4.1}
$$
\ba
      J_{44} = &J_{44}^\prime + J_{44}^{\prime\prime}
                =  (-1)^{N+P}\frac{(c_Nc_P)^{1/2}}{(NP)^{1/2}}\\
     &\times\bigg[ \frac1{4m} 
      +  O(P^{-1/3})+O(P^{-1/(2m)}) \\
       &\qquad\qquad\qquad+ O\Big(\delta^{1/2}  + \delta^{-3/4}P^{-1/2}
                            +\delta^{-3/2}P^{-1}\Big)
           \qquad\textrm{by \eqref{eqN.51.2}}\\
      &\qquad + O(N^{-1/3})\qquad\textrm{by \eqref{eqN.5.1}}\\
      &\qquad + O(\delta^{1/2}) +O(P^{-1/3}) 
            \qquad\textrm{by \eqref{eqN.17.1NN}}
     \bigg]                 \\
&=  \frac{(-1)^{N+P}}{4m}
             \frac{(c_Nc_P)^{1/2}}{(NP)^{1/2}}
     \bigg[ 1 + O(P^{-1/3})+O(P^{-1/(2m)})\\
                      &\qquad\qquad\qquad+  O\Big(\delta^{1/2} + \delta^{-3/4}P^{-1/2}
                            +\delta^{-3/2}P^{-1}\Big)
     \bigg].     
\ea                   
$$
By this estimate and \eqref{eqN.3.2}, \eqref{eqN.3.3}, \eqref{eqN.3.4},
we find finally that
\be\la{eqN.52.1}
\ba
     J_4 =   \frac{(c_Nc_P)^{1/2}}{(NP)^{1/2}}
     \bigg[  \frac{(-1)^{N+P}}{4m}
                + &\frac{(-1)^{N}}{2m}
             + O(P^{-1/3})+O(P^{-1/(2m)})\\
              &+  O\Big(\delta^{1/2}  + P^{-1/2}\delta^{-3/4}
                      + P^{-1}\delta^{-3/2}\Big)
     \bigg].     
\ea
\ee

In a similar way we obtain
\be\la{eqN.53.1}
\ba
     J_2 \equiv 
       \int_{1-\delta}^{1+\delta} &c_N\phi_N(c_Nx+d_N)\,
           \Big(\int_{X(x)}^{1+\delta} c_P\phi_P(c_Px+d_P) \Big)\,dx\\
      &=  \frac{(c_Nc_P)^{1/2}}{(NP)^{1/2}}
     \bigg[  \frac{1}{4m}
                  + O(P^{-1/3})+O(P^{-1/(2m)}) \\
        &\qquad\qquad\qquad+  O\Big(\delta^{1/2} 
                       + P^{-1/2}\delta^{-3/4}+ P^{-1}\delta^{-3/2}\Big)
     \bigg].
\ea
\ee
(Note that the integral leading to the term $\frac{(-1)^N}{2m}$
in \eqref{eqN.52.1}, is not present for $J_2$.)

It remains to consider $J_3$.
Again assume that $N,P\too\infty$.
By \eqref{eqN.2.2}
$$
\ba
  J_3 = &c_Nc_P \int_{I_3}
           \phi_N(c_Nx+d_N)\,dx
                       \int_{X(x)}^{+\infty}   \phi_P(c_Py+d_P)\,dy\\
          &=c_Nc_P \int_{I_3}
           \phi_N(c_Nx+d_N)\,dx
                       \bigg( \int_{X(x)}^{1-\delta}+\int_{I_2}
                            +\int_{I_1} \bigg)  \phi_P(c_Py+d_P)\,dy\\
          &\equiv J_{33}+J_{32}+J_{31}.
\ea
$$
By \eqref{eqI.18.1}, \eqref{eqI.17.1}, \eqref{eqI.10.1}
$$     
       |J_{32}| + |J_{31}| = c_N^{1/2} N^{-1} 
          \,O\big(  (c_P/P)^{1/2} + (c_P/P)^{1/2}
                                    \,e^{-Pc_{3}}\big) 
        =  \frac{ (c_Nc_P)^{1/2}}{ (NP)^{1/2}}\,
            O\big( N^{-1/2}\big). 
$$
To analyze $J_{33}$ we first introduce $x_{N,P}^{\pm}\in{}I_3$ such that
$$
      X(x_{N,P}^{\pm}) \equiv 
           x_{N,P}^{\pm}c_N/c_P + (d_N-d_P)/c_P = \pm(1-\delta)
$$
which in view of \eqref{eqXx}, implies that
for $N,P\too\infty$,
\be\la{eqN.57.0}
     X(x)\in I_3,\qquad x\in I_3^\prime\equiv [x_{N,P}^-,x_{N,P}^+]
                       \subset I_3.
\ee
Set $I_3^{\prime\prime}\equiv I_3\setminus I_3^\prime$ and note
that by Lemma \ref{lemprec2}
\be\la{eqN.56.1}
   x_{N,P}^{\pm} = \pm(1-\delta) + O(P^{-1}).
\ee
Then
$$
\ba
  J_{33} = &c_Nc_P\bigg( \int_{I_3^{\prime}} 
                               + \int_{I_3^{\prime\prime}}\bigg)
           \phi_N(c_Nx+d_N)\,dx
                       \int_{X(x)}^{1-\delta}   \phi_P(c_Py+d_P)\,dy\\
          &\equiv J_{33}^{\prime} + J_{33}^{\prime\prime}.
\ea
$$
Now
\be\la{eqN.56.p}
\ba
      |J_{33}^{\prime\prime}| \leq
          C \int_{I_{3}^{\prime\prime}}
                   \frac{c_N}{c_N^{1/2}} 
                   \frac{c_N^{1/2}}{N^{1/2}} \,dx = C c_N\,N^{-1/2}\,
                 |I_{3}^{\prime\prime}| 
          = \frac{ (c_Nc_P)^{1/2}}{ (NP)^{1/2}}\,
                           O\big( P^{-1/2}\big)
\ea
\ee
by \eqref{eqN.4.abs}, \eqref{eqI.19.1}, \eqref{eqN.56.1}.

Now taking into account \eqref{eqN.57.0} and using \eqref{eqI.9.1}
$$
\ba
   J_{33}^\prime =
    &\frac2\pi \sqrt{c_Nc_P}
       \int_{I_3^\prime}
          \frac{dx}{|1-x^2|^{1/4}} \Big[
          \Big(\cos f_N^+(x)\Big)\,\Big(1+O(N^{-1}\Big)
        + \Big(\sin f_N^-(x)\Big)\,O(N^{-1})\Big]\\
   &\times  \int_{X(x)}^{1-\delta}
          \frac{dy}{|1-y^2|^{1/4}} \Big[
          \Big(\cos f_P^+(y)\Big)\,\Big(1+O(P^{-1}\Big)
        + \Big(\sin f_P^-(y)\Big)\,O(P^{-1})\Big]
\ea
$$
where
\be\la{eqN.57.2}
   f_N^{\pm}(x) \equiv  \frac{N}2\int_1^x|1-y^2|^{1/2}h_N(y)dy
                       \pm \frac12\arcsin x
\ee
and $h_N$ satisfies \eqref{hestbelow} and \eqref{eq9pSTAR}.

Clearly a product of two terms, one of which 
has a factor $O(N^{-1})$ and the other $O(P^{-1})$,
 gives rise to a smaller order 
contribution $\frac{\sqrt{c_Nc_P}}{\sqrt{NP}}\,O(N^{-1})$.
Consider next the integral
$$
\ba
  \bigg|       \int_{I_3^\prime}
          &\frac{dx}{|1-x^2|^{1/4}} 
               \Big(\sin f_N^-(x)\Big)\,O(N^{-1})
           \int_{X(x)}^{1-\delta}
          \frac{dy}{|1-y^2|^{1/4}} 
          \Big(\cos f_P^+(y)\Big)\bigg|\\
     &\leq O(N^{-1})
           \int_{I_3^\prime}
          \frac{dx}{|1-x^2|^{1/4}} 
             \bigg|\int_{X(x)}^{1-\delta}
          \frac{dy}{|1-y^2|^{1/4}} 
          \Big(\cos f_P^+(y)\Big)\bigg|\\
     &\leq O(N^{-1})
           \bigg(\int_{I_3^\prime}
          \frac{dx}{|1-x^2|^{1/4}} \bigg)
           \max_{-1+\delta\leq x_1\leq1-\delta}  \bigg|\int_{x_1}^{1-\delta}
          \frac{dy}{|1-y^2|^{1/4}} 
          \Big(\cos f_P^+(y)\Big)\bigg|\\
    &=O((NP)^{-1}),
\ea
$$
where we have used \eqref{eqI.18.0} in the last step. 
Thus
$$
\ba
  \frac2\pi\sqrt{c_Nc_P}
           \int_{I_3^\prime}
          &\frac{dx}{|1-x^2|^{1/4}} 
               \Big(\sin f_N^-(x)\Big)\,O(N^{-1})
           \int_{X(x)}^{1-\delta}
          \frac{dy}{|1-y^2|^{1/4}} 
          \Big(\cos f_P^+(y)\Big)\\
    &=\frac{\sqrt{c_Nc_P}}{\sqrt{NP}}\,O(N^{-1}).
\ea
$$
Similarly, interchanging the order of integration as in \eqref{eqN.16.extra},
we find
$$
\ba
  \frac2\pi\sqrt{c_Nc_P}
          &\int_{I_3^\prime}
           \cos f_N^+(x)\,\frac{dx}{|1-x^2|^{1/4}} 
                    \int_{X(x)}^{1-\delta}
           \Big(\sin f_P^-(y)\Big)\,O(P^{-1})\,
      \frac{dy}{|1-y^2|^{1/4}}   \\
  &=\frac2\pi\sqrt{c_Nc_P}
           \int_{-1+\delta}^{1-\delta}
           \frac{\big(\sin f_P^-(y)\big)\,O(P^{-1})}{|1-y^2|^{1/4}}\,dy 
                    \int_{-1+\delta}^{X^{-1}(y)}
                            \frac{\cos f_N^+(x)}{|1-x^2|^{1/4}} \,dx
            \\
    &=\frac{\sqrt{c_Nc_P}}{\sqrt{NP}}\,O(N^{-1}).
\ea
$$

Thus
\be\la{eqN.59.1}
   J_3 =      \frac2\pi\sqrt{c_Nc_P} \int_{I_3^\prime}
       \frac{\cos f_N^+(x)}{|1-x^2|^{1/4}}\,dx 
           \int_{X(x)}^{1-\delta}
          \frac{\cos f_P^+(y)}{|1-y^2|^{1/4}} \,dy
            +\frac{\sqrt{c_Nc_P}}{\sqrt{NP}}\,O(N^{-{1/2}}).
\ee
Note that,
in view of \eqref{hestbelow}, for a fixed $\delta>0$,
 there exists $\eta_\delta>0$
such that as $N\to\infty$
\be\la{eqN.59.0}
     (f_N^{\pm}(x))^\prime 
         = \frac{N}2 |1-x^2|^{1/2}h_N(x) \pm \frac1{2|1-x^2|^{1/2}}
         \geq N\eta_\delta
\ee
uniformly for $x\in I_3^\prime$.
The integral in \eqref{eqN.59.1} is given by $\frac2\pi\sqrt{c_Nc_P}$ times
\be\la{eqN.60.1}
\ba
    \int_{I_3^\prime}
       \frac{\cos f_N^+(x)}{|1-x^2|^{1/4}}\,dx 
         \bigg[ &\frac{\sin f_P^+(y)}{(f_P^+(y))^\prime|1-y^2|^{1/4}}
                     \bigg|_{X(x)}^{1-\delta}\\
      &-   \int_{X(x)}^{1-\delta}
          (\sin f_P^+(y)) 
                \Big(\frac1{|1-y^2|^{1/4}(f_P^+(y))^\prime}\Big)^\prime \,dy\bigg]\\
    = - \int_{I_3^\prime}
       \frac{\cos f_N^+(x)}{|1-x^2|^{1/4}}
          &\frac{\sin f_P^+(X(x))}{f_P^{+\prime}(X(x)) |1-(X(x))^2|^{1/4}}
                    \,dx + O((NP)^{-1}) 
\ea
\ee
in view of \eqref{eqN.59.0}, 
since we can integrate by parts in $x$ once more for
the boundary term at $y=1-\delta$, 
and also integrate by parts again
in the integral term $\int_{X(x)}^{1-\delta}\cdots$. 
Note next that by Lemma \ref{lemprec2} we have again
uniformly for $x\in I_3^\prime$ 
$$
       |X(x)- x|=O(P^{-1}).
$$
Also, uniformly for $x\in{}I_3$,
$|f_P^{+\prime\prime}(x)|\leq\const\cdot P$ 
and $|1-x^2|^{-1}\leq c$.
Thus, as $X(I_3^\prime)\subset I_3$, we conclude that
$$
  \bigg|           \frac1{f_P^{+\prime}(X(x)) |1-(X(x))^2|^{1/4}}
     -           \frac1{f_P^{+\prime}(x) |1-x^2|^{1/4}}\bigg|
           \leq \frac{\const}{P^2}
$$                  
and hence using  \eqref{eqN.60.1}
we find
\be\la{eqN.61.1}
\ba
    J_3  = &-\frac2\pi\sqrt{c_Nc_P}
           \int_{I_3^\prime}
       \frac{\cos f_N^+(x)\, \sin f_P^+(X(x))}
                 {f_P^{+\prime}(x)|1-x^2|^{1/2}}\,dx
              + \frac{\sqrt{c_Nc_P}}{\sqrt{NP}}\,O(N^{-{1/2}}).
\ea
\ee
Recalling \eqref{eqN.59.0},
\eqref{hestbelow}
and \eqref{eq9pSTAR},
we derive from \eqref{eqN.61.1}
\be\la{p.45.1}
\ba
    J_3  = &-\frac4\pi\frac{\sqrt{c_Nc_P}}{\sqrt{NP}}
           \int_{I_3^\prime}
       \frac{\cos f_N^+(x)\, \sin f_P^+(X(x))}
                 {(1-x^2)\,h(x)}\,dx
              + \frac{\sqrt{c_Nc_P}}{\sqrt{NP}}\,O(N^{-{1/(2m)}}).
\ea
\ee
Now the integral above is given by
\be\la{eqN.62.1}
\ba
       \int_{I_3^\prime}
       \frac{\cos f_N^+(x)\, \sin f_P^+(X(x))}
                 {(1-x^2)\,h(x)}\,dx
     =&\frac12    \int_{I_3^\prime}
       \frac{\sin\big[  f_P^+(X(x)) + f_N^+(x)\big]}
                 {(1-x^2)\,h(x)}\,dx\\
  &+\frac12    \int_{I_3^\prime}
       \frac{\sin\big[  f_P^+(X(x)) - f_N^+(x)\big]}
                 {(1-x^2)\,h(x)}\,dx \equiv \spIst+\spIIst.
\ea
\ee
Integrating by parts in $\spIst$ in the same way as in \eqref{eqN.60.1}
and noting that
for sufficiently large $N$, by \eqref{eqN.59.0},
$$
      f_P^{+\prime}(X(x))\,X^\prime(x) + f_N^{+\prime}(x)
          \geq N\tilde{\eta}_\delta
$$
uniformly for $x\in{}I_3^\prime$, we conclude that 
$$
       |\spIst|\leq O(N^{-1}).
$$
Finally consider $\spIIst$.
Introduce the convenient notation
\be\la{eqtheta}
      \theta(x)\equiv \frac12\int_0^x |1-t^2|^{1/2}\,h(t)\,dt
\ee
and note that $\theta(1)=\pi/2$ by \eqref{psiN}, \eqref{eq9pSTAR},
the fact that $h(t)$ is even (see \eqref{hz}) and
the normalization condition $\int_{-1}^1\psi_N^{(\textrm{eq})}(x)\,dx=1$.
\begin{lemma}
Uniformly for $x\in{}I_3^\prime$ as $N,P\too\infty$,
\be\la{eqN.63.1}
      f_P^+(X(x)) - f_N^+(x) = 
         (N-P)\bigg[ \frac1{2m}\,x\theta^\prime(x)
           - \theta(x) + \frac\pi2\bigg]
            +O(N^{-1/(2m)}).
\ee
\end{lemma}
\begin{proof}
We have 
$$
       f_P^+(X(x)) - f_N^+(x) =     [f_P^+(X(x)) - f_P^+(x)] +
               [f_P^+(x) - f_N^+(x)] \equiv F_1 + F_2.
$$
By \eqref{eqN.57.2}
$$
    F_2 =  \frac12
             \int_1^x|1-y^2|^{1/2}[Ph_P(x)-Nh_N(y)]dy.
$$
Note that by \eqref{eq9pSTAR} with $q=2m$
and \eqref{eqNPalpha}
$$
\ba
         Ph_P(x)-Nh_N(x) &= P(h_P(x)-h_N(x)) + (P-N)h_N(x)\\
            &=P \sum_{k=1}^{2m}\Big[h_{(k)}(x)\cdot 
                     (P^{-k/(2m)}-N^{-k/(2m)})\big] + P\,O(P^{-1-1/(2m)})\\
             &\qquad+ (P-N)[h(x) + O(N^{-1/(2m)})]\\
           &= (P-N) h(x) + O(N^{-1/(2m)})
\ea
$$
uniformly for $-1\leq x\leq 1$. Hence,
uniformly for $x\in I_3^\prime$,
$$
\ba
     F_2 =  &(P-N)\frac12
             \int_1^x|1-y^2|^{1/2}h(y)dy  + O(N^{-1/(2m)})\\
         &=- (N-P)\Big[\theta(x) - \frac12
             \int_0^1|1-y^2|^{1/2}h(y)dy\Big]  + O(N^{-1/(2m)})\\
         &=- (N-P)\Big[\theta(x) -\frac\pi2\Big]  + O(N^{-1/(2m)}),
\ea
$$
since $\theta(1)=\pi/2$.
Next 
$$
\ba
  F_1 = &f_P^{+\prime}(x)\,(X(x)-x) 
              + f_P^{+\prime\prime}(\xi(x))\,(X(x)-x)^2/2
\ea
$$
for some $\xi(x)$ between $x$ and $X(x)$.
Note that by \eqref{eqN.1.1} and Lemma \ref{lemprec2}
$$
         X(x) - x = (c_N/c_P -1 )x +(d_N-d_P)/c_P 
        = \frac1{2m}\frac{N-P}P\,x + O(P^{-1-1/(2m)}).
$$
Also by \eqref{eqN.59.0} and \eqref{eq9pSTAR},
$|f_P^{+\prime\prime}(x)|\leq\const\cdot P$ uniformly for $x\in I_3$.
But
$$
     f_P^{+\prime}(x) = \frac{P}2 |1-x^2|^{1/2} h(x)\Big[
                 1+O(P^{-1/(2m)}) + O(P^{-1})\Big]
$$
and hence by \eqref{eqtheta}
$$
        F_1 = \frac12\frac{N-P}{2m}\,x|1-x^2|^{1/2} h(x) + O(P^{-1/(2m)})
          =\frac{N-P}{2m}\,x\theta^\prime(x) + O(P^{-1/(2m)})
$$
which completes the proof of the lemma.
\end{proof}
We need the following fortunate and remarkable fact.
\begin{lemma}
\la{s5propde}
   For any $V(x)=\kappa_{2m}x^{2m}+\cdots$ 
the function $\theta$ defined by \eqref{eqtheta}
solves the linear first order ODE
%(and $h(t)$ also satisfies a different linear first order ODE!)
$$
   \theta(x) -\frac1{2m}x\theta^\prime(x) =\arcsin x,\qquad-1\leq x\leq1.
$$
\end{lemma}
\begin{proof} See Remark~\ref{rmpfprop1} in Subsection~\ref{subsec72}.
\end{proof}
Thus from \eqref{eqN.62.1}, \eqref{eqN.63.1}
$$
    J_3 = -\frac2\pi \frac{\sqrt{c_Nc_P}}{\sqrt{NP}}
       \int_{I_3^\prime}
       \frac{\sin\big[ (P-N) \arcsin x +(N-P) \frac\pi2\big) \big]}
                 {(1-x^2)\,h(x)}\,dx
            +\frac{\sqrt{c_Nc_P}}{\sqrt{NP}}\,O(N^{-{1/(2m)}}).
$$
Note that by \eqref{eqN.56.1} we can now replace $I_3^\prime$
with $I_3\equiv[-1+\delta,1-\delta]$ introducing an error
$\frac{\sqrt{c_Nc_P}}{\sqrt{NP}}\,O(N^{-1})$.
Now using the formula for the sine of a sum and noting
that $\frac{\sin[(P-N) \arcsin x]}
                 {(1-x^2)\,h(x)}$ is an odd function 
we find
$$
\ba
    J_3 = &-
          \frac2\pi \frac{\sqrt{c_Nc_P}}{\sqrt{NP}}
              \,\sin\Big(\frac{(N-P)\pi}2\Big)
       \int_{-1+\delta}^{1-\delta}
       \frac{\cos\big[ (P-N) \arcsin x  \big]}
                 {(1-x^2)\,h(x)}\,dx\\
           & +\frac{\sqrt{c_Nc_P}}{\sqrt{NP}}\,O(N^{-{1/(2m)}}).
\ea
$$
Assume $N-P$ is even. 
Then $J_3=\frac{\sqrt{c_Nc_P}}{\sqrt{NP}}\,O(N^{-{1/(2m)}})$.
Assume $N-P$ is odd. Then 
$$
     \frac{\cos\big[ (P-N) \arcsin x  \big]}
                 {(1-x^2)\,h(x)} = O\Big(\frac1{|1-x^2|^{1/2}}\Big),
            \qquad x\to\pm(1-0)
$$
and hence for $N-P$ odd
\be\la{eqN.65.1}
\ba
    J_3 = &-
          \frac2\pi \frac{\sqrt{c_Nc_P}}{\sqrt{NP}}
              \,\sin\Big(\frac{(N-P)\pi}2\Big)
       \int_{-1}^{1}
       \frac{\cos\big[ (N-P) \arcsin x  \big]}
                 {(1-x^2)\,h(x)}\,dx\\
           & +\frac{\sqrt{c_Nc_P}}{\sqrt{NP}}\,\Big(O(N^{-{1/(2m)}})
                        +O(\delta^{1/2})\Big).
\ea
\ee

By \eqref{s5eq29}, \eqref{s5eq31}, \eqref{eqN.2.2}, as $N,P\too\infty$,
\be\la{eqN.67.1}
\ba
     (\phi_N,\eps\phi_P)  = &\frac12 
            \Big(\int_{-\infty}^{+\infty}\phi_N(x)\,dx\Big)
            \Big(\int_{-\infty}^{+\infty}\phi_P(y)\,dy\Big)\\
     &- \bigg[c_Nc_P 
          \Big(\int_{I_1}+\int_{I_5}\Big) \phi_N(c_Nx+d_N)\,dx
                       \int_{X(x)}^{+\infty}   \phi_P(c_Py+d_P)\,dy \\
   &\qquad\qquad\qquad+\Big(J_4+J_2\Big)  + J_3\bigg]\\
=\frac{\sqrt{c_Nc_P}}{\sqrt{NP}}
          &\bigg\{
     \frac12 \,\frac1{2m}\,(1+(-1)^N+O(N^{-1/2})+O(N^{-1/(2m)}))\\
           &\qquad\qquad \times(1+(-1)^P+O(P^{-1/2})+O(P^{-1/(2m)}))
                  \qquad\textrm{by \eqref{s5eq30}}\\
     &\qquad   -\bigg[ O(e^{-c_3N})
           \qquad\textrm{by \eqref{eqN.2.1}}\\
      &\qquad\qquad +\frac{(-1)^{N+P}}{4m} 
                     +\frac{(-1)^{N}}{2m} +\frac{1}{4m}  \\
      &\qquad\qquad\qquad
                     +O(P^{-1/3})+O(P^{-1/(2m)})\\
          &\qquad\qquad\qquad+ O\Big(\delta^{1/2} 
                       + P^{-1/2}\delta^{-3/4}+ P^{-1}\delta^{-3/2}\Big)
                 \quad\textrm{by \eqref{eqN.52.1}, \eqref{eqN.53.1}}\\
      &\qquad\qquad   -
            \sin\Big(\frac{(N-P)\pi}2\Big)\,\frac2\pi
       \int_{-1}^{1}
       \frac{\cos\big[ (N-P) \arcsin x  \big]}
                 {(1-x^2)\,h(x)}\,dx\\
       &\qquad\qquad\qquad
               +O(N^{-{1/(2m)}})     +O(\delta^{1/2}) \bigg] \bigg\}      
                    \quad\textrm{by \eqref{eqN.65.1}}
            \\
   =\frac{\sqrt{c_Nc_P}}{\sqrt{NP}}
           &\bigg[
               \frac{(-1)^P-(-1)^N}{4m} + I(N-P)
\\     &\qquad
+ O(P^{-1/3})+O(P^{-1/(2m)})+ O\Big(\delta^{1/2} 
                       + P^{-1/2}\delta^{-3/4}+ P^{-1}\delta^{-3/2}\Big)
            \bigg]
\ea                   
\ee
where
$$
   I(q)\equiv\begin{cases}\sin\Big(\frac{q\pi}2\Big)\,\frac2\pi
       \int_{-1}^{1}
       \frac{\cos(q\arcsin x)}
                 {(1-x^2)\,h(x)}\,dx, &q\textrm{ odd}\cr
         0, &q\textrm{ even.}
                   \end{cases}
$$
As $(\phi_N,\eps\phi_P)=-(\eps\phi_N,\phi_P)=-(\phi_P,\eps\phi_N)
       $ we see that \eqref{eqN.67.1}
is true also if $N<P$. 
Hence as $N,P\too\infty$
$$
      (\eps\phi_N,\phi_P) =
         \frac{\sqrt{c_Nc_P}}{\sqrt{NP}}
               \begin{cases}
                     \frac{(-1)^N}{2m} - I(N-P) + O(\delta^{1/2})+o_\delta(1),
                            &N-P\textrm{ odd}\cr
                         O(\delta^{1/2})+o_\delta(1),
                            &N-P\textrm{ even},
         \end{cases}
$$
where $o_\delta(1)\equiv O(P^{-1/3})+O(P^{-1/(2m)})+
 O(P^{-1/2}\delta^{-3/4}+ P^{-1}\delta^{-3/2})$.
Note that some of the smaller order terms in $o_\delta(1)$ are
  proportional to $\frac1{\delta^\alpha}$ for some
$\alpha>0$. Nevertheless
we see that e.g.~for $N-P$ odd
$$
\ba
  -O(\delta^{1/2}) \leq &\liminf_{N,P\too\infty}
                  \bigg[  \frac{\sqrt{NP}}{\sqrt{c_Nc_P}}
                         (\eps\phi_N,\phi_P) -
                     \bigg(\frac{(-1)^N}{2m} - I(N-P) \bigg)    \bigg]\\
          &\leq\limsup_{N,P\too\infty}
                 \bigg[   \frac{\sqrt{NP}}{\sqrt{c_Nc_P}}
                        (\eps\phi_N,\phi_P) -
                    \bigg(  \frac{(-1)^N}{2m} - I(N-P) \bigg)    \bigg] 
           \leq O(\delta^{1/2}).
\ea
$$
Letting $\delta\to0$ we conclude that
$\liminf=\limsup$ and hence
\be\la{eqN.68.1}
      (\eps\phi_N,\phi_P) =
         \frac{\sqrt{c_Nc_P}}{\sqrt{NP}}
                \bigg[     \frac{(-1)^N}{2m} - I(N-P) +o(1) \bigg]
\ee
as $N,P\too\infty$, $N-P$ odd. 
Similarly
\be\la{eqN.69.1}
      (\eps\phi_N,\phi_P) =
         \frac{\sqrt{c_Nc_P}}{\sqrt{NP}}
                \,o(1)
\ee
as $N,P\too\infty$, $N-P$ even.
This completes the proof of Theorem \ref{thm2}. 
\begin{remark}\la{rem4.139p}
As noted above, the proof of Theorem \ref{thm2}
involves a ${\liminf}$/${\limsup}$ argument as $N,P\too\infty$,
followed by a limit as $\delta\to0$.
However, as noted at the beginning of Subsection \ref{4.2.1},
 the error estimates from \cite{DKMVZ2}
in \eqref{eqdoms} et seq., are uniform only for $\delta$
in compact subsets of $(0,\delta_0]$,
and the reader may be concerned that the $O(\delta^{1/2})$
term in the ${\liminf}$/${\limsup}$ argument above, 
in fact depends on constants, say, that
blow up as $\delta\to0$.
But the reader may easily check that all the error terms that arise in the
proof of Theorem \ref{thm2}, and that could blow up as $\delta\to0$,
are always multiplied by negative power of $N$ or $P$,
and hence vanish in the ${\liminf}$/${\limsup}$ argument:
all that remains are $O(\delta^{1/2})$ terms that arise
from the evaluation of integrals with explicit $\delta$-independent
integrands (see e.g.~the proof of \eqref{eqN.65.1} above),
and do not blow up as $\delta\to0$.
\end{remark}
\begin{remark}
%The reader might worry that we could have made an arithmetic
%error on the way. Here are some aruments for a check up.
%First, the formula changes sign when we interchange $N$ and $P$
%as it should.
In the case $m=1$ the OP's are just the Hermite polynomials.
They have the exceptional property that if $N$ is odd then $\eps\phi_N$
is a polynomial of degree $N-1$ times $e^{-x^2/2}$,
and hence $(\eps\phi_N,\phi_P)$ is identically
zero for $N$ odd and $N<P$. 
However, we have 
$$
    \int_{-1}^1 \frac{\cos\big(q\arcsin x\big)}{1-x^2}\,dx 
    = \begin{cases}
            \pi,&q=1,5,9,\cdots\cr
            -\pi, &q=3,7,\cdots,
       \end{cases}
$$
and for $m=1$, $h(x)\equiv4$ by \eqref{hz}.
A simple calculation now shows that the leading coefficient
in \eqref{eqN.68.1} is zero
for $N$ odd, $N<P$, $P$ even, so that the calculations match.
\end{remark}
\begin{remark}
The following (nonrigorous) argument 
is consistent with our asymptotic formulae. 
We know by Theorem~\ref{thm1}
that for any polynomial $V$
 the matrix $D_\infty$ is banded and 
looks asymptotically 
like the product of a Toeplitz matrix, say $\tilde{D}_\infty$,
 whose diagonals
are given by certain binomial coefficients
times the diagonal matrix
 $\tilde{T}_\infty\equiv\diag((m\kappa_{2m}b_j^{2m-1})_{j\geq0})$,
$b_0\equiv1$ (see \eqref{Das}).
Take a large enough even $N$
and let $\tilde{D}_N$ be an $N\times{}N$
section 
of $\tilde{D}_\infty$, i.e.{} $(\tilde{D}_N)_{i,j}=(\tilde{D}_\infty)_{i,j}$,
$0\leq i,j\leq N-1$.
Also set
$$
          \tilde{T}_N\equiv\diag((m\kappa_{2m}b_j^{2m-1})_{j=0}^{N-1},
              \qquad b_0\equiv1.
$$
Consider a submatrix
in the middle of the matrix $(\tilde{D}_N\tilde{T}_N)^{-1}$
near its diagonal which is small compared to $N$.
For large $N$, in view of Theorem \ref{thm6}(i),
we would expect this small submatrix 
to look like the corresponding submatrix 
of the matrix $\eps_\infty$.

Let us take, e.g., $m=2$
and $V(x)=x^4$. Then $n=2m-1=3$.
In this case the Toeplitz matrix $\tilde{D}_N$
has diagonals (cf. Theorem \ref{thm1})
$$(\cdots\quad 0\quad1\quad0\quad3\quad0\quad-3\quad0\quad-1
  \quad0\quad\cdots).
$$
In view of the preceding discussion, we would
expect that for large enough even $N$
 the submatrix of $\tilde{T}_N^{-1}\tilde{D}_N^{-1}$
formed by the rows $M-n,\cdots,M-1$ and the columns
$M,\cdots,M+n-1$, $M$ even and $M\sim {}N/2$,
to look like the submatrix of $\eps_\infty$ 
formed by the same rows and columns.
In other words, we would expect that this small submatrix
of $\tilde{D}_N^{-1}$ would look like the small submatrix
of $\tilde{T}_N\eps_N$ located at the same position.
(Recall that the
rows and columns in $\tilde{D}_N^{-1}$
and $\tilde{T}_N\eps_N$ are enumerated from $0$ to $N-1$
so that the submatrix chosen above
corresponds to  $B_{12}$ defined in \eqref{B}
about the center $(M,M)$.)

The overall coefficient
multiplying the elements in the middle of the matrix $T_N\eps_N$
is the same as \eqref{s5eq80}, that is 
$
    m\kappa_{2m}b_{M}^{2m-1}\,c_MM^{-1} 
     = \frac{2(m!)^2}{(2m)!}(1+o(1))$,
for $M$ large enough.
Using Maple software we have computed
the matrix $\tilde{D}_N^{-1}$ for $N=20$. 
It turns out that the submatrix of $\tilde{D}_N^{-1}$
 with rows $7,8,9$ and columns $10,11,12$ 
(here $M=N/2=10$, $n=3$)
 and the corresponding submatrix
of $\tilde{T}_N\eps_N$ computed using 
Theorem~\ref{thm2}
(here $h(x)=\frac83(1+2x^2)$ by \eqref{hz}) are given by
$$
\Bigg(\begin{matrix}
       -0.01630&0&0.00435\\
                0&0.15078&0\\
                         0.06113&0&-0.01630
\end{matrix}\Bigg),\qquad
\Bigg(\begin{matrix}
       -0.01635&0&0.00438\\
                0&0.15032&0\\
                         0.06100&0&-0.01635
\end{matrix}\Bigg),
$$
respectively.
We see that the corresponding nonzero elements differ by at most $0.8\%$.
For $N=40$, $M=N/2$, the maximal difference is already less than $0.002\%$.
We have also done similar computations for $m=3,4$, $n=2m-1$
again for $N=40$, $M=N/2$.
It turns out that the nonzero elements in the 
small blocks 
computed using the matrix inversion and
Theorem~\ref{thm2}
as described above differ by at most
$0.005\%$, $0.02\%$ for $m=3,4$, respectively. 
\end{remark}
\subsection{Convergence of derivatives and integrals of the Christoffel--Darboux
kernel for weights $e^{-V(x)}$,
 $V(x)=\kappa_{2m}x^{2m}+\cdots$, $\kappa_{2m}>0$}
\la{subsec_for_sec_3}
We start with the convergence of the derivatives 
in the $12$ entries
of the kernels $K_{N,\beta}$, $\beta=1,4$.
Our main results in this direction are \eqref{M.4.5}, \eqref{M.5.1}
and their Corollaries \eqref{M.6.2}, \eqref{M.6.3} below.
After that we prove the convergence of the integrals 
in the $21$ entries
of the kernels $K_{N,\beta}$, $\beta=1,4$.
The main result for the integrals can be found in \eqref{M.13.0} below.
\subsubsection{Derivatives}
Fix $L_0>0$.
For $r$, $\xi$, $\eta$ in the compact set $|\xi|$, $|\eta|$, $|r|\leq2L_0$,
define
\be\la{M.1.1}
    D_{r,N}(\xi,\eta) \equiv \frac1{q_N}
              K_N\bigg(r+\frac\xi{q_N},r+\frac\eta{q_N}\bigg).
\ee
In what follows, $q_N$ is {\em any\ }sequence of numbers with the property
\be\la{M.1.2}
    q_N = \frac{Nh_N(0)}{2\pi c_N}\,(1+O(N^{-\alpha}))
\ee
for some $0<\alpha\leq1$, as $N\ra\infty$. 
\begin{remark}\la{rem4.141p}
In the calculations that follow we use formulae for $D_{r,N}(\xi,\eta)$
based on \eqref{eq_KN} that holds for $\xi\neq\eta$.
But $D_{r,N}(\xi,\eta)$ is continuous in $\xi,\eta$
(see e.g.~\eqref{K}), and it will be clear from the calculations
that we can obtain analogous
formulae for $D_{r,N}$ on the diagonal
by taking the limit $\eta\to\xi$ in the formulae below.
\end{remark}
Eventually in \eqref{M.1.1}, \eqref{M.1.2}
 we will take $q_N=R_{N,1,1}(r),R_{N/2,1,4}(r)$
for $\beta=1,4$ respectively,
but at this stage asymptotics for $R_{N,1,1}(r),R_{N/2,1,4}(r)$
 of type \eqref{M.1.2}
must still be proved, see \eqref{M.5.2}, \eqref{eqsc1}, \eqref{eqsc4} below
 (cf.~\cite[p.~240 et seq.]{D} where the analog
of \eqref{M.1.2} is proved for $R_{N,1,2}(0)=K_N(0,0)$
 in the case $V(x)=x^{2m}$
without lower order terms). 
By \eqref{cN},  $q_N\sim N^{1-1/(2m)}$ to leading order.

We use the notation in \cite{DKMVZ2}. In particular set 
$$
      g(z)\equiv g_N(z)=\int_{-1}^1 \log(z-x)\,d\mu^{\textrm{(eq)}}(x)
                =\int_{-1}^1 \log(z-x)\,\frac1{2\pi}|1-x^2|^{1/2}\,h_N(x)\,dx,
$$
$z\in\C\setminus(-\infty,-1]$, and for $z\in(-1,1)$ let
\be\la{M.2.1}
\ba
 S_+(z)=&\twotwo{c_N^{-N}}{0}{0}{c_N^{N}}
                        \,e^{-\frac{Nl}{2}\sigma_3}\, Y_+(c_Nz+d_N)\\
              &\,\times e^{-N(g_+(z)-\frac{l}{2})\sigma_3}\,
                 \twotwo{1}{0}{-e^{-N(g_+(z)-g_-(z))}}{1},
\ea
\ee
here $\pm$ refer to the boundary values from above/below the real axis,
respectively (cf. \cite[(4.22)]{DKMVZ2}).
Here $Y$ solves the Fokas--Its--Kitaev Riemann--Hilbert problem
fot the polynomials orthogonal with respect to the weight $e^{-V(x)}dx$
(see \cite[Thm.~3.1]{DKMVZ2}, and the constant $l\equiv l_N$
is given by (5.35), loc. cit.).
Finally, for $z\in(-1,1)$ let
\be\la{M.2.2}
   \xi_N(z) \equiv g_+(z)-g_-(z) = i\int_z^1 |1-x^2|^{1/2} h_N(x)\,dx
\ee
and set 
$$
            r_N\equiv (r-d_N)/c_N.
$$
Note that $\xi_N(z)\in i\R_+$
and $r_N=O(N^{-1/(2m)})$ by \eqref{cN}, \eqref{dN}.
We need the following properties of $S_+$ which are proved in
\cite[Sec.~7]{DKMVZ2}:

(i) For $-1<z<1$, 
\be\la{M.2.3}
\ba
   \det S_+(z) = 1.
\ea
\ee

(ii)
Let $a_+(z)=\big(\frac{z-1}{z+1}\big)_+^{1/4}$
and set
$$
  S_+^{(\infty)}(z)\equiv
    \frac12\twotwo{a_+(z)+a_+(z)^{-1}}{i(a_+(z)^{-1}-a_+(z))}
              {i(a_+(z)-a_+(z)^{-1})}{a_+(z)+a_+(z)^{-1}},
      \qquad -1<z<1.
$$
Then as $N\to\infty$, $S_+(z)$ converges with all its derivatives
to $S_+^{(\infty)}(z)$ for all $z$ in compact subsets of $(-1,1)$.
In particular, for any $0<\delta<1$,
\be\la{M.3.1}
   \sup_N \max_{-1+\delta\leq z\leq 1-\delta}
         \Big|\frac{d^k}{dz^k} S_+(z)\Big| \leq c_k<\infty
\ee
for $k=0,1,2,\cdots$.

Now a simple calculation (cf.~\cite[p.~24]{D})
shows that
\be\la{M.3.2}
\ba
   \null&D_{r,N}(\xi,\eta)\\
  \\&= \bigg[
        \bigg(   S_{+11}   \Big(r_N+\frac{\eta}{q_Nc_N}\Big)\,
           e^{\frac{N}2\xi_N\big(r_N+  \eta/(q_Nc_N)\big)}
    +S_{+12}   \Big(r_N+\frac{\eta}{q_Nc_N}\Big)\,
           e^{-\frac{N}2\xi_N\big(r_N+  \eta/(q_Nc_N)\big)}
   \bigg)\\
   &\quad\times\bigg(   S_{+21}   \Big(r_N+\frac{\xi}{q_Nc_N}\Big)\,
           e^{\frac{N}2\xi_N\big(r_N+  \xi/(q_Nc_N)\big)}
    +S_{+22}   \Big(r_N+\frac{\xi}{q_Nc_N}\Big)\,
           e^{-\frac{N}2\xi_N\big(r_N+  \xi/(q_Nc_N)\big)}
   \bigg)\\
   &\qquad-(\xi\leftrightarrow \eta)
   \bigg]/\big(2\pi i(\xi-\eta)\big)
\ea
\ee
where $(\xi\leftrightarrow \eta)$ indicates the same terms with $\xi$
and $\eta$ interchanged.
Consider the terms
\be\la{M.3.3}
\ba
   \bigg[
        &   S_{+11}   \Big(r_N+\frac{\eta}{q_Nc_N}\Big)\,
      S_{+22}   \Big(r_N+\frac{\xi}{q_Nc_N}\Big)\,
                e^{\frac{N}2\big(\xi_N(r_N+  \eta/(q_Nc_N))
                   -\xi_N(r_N+  \xi/(q_Nc_N))\big)}\\
     & -  S_{+11}   \Big(r_N+\frac{\xi}{q_Nc_N}\Big)\,
      S_{+22}   \Big(r_N+\frac{\eta}{q_Nc_N}\Big)\,
                e^{\frac{N}2\big(\xi_N(r_N+  \xi/(q_Nc_N))
                   -\xi_N(r_N+  \eta/(q_Nc_N))\big)} \bigg]\\
    &/\big(2\pi i(\xi-\eta)\big)
            \equiv \specIthree + \specIIthree + \specIIIthree
\ea
\ee
where
$$
\ba
   \specIthree\equiv &\frac
        {   S_{+11}   \Big(r_N+\frac{\eta}{q_Nc_N}\Big) -
      S_{+11}   \Big(r_N+\frac{\xi}{q_Nc_N}\Big)} {2\pi i(\xi-\eta)}\\
           &\times S_{+22}   \Big(r_N+\frac{\xi}{q_Nc_N}\Big)
   \,  e^{\frac{N}2\big(\xi_N(r_N+  \eta/(q_Nc_N))
                   -\xi_N(r_N+  \xi/(q_Nc_N))\big)},
\ea
$$
and
$$
\ba
   \specIIthree\equiv &S_{+11}   \Big(r_N+\frac{\xi}{q_Nc_N}\Big)
   \,  e^{\frac{N}2\big(\xi_N(r_N+  \eta/(q_Nc_N))
                   -\xi_N(r_N+  \xi/(q_Nc_N))\big)}\\
           &\times\frac
        {   S_{+22}   \Big(r_N+\frac{\xi}{q_Nc_N}\Big) -
               S_{+22}   \Big(r_N+\frac{\eta}{q_Nc_N}\Big)} {2\pi i(\xi-\eta)},
\ea
$$
and also
$$
\ba
   \specIIIthree\equiv &S_{+11}   \Big(r_N+\frac{\xi}{q_Nc_N}\Big)
            \,S_{+22}   \Big(r_N+\frac{\eta}{q_Nc_N}\Big)\\
     &\times\frac
      { e^{\frac{N}2\big(\xi_N(r_N+  \eta/(q_Nc_N))
                   -\xi_N(r_N+  \xi/(q_Nc_N))\big)}
           -e^{\frac{N}2\big(\xi_N(r_N+  \xi/(q_Nc_N))
                   -\xi_N(r_N+  \eta/(q_Nc_N))\big)}
       } {2\pi i(\xi-\eta)}.
\ea
$$
Note that 
\be\la{M.4.1}
\ba
   \frac
        {   S_{+11}   \Big(r_N+\frac{\eta}{q_Nc_N}\Big) -
      S_{+11}   \Big(r_N+\frac{\xi}{q_Nc_N}\Big)} {2\pi i(\xi-\eta)}
 =-\frac1{2\pi i}
     \int_0^1 S_{+11}^\prime  
              \Big(r_N+\frac{\eta+\tau(\xi-\eta)}{q_Nc_N}\Big)
        \, \frac{d\tau}{q_Nc_N}
\ea
\ee
and
\be\la{M.4.2}
\ba
   \frac{N}2\bigg(\xi_N\big(r_N+  \frac{\eta}{q_Nc_N}\big)
                   -\xi_N(r_N+  \frac{\xi}{q_Nc_N})\bigg)
      &=i\pi(\xi-\eta)\,G_N(\xi,\eta)\\
      &=i\pi(\xi-\eta) + O(N^{-\alpha})
\ea
\ee
as $N\to\infty$,
where we have denoted
\be\la{GN}
 G_N(\xi,\eta)\equiv
  \frac{N}{2\pi q_Nc_N} 
       \int_{0}^{1}
          \sqrt{ 1-\Big( \frac{\eta+t(\xi-\eta)}{q_Nc_N}\Big)^2 }
            h_N\Big( \frac{\eta+t(\xi-\eta)}{q_Nc_N}\Big)\,dt.
\ee
and observed that by \eqref{M.1.2},
\be\la{GNone}
           G_N(\xi,\eta) = \frac{Nh_N(0)}{2\pi q_Nc_N}\,(1+O(N^{-1})) 
         = 1+O(N^{-\alpha}),\qquad N\to\infty.
\ee
From the formulae above 
and the relation
 $\frac{iN(\xi-\eta)}{2q_Nc_N}
  =\frac{i\pi(\xi-\eta)}{h_N(0)+o(1)} $,
we see that $\specIthree$, and similarly $\specIIthree$, and all their $\xi,\eta$
derivatives converge to zero uniformly for $|\xi|,|\eta|,|r|\leq2L_0$,
with errors of order $O(N^{-1})$.

We also have the formula
\be\la{M.4.4}
\ba
   \specIIIthree=S_{+11}   \Big(r_N+\frac{\xi}{q_Nc_N}\Big)
            \,&S_{+22}   \Big(r_N+\frac{\eta}{q_Nc_N}\Big)
\\     &\times
           G_N(\xi,\eta)
                     \int_0^1 \cos\big(\tau\pi(\xi-\eta)\,G_N(\xi,\eta)\big)\,d\tau.
\ea
\ee
Together with similar formulae and
 calculations for the other terms in \eqref{M.3.2}
we see 
that $D_{r,N}(\xi,\eta)=\frac{\sin\big((\xi-\eta)
\frac{Nh_N(0)}{2\pi q_Nc_N}\big)}{\pi(\xi-\eta)}+O(N^{-1})
=K_\infty(\xi,\eta)+O(N^{-\alpha})+O(N^{-1})
$ 
where
$K_\infty(\xi-\eta)$ is the sine kernel $\frac{\sin\pi(\xi-\eta)}{\pi(\xi-\eta)}$.
The same is true for all the $\xi,\eta$ derivatives.
More precisely we have for any $j,k\geq0$,
\be\la{M.4.5}
         \frac{\partial^j}{\partial\xi^j}\,
            \frac{\partial^k}{\partial\eta^k}\,
            \frac1{q_N}
                 K_N\bigg(r+\frac\xi{q_N},r+\frac\eta{q_N}\bigg)
   \to  
            \frac{\partial^j}{\partial\xi^j}\,
                 \frac{\partial^k}{\partial\eta^k}\,
                     K_\infty(\xi-\eta)
\ee
uniformly for $|\xi|,|\eta|,|r|\leq2L_0$
as $N\to\infty$. 
% with the error $O(N^{-\alpha})$.
%
Note that  the terms of the form $S_{+11}(\cdots)S_{+21}(\cdots)$
and $S_{+12}(\cdots)S_{+22}(\cdots)$ do not contribute to leading
order, and also 
$S_{+11}(r_N)S_{+22}(r_N)-S_{+12}(r_N)S_{+21}(r_N)=1$
by \eqref{M.2.3}. 
Moreover, keeping track of the estimates, we find that
\be\la{M.5.1}
  \textrm{the error term in \eqref{M.4.5} is }O(N^{-\alpha})
  \textrm{ uniformly for }|\xi|,|\eta|,|r|\leq2L_0.
\ee
We see from the above that the largest error in \eqref{M.5.2}
arises purely from the asymptotic evaluation of $\frac{Nh_N(0)}{2\pi q_Nc_N}$
using \eqref{M.1.2} (cf.{} \eqref{GNone}):
 apart from this term the error in
\eqref{M.5.2} is $O(N^{-1})$, rather than $O(N^{-\alpha})$.
\subsubsection{}%An aside}
We can now verify asymptotics for $K_N(r,r)$ of type \eqref{M.1.2}.
Indeed from the uniform convergence
above, taking $q_N\equiv\frac{Nh_N(0)}{2\pi c_N}$,
for which we may, in turn, take $\alpha=1$ in \eqref{M.1.2},
we have for $|r|\leq2L_0$, as $N\to\infty$,
\be\la{M.5.2}
\ba
   K_N(r,r) &= q_N\,D_{r,N}(0,0) = q_N\lim_{\xi\to0}D_{r,N}(\xi,0) 
     = q_N\bigg(\lim_{\xi\to0}\frac{\sin\pi(\xi-0)}{\pi(\xi-0)} + O(N^{-1})\bigg)\\
     &= q_N\big(1 + O(N^{-1})\big)
   =\frac{Nh_N(0)}{2\pi c_N}\big(1 + O(N^{-1})\big).
%      =\frac{N}{2\pi c_N}\big(h_N(0)+ O(N^{-1})\big).
 \ea
\ee
This means in particular that \eqref{M.4.5}, \eqref{M.5.1}
are true for $q_N=K_N(r,r)$ with $\alpha=1$ for any $|r|\leq2L_0$.
Hence
\be\la{eqx0}
            K_N(r,r)  \sim N^{1-1/(2m)},\qquad\textrm{for }|r|\leq2L_0.
\ee
Estimating the correction terms for $S_{N,1}(r,r)$ 
as in \eqref{M.14.2}, we find for any $|r|\leq2L_0$
\be\la{eqsc1}
\ba
    \frac{S_{N,1}(r,r)}{K_N(r,r)} 
       &=  1+O(N^{-1/2}).
\ea
\ee
Similarly for any $|r|\leq2L_0$
\be\la{eqsc4}
    \frac{S_{N/2,4}(r,r)}{K_N(r,r)} = 1+O(N^{-1/2}).
\ee
Note that \eqref{eqsc1}, \eqref{eqsc4} together with
 \eqref{eqx0} prove \eqref{eqscsec1}, and hence \eqref{eqSTARSTAR}.
\subsubsection{Derivatives, continued}
We now consider the term
\be\la{M.6.1}
          X_N(\xi,\eta)\equiv \frac1{q_N^2}(K_ND)
                       \bigg(r+\frac\xi{q_N},r+\frac\eta{q_N}\bigg),\qquad
           q_N\textrm{ as in \eqref{M.1.2}}.
\ee
Terms of this form arise in the $12$ entry of 
$\frac1{q_N}K_{N,1}^{(\lambda_{N,1})}
       \big(r+\frac\xi{q_N},r+\frac\eta{q_N}\big)$.
Clearly $X_N(\xi,\eta)=-\frac{\partial}{\partial\eta}D_{r,N}(\xi,\eta)$
(cf.~\eqref{M.1.1}) and so by \eqref{M.4.5} we see that
\eqref{M.6.1} satisfies
\be\la{M.6.2}
          X_N(\xi,\eta)\to
      -\frac{\partial}{\partial\eta}K_{\infty}(\xi-\eta)=
       \frac{\partial}{\partial\xi}K_\infty(\xi-\eta)
\ee
together with all its $\xi,\eta$ derivatives,
uniformly for $|\xi|,|\eta|,|r|\leq2L_0$.
Moreover as before
\be\la{M.6.3}
  \textrm{the error term in \eqref{M.6.2} is }O(N^{-\alpha})
  \textrm{ uniformly for }|\xi|,|\eta|,|r|\leq2L_0.
\ee
\subsubsection{Integrals}
Finally, consider the term
\be\la{M.6.4}
  Y_N(\xi,\eta)\equiv \frac12 \int_{-\infty}^\infty \sgn
         \bigg(r+\frac\xi{q_N} -s\bigg)
         \,K_N\bigg(s,\, r+\frac\eta{q_N}\bigg) \,ds,\qquad
           q_N\textrm{ as in \eqref{M.1.2}}.
\ee
Terms of this form arise in the $21$ entry of 
 $\frac1{q_N}K_{N,1}^{(\lambda_{N,1})}
       \big(r+\frac\xi{q_N},r+\frac\eta{q_N}\big)$.
We want to show that
 $Y_N(\xi,\eta)\to\int_0^{\xi-\eta}K_\infty(t)\,dt$ as $N\to\infty$.
Changing variables $s=r+(\eta+t)/q_N$ we write $Y_N$ as
\be\la{M.6.5}
   Y_N(\xi,\eta) = \specIsix+\specIIsix+\specIIIsix
\ee
where (cf.~\eqref{M.1.1})
$$
\ba
   \specIsix &\equiv\int_0^{\xi-\eta} D_{r,N}(t+\eta,\eta)\,dt\\
   \specIIsix &\equiv\frac12\int_{-\infty}^0 D_{r,N}(t+\eta,\eta)\,dt\\
   \specIIIsix &\equiv-\frac12\int_0^{\infty} D_{r,N}(t+\eta,\eta)\,dt.
\ea
$$
Again we consider $\xi,\eta,r$ in the compact set $[-2L_0,2L_0]$.
By \eqref{M.4.5},  $\specIsix$ converges, with all its derivatives, to
$
   \int_0^{\xi-\eta}K_\infty(t+\eta-\eta)\,dt 
        = \int_0^{\xi-\eta}K_\infty(t)\,dt
$
as $N\to\infty$, and so we must show that $\specIIsix+\specIIIsix$
converges to $0$.

To analyze $\specIIsix,\specIIIsix$ it is convenient to consider the cases $|t|\leq1$
and $|t|>1$ separately.
By the above calculations, we have as $N\to\infty$
\be\la{M.7.1}
\ba
  \frac12\bigg(&\int_{-1}^0 
                             -\int_0^1 \bigg) D_{r,N}(t+\eta,\eta)\,dt\\
  &=\frac12\bigg(\int_{-1}^0 
                             -\int_0^1 \bigg) K_\infty(t+\eta-\eta)\,dt+ O(N^{-\alpha})
     = O(N^{-\alpha}),
\ea
\ee
as $K_\infty(t)$ is even.
As before the term $O(N^{-\alpha})$ is uniform
for $|r|,|\eta|\leq2L_0$.
Now we must show
$\specIseven-\specIIseven\to0$ as $N\to\infty$, where 
$$
  \specIseven \equiv\int_{-\infty}^{-1} D_{r,N}(t+\eta,\eta)\,dt
   \qquad\textrm{and}\qquad
 \specIIseven \equiv \int_1^{\infty} D_{r,N}(t+\eta,\eta)\,dt.
$$

Consider $\specIIseven$. We utilize the change of variables
\be\la{M.7.2}
         r+(t+\eta)/q_N = c_Nz+d_N
\ee
i.e.
\be\la{M.7.3}
\ba
       t &= -\eta +(d_N-r)q_N + q_Nc_Nz\\
       z &= (r-d_N)/c_N + (t+\eta)/(q_Nc_N).
\ea
\ee
Fix $\delta>0$ small as in Subsection \ref{ssec42},
set 
$t_{\pm}\equiv-\eta +(d_N-r)q_N + q_Nc_N(1\pm\delta)$,
and write $\specIIseven\equiv \specIeight+\specIIeight+\specIIIeight$
where
$$
    \specIeight\equiv\int_1^{t_-},\qquad
       \specIIeight\equiv\int_{t_-}^{t_+},\qquad
         \specIIIeight\equiv\int_{t_+}^\infty.
$$
Note that $t_{\pm}\sim  N$ as $N\to\infty$.

We have by \eqref{M.1.1}, \eqref{eq_KN}
$$
   \specIIIeight = \int_{t_+}^\infty
        b_{N-1}\bigg[ 
     \phi_N \bigg(r+\frac{t+\eta}{q_N}\bigg)
       \phi_{N-1} \bigg(r+\frac{\eta}{q_N}\bigg)
   -     \phi_{N-1} \bigg(r+\frac{t+\eta}{q_N}\bigg)
       \phi_{N} \bigg(r+\frac{\eta}{q_N}\bigg)
   \bigg]\,\frac{dt}{t}.
$$
Changing variables $t\to z$ as in \eqref{M.7.3},
the first term takes the form
$$
 \int_{1+\delta}^\infty
        b_{N-1}
     \phi_N(c_Nz+d_N)\,
       \phi_{N-1}(c_Nz_0+d_N)
           \,\frac{q_Nc_N\,dz}{q_Nc_Nz+q_N(d_N-r)-\eta}
$$
where $z_0=z(t=0)=(r-d_N)/c_N + \eta/(q_Nc_N) =O(N^{-1/(2m)})$.
As $b_{N-1}\sim  c_N/2$,
using \eqref{eqthm10}, \eqref{eqI.10.1}
we see that the above term is bounded by
$$
 \const \cdot ((c_N/N)^{1/2}e^{-cN})\cdot (1/c_N^{1/2})
      \leq \const \cdot e^{-cN}
$$
for some $c=c(\delta)>0$. There is a similar estimate
 for the second term in $\specIIIeight$, and we have
\be\la{M.8.1}
   |\specIIIeight|=O(e^{-cN}),\qquad c=c(\delta)>0,
\ee
uniformly for $|r|,|\eta|\leq2L_0$.

Changing variables $t\to z$ as in \eqref{M.7.3}
in the first term of $\specIIeight$, and then integrating by parts, we obtain
$$
\ba
        b_{N-1} \phi_{N-1}(c_Nz_0+d_N)
  \bigg[&\bigg( \int_{1-\delta}^{1+\delta}\phi_{N}(c_Nw+d_N)\,dw\bigg)
     \,\frac1{1+\delta+\frac{d_N-r}{c_N}-\frac{\eta}{q_Nc_N}}\\
  &+\int_{1-\delta}^{1+\delta}
  \bigg( \int_{1-\delta}^{z}\phi_{N}(c_Nw+d_N)\,dw\bigg)
     \,\frac{dz}{(z+\frac{d_N-r}{c_N}-\frac{\eta}{q_Nc_N})^2}
  \bigg]
\ea
$$
which is bounded by
$$
 \const \cdot c_N \cdot c_N^{-1/2} \cdot (c_N^{-1/2}N^{-1/2})
      = O(N^{-1/2})
$$
in view of \eqref{eqthm10}, \eqref{eqI.17.1} (of course $z\sim1$ 
in the above integrals, so
there is no singularity in the integrand).
Again there is a similar estimate
 for the second term in $\specIIeight$, and so
\be\la{M.9.1}
   |\specIIeight|=O(N^{-1/2})
\ee
uniformly for $|r|,|\eta|\leq2L_0$.

Let $0<\alpha^\prime<\min\big(\frac12,\alpha\big)$
 and set $t_N\equiv N^{\alpha^\prime}$.
Thus $1<t_N<t_-$ as $N\to\infty$. 
Let $\specIeight=\specIeight^\prime+\specIeight^{\prime\prime}$ where
$$
  \specIeight^{\prime} \equiv\int_1^{t_N} D_{r,N}(t+\eta,\eta)\,dt
   \qquad\textrm{and}\qquad
 \specIeight^{\prime\prime} \equiv \int_{t_N}^{t_-} D_{r,N}(t+\eta,\eta)\,dt.
$$
First we consider $\specIeight^{\prime}$. From \eqref{M.3.2} we again see that
$D_{r,N}(t+\eta,\eta)$ is a sum
of terms of the form $S_{+11}(\cdots)S_{+22}(\cdots)$,
$S_{+12}(\cdots)S_{+21}(\cdots)$, $\cdots$.
Express the term $S_{+11}S_{+22}$ as a sum 
$$
   \specIthree(t+\eta,\eta)+\specIIthree(t+\eta,\eta)+\specIIIthree(t+\eta,\eta)
$$
as in \eqref{M.3.3}. As in the proof of \eqref{M.4.5}, we have
$|\specIthree(t+\eta,\eta)|,|\specIIthree(t+\eta,\eta)|=O(N^{-1})$ as $N\to\infty$
uniformly for $1<t<t_N$ and $|r|,|\eta|\leq2L_0$.
Thus the contribution of $\specIthree(t+\eta,\eta),\specIIthree(t+\eta,\eta)$
to $\specIeight^\prime$ is $O(t_N N^{-1})=O(N^{-1+\alpha^\prime})$
uniformly for $|r|,|\eta|\leq2L_0$.
From \eqref{M.4.4}
$$
  \specIIIthree(t+\eta,\eta)  = S_{+11}   \Big(r_N+\frac{t+\eta}{q_Nc_N}\Big)
            \,S_{+22}   \Big(r_N+\frac{\eta}{q_Nc_N}\Big)
              \, \frac{\sin(\pi t\,G_N(t+\eta,\eta))}{\pi t}
$$
where $G_N$ is as in \eqref{GN}.
By \eqref{M.3.1}
$$
\ba
  \specIIIthree(t+\eta,\eta)  &=
       S_{+11} (r_N)
            \,S_{+22}(r_N)\,
               \frac{\sin(\pi t\,G_N(t+\eta,\eta))}{\pi t} 
        + O\bigg(\frac1N \,\max_{t\geq1}\Big|\frac{t+|\eta|}{t}\Big|\bigg)\\
    &=
       S_{+11} (r_N)
            \,S_{+22}(r_N)\,
               \frac{\sin \pi t}{\pi t}\,
                     \cos \big[\pi t (G_N(t+\eta,\eta)-1) \big]\\
     &\quad+ S_{+11} (r_N)
            \,S_{+22}(r_N)\,
               \frac{\cos\pi t}{\pi t}\,
                     \sin \big[\pi t (G_N(t+\eta,\eta)-1) \big]
        + O(N^{-1}).
\ea
$$
But by \eqref{M.1.2}, \eqref{M.5.2},
we find from \eqref{GN}
$$
   G_N(t+\eta,\eta) = \frac{Nh_N(0)}{2\pi q_Nc_N}\,(1+O(t/N))
                               = 1+ O(t/N) + O(N^{-\alpha})
$$
uniformly for $1\leq t\leq t_N$, $|\eta|\leq2L_0$.
Thus
$$
   \bigg| \frac{ \sin \big[\pi t (G_N(t+\eta,\eta)-1) \big]}{\pi t} \bigg| 
      \leq |G_N(t+\eta,\eta)-1| = O(t/N) + O(N^{-\alpha})
$$
and similarly
$$
   \big| \cos \big[\pi t (G_N(t+\eta,\eta)-1) \big] -1  \big| 
      = O(t^4/N^2) + O(t^2/N^{2\alpha}).
$$
Thus as $\alpha\leq1$,
$$
  \specIIIthree(t+\eta,\eta)  =
       S_{+11} (r_N)
            \,S_{+22}(r_N)\,
               \frac{\sin\pi t}{\pi t} 
        + O\bigg(\frac{t^3}{N^2} + \frac{t}N 
                     + \frac{t}{N^{2\alpha}} + \frac1{N^\alpha} \bigg)
$$
and so
$$
\ba
  \int_{1}^{t_N} \specIIIthree(t+\eta,\eta)\,dt  =
       &S_{+11} (r_N)
            \,S_{+22}(r_N)\,
               \int_{1}^{t_N} \frac{\sin\pi t}{\pi t} \,dt\\
        &+ O\bigg( \frac{N^{4\alpha^\prime}}{N^2} 
                              + \frac{N^{2\alpha^\prime}}N
               + \frac{N^{2\alpha^\prime}}{N^{2\alpha}}
               +  \frac{N^{\alpha^\prime}}{N^{\alpha}} \bigg).
\ea
$$
Assembling the above estimates we find for $0<\alpha^\prime<\min\big(\frac12,
\alpha\big)$
as $N\to\infty$, the contribution of the $ S_{+11}S_{+22}$
term to $\specIeight^\prime$ is given by
$$
    S_{+11} (r_N)
            \,S_{+22}(r_N)\,
               \int_{1}^{t_N} \frac{\sin\pi t}{\pi t} \,dt
        + O(N^{-1+2\alpha^\prime})+ O(N^{-\alpha+\alpha^\prime})
$$
uniformly for $|r|,|\eta|\leq2L_0$. 
Analyzing the other contributions
$ S_{+12}S_{+21}$, $\cdots$ we obtain as $N\to\infty$
\be\la{M.11.1}
   \specIeight^{\prime} =\int_1^{t_N} D_{r,N}(t+\eta,\eta)\,dt
        = \int_{1}^{t_N} \frac{\sin\pi t}{\pi t} \,dt
              + O(N^{-1+2\alpha^\prime})+ O(N^{-\alpha+\alpha^\prime})
\ee
uniformly for $|r|,|\eta|\leq2L_0$. Again we have used \eqref{M.2.3}.

Finally we consider 
$$
  \specIeight^{\prime\prime}  
    = \int_{t_N}^{t_-}
        b_{N-1}\bigg[ 
     \phi_N \bigg(r+\frac{t+\eta}{q_N}\bigg)
       \phi_{N-1} \bigg(r+\frac{\eta}{q_N}\bigg)
   -     \phi_{N-1} \bigg(r+\frac{t+\eta}{q_N}\bigg)
       \phi_{N} \bigg(r+\frac{\eta}{q_N}\bigg)
   \bigg]\,\frac{dt}{t}.
$$
Changing variables $t\to z$ as in \eqref{M.7.3} and then
integrating by parts, we obtain for the first term in $\specIeight^{\prime\prime}$
$$
\ba
        b_{N-1} \phi_{N-1}(r&+\eta/q_N)
  \bigg[\bigg( \int_{z(t_N)}^{1-\delta}
            \phi_{N}(c_Nw+d_N)\,dw\bigg)
     \,\frac{q_Nc_N}{q_Nc_N(1-\delta) + (d_N-r)q_N-\eta}\\
  &+\int_{z(t_N)}^{1-\delta}
  \bigg( \int_{z(t_N)}^{z}
             \phi_{N}(c_Nw+d_N)\,dw\bigg)
    \,\frac{(q_Nc_N)^2\,dz}{(q_Nc_Nz + (d_N-r)q_N-\eta)^2}
  \bigg].
\ea
$$
Note that $z(t_N)\sim N^{-1+\alpha}$
and recall \eqref{eqthm10}, \eqref{eqI.18.1}
%\eqref{eqN.4.1half}, \eqref{eqI.16.2}
to conclude that the first term in  $\specIeight^{\prime\prime}$ is bounded by
$$
 \const \cdot b_{N-1} \cdot c_N^{-1/2} \cdot
       (c_N^{-1/2}N^{-1}) \cdot \bigg(\const +
            q_Nc_N\int_{t_N}^{t_-}\frac{dt}{t^2} \bigg)
      =  O(N^{-\alpha^\prime}).
$$
The same is true for the second term in $\specIeight^{\prime\prime}$,
and so together with \eqref{M.11.1} we find
\be\la{M.12.1}
   \specIeight = \int_1^{t_-} D_{r,N}(t+\eta,\eta)\,dt
        = \int_{1}^{t_N} \frac{\sin\pi t}{\pi t} \,dt
              + O(N^{-1+2\alpha^\prime} + N^{-\alpha^\prime}
               + N^{-\alpha+\alpha^\prime})
\ee
and hence, as $\alpha^\prime<1/2$
$$
\ba
  \specIIseven &= \int_1^{\infty} D_{r,N}(t+\eta,\eta)\,dt\\
        &= \int_{1}^{t_N} \frac{\sin\pi t}{\pi t} \,dt
              + O( N^{-1+2\alpha^\prime} + N^{-\alpha^\prime}+N^{-1/2}
               + N^{-\alpha+\alpha^\prime}+e^{-cN})\\
           &=\int_{1}^{t_N} \frac{\sin\pi t}{\pi t} \,dt
              + O( N^{-1+2\alpha^\prime} + N^{-\alpha^\prime}
               + N^{-\alpha+\alpha^\prime}).
\ea
$$
The best error estimate is clearly obtained for $\alpha^\prime=\alpha^*\equiv
\min\big(\frac\alpha{2},\frac13\big)<\alpha$.
Thus
\be\la{M.12.2}
    \int_1^\infty D_{r,N}(t+\eta,\eta)\,dt
        = \int_{1}^{t_N} \frac{\sin\pi t}{\pi t} \,dt
              + O(N^{-\alpha^*})
\ee
and similarly
\be\la{M.12.3}
    \int_{-\infty}^{-1} D_{r,N}(t+\eta,\eta)\,dt
        = \int_{-t_N}^{-1} \frac{\sin\pi t}{\pi t} \,dt
              + O(N^{-\alpha^*}).
\ee
But $\frac{\sin\pi t}{\pi t}$ is even, and together with \eqref{M.7.1}
we see that 
\be\la{M.12.4}
   \specIIsix+\specIIIsix =\frac12\bigg( \int_{-\infty}^{0} D_{r,N}(t+\eta,\eta)\,dt
          -  \int_0^{\infty} D_{r,N}(t+\eta,\eta)\,dt\bigg)
        = O(N^{-\alpha^*})
\ee
as $N\to\infty$ uniformly for $|r|,|\eta|\leq2L_0$.
We have proved the following (cf.~\eqref{M.6.4}):
\bq\la{M.13.0}
\ba
      Y_N(\xi,\eta) &= A_N(\xi,\eta) + B_N(\eta),\qquad
                 \textrm{where as }N\to\infty\\
     \textrm{(a)}\quad &A_N(\xi,\eta)\textrm{ converges with all its $\xi,\eta$
            derivatives to $\int_0^{\xi-\eta}K_\infty(t)\,dt$}\\
              &\qquad\qquad\textrm{with error $O(N^{-\alpha})$ uniformly for 
                                           $\xi,\eta,r\in[-2L_0,2L_0]$;}\\   
  \textrm{(b)}\quad &B_N(\eta)=O(N^{-\alpha^*})
             \textrm{ uniformly for $\eta,r\in[-2L_0,2L_0]$}.
\ea
\ee
Here $\alpha,\alpha^*$ are as in \eqref{M.1.2}, \eqref{M.12.2},
respectively.
\section{Proofs of Theorems~\ref{thm6}, 
~\ref{thm4} and~\ref{thm5}}
\la{sectfive}
\subsection{Proof of Theorem~\ref{thm6}}
We know by Lemma~\ref{lembor}
that $D\phi_k$ is a (finite) linear 
combination of $\phi_{\max(0,k-n)},\cdots,\phi_{k+n}$, $n=\deg{}V^\prime$.
Hence
\be\la{s6e1}
    D\phi_k=\sum_{j=\max(0,k-n)}^{k+n}(D\phi_k,\phi_j)\,\phi_j.
\ee
Note next that since $\phi_k$ and its derivative are 
rapidly decaying, 
\be\la{s6e2}
      \eps{}D\phi_k=\phi_k,\qquad k\in\ZP.
\ee
Apply $\eps$ to \eqref{s6e1} and take into account \eqref{s6e2}
to find
\be\la{s6e3}
       \phi_k = \eps D\phi_k = \sum_{j=\max(0,k-n)}^{k+n}(D\phi_k,\phi_j)\,\eps\phi_j.
\ee
Finally, take the inner product of \eqref{s6e3} with any $\phi_l$
to find
\be\la{s6e4}
        \delta_{k,l} =  \sum_{j=\max(0,k-n)}^{k+n}
              (D\phi_k,\phi_j)\,(\eps\phi_j,\phi_l)
\ee
which shows that
  $I_\infty=D_\infty\eps_\infty$ (recall that $(D\phi_k,\phi_j)=0$
for $|j-k|>n$).
Taking the transposes and using the fact
that $D_\infty$ and $\eps_\infty$
are both skew symmetric gives the second statement in Theorem~\ref{thm6}(i).

Now for the proof of Theorem~\ref{thm6}(ii).
Denote the $N\times(N+n)$ section
of $\eps_\infty$ by $\eps_{N,N+n}$
and denote the last $n$ columns of  $\eps_{N,N+n}$ by
\be\la{epsstar}        
  \bigg(\begin{matrix} \eps_{N-n,n}^*\\
                                  B_{12} 
\end{matrix}\bigg)
\ee
($\eps_{N-n,n}^*$ is the same matrix as in \eqref{W}).
Denote the $(N+n)\times{}N$ section of $D_\infty$ by $D_{N+n,N}$.
Note that the last $n$ rows of this matrix are 
$       
  \big(\begin{matrix} 
                         0_{n,N-n} & D_{21} 
\end{matrix}\big).
$
By Theorem~\ref{thm6}(i)
$$
\begin{aligned}
      \twotwo{I_{N-n}}{0}{0}{I_n} &= \eps_{N,N+n} D_{N+n,N}
% \\  &
= \eps_{N} D_N 
       + \Bigg(0_{N,N-n}\Bigg| 
     \genfrac{}{}{0pt}{}{\eps_{N-n,n}^*D_{21}}{B_{12}D_{21}}\Bigg)
\end{aligned}
$$
which proves Theorem~\ref{thm6}(ii).
\subsection{Proof of Theorem~\ref{thm4}}
Recall the definitions of the block matrices $A$, $B$, and $C$.
Note first that
$$
       BA = \twotwo{B_{12}A_{21}}{B_{11}A_{12}}
                              {B_{22}A_{21}}{B_{21}A_{12}}
              = \twotwo{-B_{12}D_{21}}{B_{11}D_{12}}
                              {-B_{22}D_{21}}{B_{21}D_{12}}.
$$
Recall from Section \ref{secttwo}
 that $R$ denotes the $n\times{}n$ matrix
with all zeros and ones on the anti-diagonal.
Let $G^T$ denote the transpose of $G$
and $G^\perp$ be the transpose with respect to the anti-diagonal.
An application of $R$ to a matrix from both sides
interchanges the order of all rows and columns in the matrix.
We have for any $G$
\be\la{eqRR}
  RGR=(G^\perp)^T=(G^T)^\perp\quad\textrm{and also}\quad RR=I_n.
\ee
%Below when we write ``$o(1)$'' or ``modulo $o(1)$'' we let $N\to\infty$, $N$ even.
It is convenient to denote for $j,k=1,2$
$$
      \tilde{B}_{jk}\equiv N^{1-1/(2m)}\,B_{jk},\qquad
        \tilde{D}_{jk}\equiv N^{-1+1/(2m)}\,D_{jk},\qquad
         \tilde{A}_{jk}\equiv N^{-1+1/(2m)}\,A_{jk}.
$$
Then by Theorem \ref{thm1} and \ref{thm2} the matrices
with the tilde tend to certain explicit limiting matrices, as $N\ra\infty$,
see e.g.~\eqref{BAformula}, \eqref{cN}. 
Note that \eqref{BAformula} implies that as $N\to\infty$, $N$ even,
\be\la{eqperpprop}
    \tilde{B}_{12}^\perp = \tilde{B}_{12}+o(1),
          \qquad  \tilde{D}_{21}^\perp= \tilde{D}_{21}+o(1)
\ee
and also from the skew symmetry of $\eps_\infty$ and $D_\infty$
$$ 
    B_{12}^T = -B_{21},\qquad D_{21}^T=-D_{12}.
$$

We prove now the first statement of Theorem~\ref{thm4}:
\be\la{s6eq20}
\begin{aligned}
  -R(BA)_{11}R &= RB_{12}D_{21}R = RB_{12}RRD_{21}R
       = (B_{12}^\perp)^T(D_{21}^\perp)^T
       = (\tilde{B}_{12}^\perp)^T(\tilde{D}_{21}^\perp)^T\\
        &  = (\tilde{B}_{12}^T+o(1))\,( \tilde{D}_{21}^T+o(1))
          =(\tilde{B}_{21}+o(1))\,(\tilde{D}_{12} +o(1))\\
        &=\tilde{B}_{21}\tilde{D}_{12} +o(1)  
           ={B}_{21}{D}_{12} +o(1) = (BA)_{22}+o(1).
\end{aligned}
\ee

Next we prove the second statement of Theorem~\ref{thm4}.
First note that 
$$
       BAC = \twotwo
        {(BAC)_{11}}{(BAC)_{12}}{(BAC)_{21}}{(BAC)_{22}}
$$
where
\be\la{allfour}
\begin{aligned}
        (BAC)_{11}&= B_{12}A_{21}+B_{12}A_{21}B_{12}A_{21}
                                   +B_{11}A_{12}B_{22}A_{21} \\
        (BAC)_{12}&= B_{12}A_{21}B_{11}A_{12}
                                   +B_{11}A_{12}B_{21}A_{12}\\
        (BAC)_{21}&= B_{22}A_{21} + B_{22}A_{21}B_{12}A_{21} 
                                     + B_{21}A_{12}B_{22}A_{21}\\
        (BAC)_{22}&= B_{22}A_{21}B_{11}A_{12}
                                     +B_{21}A_{12}B_{21}A_{12}.
\end{aligned}
\ee
Let us prove first that $(BAC)_{12}=0$. Since $\det{A_{12}}\neq0$
(by Theorem \ref{thm1} and since $\det\tilde{A}_{12}\neq0$)
this is the same as proving
\be\la{s6eq10}
  B_{12}A_{21}B_{11} +B_{11}A_{12}B_{21} = 0.
\ee
We claim that to prove \eqref{s6eq10} is in turn the same as to prove
that $B_{12}A_{21}B_{11}$ is skew symmetric. 
This is because 
$$
       B_{11} A_{12} B_{21}
        = B_{11}^T A_{21}^T B_{12}^T =  (B_{12}A_{21}B_{11})^T
$$
(recall $A_{12}=A_{21}^T$).
Now since $B_{11}$ is skew, we note that $B_{12}A_{21}B_{11}$ is skew if and only if
$$
    (I_n+B_{12}A_{21})B_{11} = (I_n-B_{12}D_{21})B_{11} 
$$
is skew. But the last matrix is skew because it is the lower right $n\times n$
corner of the skew matrix $\eps_N D_N\eps_N$.
 (Here we used \eqref{W}
and the fact that $B_{11}$ is the lower right $n\times n$
corner of $\eps_N$.)

Next we show that $(BAC)_{11}=0$. Since $\det{A_{21}}\neq0$
this is the same as proving that 
\be\la{s6eq22}
   (I_n+B_{12}A_{21}) B_{12}  +B_{11}A_{12}B_{22}  = 0.
\ee
Recall that $A_{12}=D_{12}$, $A_{21}=-D_{21}$
 and note that we have to prove
\be\la{s6eq14}
        (I_n-B_{12}D_{21})B_{12} =  -B_{11}D_{12}B_{22}
\ee
The matrix $(I-B_{12}D_{21})$ as we know is the lower right $n\times n$
corner of $\eps_N D_N$. In view of \eqref{W} and \eqref{epsstar},
$(I-B_{12}D_{21})B_{12}$ equals the product of
the last $n$ rows of $(\eps_N D_N)$ with $\bigg(\begin{matrix} \eps_{N-n,n}^*\\
                                  B_{12}
\end{matrix}\bigg)$. 
But this is the 
same as the product of the  last $n$ rows of $\eps_N$ with 
$
   D_N\bigg(\begin{matrix} \eps_{N-n,n}^*\\
                                  B_{12}
\end{matrix}\bigg)
$.
We now derive a different expression for the latter product.
To this end we consider the product
 of the rows $0,\cdots,N-1$ in $D_\infty$ with the columns
$N,\cdots,N+n-1$ in $\eps_\infty$.
Since $D_\infty\eps_\infty=I_\infty$ this product equals $0_{N,n}$.
But using the fact that $D_\infty$ is banded, and 
the expression \eqref{epsstar} for $\eps_{N,N+n}$,
 this relation becomes
$$
      D_{N}\, \bigg(\begin{matrix} \eps_{N-n,n}^*\\
                                  B_{12}
           \end{matrix}\bigg) = \bigg(\begin{matrix} 0_{N-n,n}\\
                                   -D_{12}B_{22}
           \end{matrix}\bigg).
$$
Multiplying the latter expression 
 from the left by the last $n$ rows of $\eps_N$ 
we see that the product is $-B_{11}D_{12}B_{22}$, and so \eqref{s6eq14}
is established.

Our next goal is to show $(BAC)_{22}=(BA)_{22}+o(1)$.
 In view of \eqref{s6eq20}
we must prove that $(BAC)_{22}$ equals 
$$\begin{aligned}
      -R(BA)_{11}R+o(1) 
      &= -RB_{12}A_{21}R+o(1)
                         =  -(RB_{12}R)(RA_{21}R)+o(1)\\
         &=  - (RB_{12}R)(A_{21}^\perp)^T +o(1)
                  = -(RB_{12}R) A_{21}^T +o(1)\\
      &= - (RB_{12}R)A_{12}+o(1).
\end{aligned}
$$
Now using \eqref{allfour} and inserting tildes,
we see that (recall $\det\tilde{A}_{12}\neq0$)
 we must prove
$$
     \tilde{B}_{22}\tilde{A}_{21}\tilde{B}_{11} 
       + \tilde{B}_{21}\tilde{A}_{12}\tilde{B}_{21} 
                    + R\tilde{B}_{12} R =  o(1).
$$
After noting that $R\tilde{B}_{12}R=-\tilde{B}_{21}+o(1)$,
and then taking transposes, we see that we must prove
$$
     \tilde{B}_{12} + \tilde{B}_{11}\tilde{A}_{12}\tilde{B}_{22} + \tilde{B}_{12}\tilde{A}_{21}\tilde{B}_{12}= o(1).
$$
But this follows from \eqref{s6eq22} (which holds exactly) 
and which we have already proved in the $(BAC)_{11}$ part.

Finally we show $(BAC)_{21}=(BA)_{21}+o(1)$ 
(which is not really needed in this paper;
our proof of the fact that $\lim_{N\to\infty}\det( I_{2n}+BAC)\neq0$
in Section \ref{secttwo} only uses $(BAC)_{22}=(BA)_{22}+o(1)$). 
By \eqref{allfour}, since $(BA)_{21}=B_{22}A_{21}$
and again $\det{\tilde{A}_{21}}\neq0$,
we must show
$$
        \tilde{B}_{22}\tilde{A}_{21}\tilde{B}_{12}   
              + \tilde{B}_{21}\tilde{A}_{12}\tilde{B}_{22} = o(1).
$$
Since $\tilde{B}_{22}=\tilde{B}_{11}^\perp+o(1)$,
this reduces to showing
$$
        \tilde{B}_{11}^\perp\tilde{A}_{21}\tilde{B}_{12}   
             + \tilde{B}_{21}\tilde{A}_{12}\tilde{B}_{11}^\perp 
                   = o(1).
$$
Now we apply the transposition across the anti-diagonal
and use the property $(GH)^\perp=H^\perp G^\perp$ 
together with \eqref{eqperpprop} to find
$$
        \tilde{B}_{11}\tilde{A}_{12}\tilde{B}_{21}   
             + \tilde{B}_{12}\tilde{A}_{21}\tilde{B}_{11} 
                   = o(1).
$$
But this follows from \eqref{s6eq10} (the latter holds exactly)
which we proved in the $(BAC)_{12}$ part.
The proof of Theorem~\ref{thm4} is now complete.
\subsection{Proof of Theorem~\ref{thm5}}
Assume $N$ is even.
Let us consider first the case when $V(x)$ is even. Then all the even-numbered
OP's and also the functions $\phi_j$ are even, and all the odd-numbered ones
are odd functions on $\R$. 
As the maps $f\to\eps f$, $f\to f^{\prime}$ reverse parity,
it follows that the $(i,j)$ entries of $\eps_N$ and $D_N$
are zero if $i$ and $j$ have the same parity.
In turn we see that the $(i,j)$ entries of $W_N=\eps_N D_N$
are zero if $i$ and $j$ have the opposite parity.
Let $W_N^{(1)}$, $W_N^{(2)}$ be the $N/2\times N/2$
matrices
constructed from the rows and columns $1,3,\cdots,N-1$ (respectively,
$0,2,\cdots,N-2$)
 of $W_N$.
Clearly $\det W_N=\det W_N^{(1)}\det W_N^{(2)}$.
But more is true: by \eqref{W} and
the asymptotics of $T_m^\prime$, $T_{m-1}$
in \eqref{Tm}, \eqref{Tm1} we must have that  
$\det{W_N^{(1)}}=\det T_m^{\prime}+o(1)$
and $\det W_N^{(2)}=\det T_{m-1}+o(1)$.
Now $W_N^{(1)}=\eps_N^{(1)}D_N^{(1)}$
and $W_N^{(2)}=\eps_N^{(2)}D_N^{(2)}$
where $\eps_N^{(i)}$, $D_N^{(i)}$, $i=1,2$,
are the following $N/2\times N/2$ matrices:
\begin{itemize}
\item $\eps_N^{(1)}$, $\eps_N^{(2)}$ are formed from rows
$1,3,\cdots,N-1$ (respectively, $0,2,\cdots,N-2$)
and columns $0,2,\cdots,N-2$ (respectively, $1,3,\cdots,N-1$)
of $\eps_N$
\item $D_N^{(1)}$, $D_N^{(2)}$ are formed from rows
$0,2,\cdots,N-2$ (respectively, $1,3,\cdots,N-1$)
and columns $1,3,\cdots,N-1$ (respectively, $0,2,\cdots,N-2$)
of $D_N$.
\end{itemize}
Thus 
$\det T_m^\prime = \det \eps_N^{(1)}\det D_N^{(1)} + o(1)$
and $\det T_{m-1} = \det \eps_N^{(2)}\det D_N^{(2)}+o(1)$.
But as $\eps_N$, $D_N$ are skew symmetric, 
a simple calculation shows that
$$
         \eps_N^{(1)} = -(\eps_N^{(2)}\big)^T,\qquad
            D_N^{(1)} = -(D_N^{(2)}\big)^T.
$$
Letting $N\to\infty$,
it follows that $\det T_m^{\prime}=\det T_{m-1}$.

Now for the general case when $V(x)$ is not assumed to be even.
Note that the $N$-independent matrices $T_m^{\prime}$ and $T_{m-1}$
depend only on the degree $2m$ of $V(x)$ (and not on $\kappa_{2m}>0$).
But this means that $\det T_m^{\prime}$ and $\det T_{m-1}$
are equal to the corresponding 
determinants for the {\em even\ }weight
$\kappa_{2m}x^{2m}$.
By the above argument, $\det T_m^{\prime}=\det T_{m-1}$.
This completes the proof of Theorem~\ref{thm5}.
\section{Differential equations for $h(x)$, $y_m(\theta)$,
and proof of Theorem~\ref{thm7}}
\la{sectsix}
\subsection{Plan of the proof}
\la{subsec70}
The goal in this Section is to prove 
that for all $m\geq2$,
$\det{}T_{m-1}\neq0$, where $T_{m-1}$ is defined in \eqref{Tm1}.
The key object to understand is $I(q)$ in \eqref{I}
for $q=3,5,\cdots,2n-3=4m-5$.
After introducing some convenient notation we proceed to the proof.
As noted in Remark \ref{rem2star}, we
use different arguments for $m$ ``small'' and ``large,'' 
and for a given $m$, we must estimate $I(q)$
for different ranges of $q$.
More precisely, in view of Proposition \ref{s7prop1}
below, what we really need is a bound on $|1+\tilde{I}(q)|$,
where $\tilde{I}(q)\equiv mI(q)-\frac12$ (see \eqref{Itilde1} and 
\eqref{Itilde} below).
For some values of $q$ and $m$,
we estimate $|\tilde{I}(q)|$, and hence 
$|1+\tilde{I}(q)|\leq1+|\tilde{I}(q)|$, but in other ranges
we must consider the absolute value of
the combination $1+\tilde{I}(q)$.
Of course if we could show that 
\be\la{Itineq}
      |\tilde{I}(q)|<1,
\ee
for any $m\geq2$ and all $q=3,5,\cdots,4m-5$,
then Theorem \ref{thm7} would follow immediately
from  Proposition \ref{s7prop1}:
unfortunately we are unable to prove \eqref{Itineq} for 
all $q,m$ as above, and in fact it may not even be true
for all $q,m$ in the range.

In Subsection \ref{subsec72}, we prove a crude
 a priori bound on the absolute value of $\tilde{I}(q)$
(see \eqref{eqcrude} below)
that is valid for all $m\geq2$ and all $q=3,5,\cdots,4m-5$.
The proof of this bound follows from 
the hypergeometric ODE for $h (x)$ mentioned above
(cf.{} \eqref{eqhz} below).

In Subsection \ref{subsec73},
we use a more refined argument
to show that the absolute value of $\tilde{I}(q)$ is bounded by $1$ 
for all $m\geq2$ and for all $q$ in a region
$O(\sqrt{m})\leq{}q\leq{}O(m)$, see \eqref{eqqestim} below. 
To this end we use the properties of $y_m(\theta)$
(see \eqref{ym})
that follow from the fact that it satisfies a certain
Riccati equation (see Proposition \ref{s7prop4}):
as noted in Remark \ref{rem2star}
this Riccati equation in turn is a consequence of the
hypergeometric ODE for $h(x)$ mentioned above.

Finally, in Subsection \ref{subsec74} we use a certain integral representation
for $h (x)$ to find, for large enough $m$, 
an accurate approximation for $\tilde{I}(q)$
which yields a bound (see \eqref{eq301} below)
on $|1+\tilde{I}(q)|$
for $q$ in the region $3\leq{}q\leq{}O(\sqrt{m})$. 

The above estimates are sufficient to prove $\det{T_{m-1}}\neq0$
for all $m\geq2$.
The crude bound in Subsection \ref{subsec72}
is sufficient to give a simple proof of the result for $m\leq51$
(see Subsection \ref{subsec72}).
The bound in Subsection \ref{subsec73} for $O(\sqrt{m})\leq{}q\leq{}O(m)$,
together with the crude bound in Subsection \ref{subsec72}
applied to $q$ in the range $3\leq{}q\leq{}O(\sqrt{m})$,
gives a proof of the result for $m\leq99$ (see Subsection \ref{subsec73}).
Finally the bound in Subsection \ref{subsec74}
for large $m$ and $3\leq{}q\leq{}O(\sqrt{m})$,
together with the bound in Subsection \ref{subsec73}
for $O(\sqrt{m})\leq{}q\leq{}O(m)$,
proves $\det{T_{m-1}}\neq0$ for $m\geq38$
(see Subsection \ref{subsec74}).
In this way the estimates in Subsections \ref{subsec72},
\ref{subsec73} and \ref{subsec74}
cover the entire range $m\geq2$, with overlap
and some redundancy.
\subsection{Notation}
\la{subsec71}
Recall the definition of $h (x)$ in \eqref{hz}.
We will repeatedly use the change of variable $x=\sin\theta$
in \eqref{I}.
We find
\be
\la{6.1p}
   I(q)=\frac1{2m}
  +\frac2{m\pi}\sin\frac{q\pi}2\int_{-\pi/2}^{\pi/2}
         \cos(q\theta)\,y_m(\theta)\,d\theta
\ee
where $y_m$ is defined in \eqref{ym},
and as indicated above we introduce the notation 
\be
\la{Itilde}
   \tilde{I}(q)\equiv{}m I(q)-\frac12 =
  \frac2\pi\sin\frac{q\pi}2\int_{-\pi/2}^{\pi/2}
         \cos(q\theta)\,y_m(\theta)\,d\theta,\qquad q=3,5,\cdots 4m-5.
\ee
In \eqref{6.1p}
we have used the following, simple identities for $q$ in the above range,
\be\la{eqcos}
  \int_{-\pi/2}^{\pi/2}\cos(q\theta)\cos\theta\,d\theta=0,\qquad
  \frac2\pi\sin\frac{q\pi}{2}
        \int_{-\pi/2}^{\pi/2}\frac{\cos(q\theta)}{\cos\theta}\,d\theta=2.
\ee
Set
$$
    \gamma_m\equiv \frac{2(m!)^2}{m(2m)!}
$$
and let $Y_{m-1}\equiv(Y_{jk})_{j,k=1}^{m-1}$ 
denote the very last matrix (of binomial coefficients) 
in \eqref{Tm1}. Set
$$
  X_{m-1}\equiv(X_{jk})_{j,k=1}^{m-1}\equiv
       \begin{pmatrix}
& 1 &1&\cdots&1 \\
& 1 &1&\cdots&1\\ 
&\cdots\\ 
&1&1&\cdots&1
\end{pmatrix} 
+ 
\begin{pmatrix}
&\tilde{I}(n) &\tilde{I}(n+2)&\cdots&\tilde{I}(2n-3) \\
&\tilde{I}(n-2) &\tilde{I}(n)&\cdots&\tilde{I}(2n-5)\\ 
&\cdots\\ 
&\tilde{I}(3) &\tilde{I}(5)&\cdots&\tilde{I}(n) 
\end{pmatrix}.
$$
Then \eqref{Tm1} becomes
\be\la{eqTm1}
   T_{m-1} = I_{m-1} - \gamma_m X_{m-1} Y_{m-1}.
\ee
We treat the second term as a perturbation
and it turns out to be important to choose the
norm $\|\cdot\|$ on $\R^{m-1}$ appropriately.
We take $\|\cdot\|$ to be the maximum norm
on $\R^{m-1}$,
$\|x=(x_1,\cdots,x_{m-1})\|=\max_{1\leq j\leq m-1}|x_j|$.
We show eventually that acting as an operator on $(\R^{m-1},\|\cdot\|)$,
the second term in \eqref{eqTm1} has norm $<1$ for all $m$.
This of course implies that $\det{}T_{m-1}\neq0$
for all $m\geq2$.
\begin{proposition}
\la{s7prop1} Assume that for some $C>0$,
$\max_{j,k=1,\cdots,{m-1}}|X_{j,k}|\leq C$.
Then 
\be\la{eqnormest} 
      \big\| \gamma_m{}X_{m-1}Y_{m-1}\big\|
           \leq C\Big(\frac12-\frac{(m!)^2}{m(2m)!}\cdot2^{2m-2}\Big).
\ee
\end{proposition}
\begin{proof}
By the definition of the norm $\|\cdot\|$ on $\R^{m-1}$
$$
\ba
     \|X_{m-1}Y_{m-1}\| 
    \leq 
         &\max_{i=1,\cdots,m-1} 
       \sum_{j=1}^{m-1} |(X_{m-1}Y_{m-1})_{ij}|\\
   &\leq 
         \max_{i=1,\cdots,m-1} 
       \sum_{j,k=1}^{m-1} |X_{ik}|\,|Y_{kj}|
    \leq C
          \sum_{j,k=1}^{m-1} |Y_{kj}| \\
   &= C \sum_{i=0}^{m-2} \sum_{l=0}^{m-2-i} \binom{2m-1}{l}
  = C \sum_{l=0}^{m-1} \binom{2m-1}{l} (m-l-1).
\ea
$$
Substituting $\sum_{l=0}^{m-1} \binom{2m-1}{l}=2^{2m-2}$
and 
$$
   \sum_{l=0}^{m-1} \binom{2m-1}{l} (m-l)
      = \frac{m}2\binom{2m-1}{m-1} + 2^{2m-3}
$$
which follows from the identity 
\be\la{eqcombid}
    2\sum_{l=0}^q\binom{p}{l}l = p \sum_{l=0}^{q-1}\binom{p}{l}
                    + q \binom{p}{q},\qquad q\leq p,
\ee
with $q=m-1$, $p=2m-1$,
we arrive at \eqref{eqnormest}. 
It remains to prove \eqref{eqcombid}. Note that
$$
\ba
    \sum_{l=0}^q\binom{p}{l}l 
     &= \sum_{l=1}^q \frac{p(p-1)\cdots(p-(l-2))}{(l-1)!}\,(p-(l-1))\\
    &=  \sum_{l=1}^q\binom{p}{l-1}(p-(l-1))
    = p \sum_{l=0}^{q-1}\binom{p}{l}
                    -\sum_{l=0}^{q-1}\binom{p}{l}l
\ea
$$
from which \eqref{eqcombid} follows.
\end{proof}
\subsection{%The case $m\leq51$. 
ODE for $h (x)$}
\la{subsec72}
The {\em final\ }estimate \eqref{6.13p} in this Subsection 
is completely superceded by the {\em final\ }estimate
 \eqref{eq56star}
 in the next Subsection,
which gives the desired result for $m$ up till $99$.
However both \eqref{6.13p} and \eqref{eq56star}
utilize the estimate \eqref{eqcrude} 
 that is proved below in this Subsection.
\begin{proposition}
\la{s7prop2}
For any $m\in\N$ the (even) function $h (x)$ satisfies 
the ODE
\be
\la{eqhz}
 x(x^2-1)h^{\prime} + ((2m-1)-2(m-1)x^2)h  = 4m,
 \qquad 0\leq x\leq1
\ee
Also 
\be
\la{eqhzic}
     h(0) =\frac{4m}{2m-1},\qquad h (1)=4m
\ee
which together with \eqref{eqhz} implies
\be
\la{eqsohz}
  h (x) = \frac{4m}{x\sqrt{1-x^2}}\int_x^1 
       \Big(\frac{x}t\Big)^{2m}\frac{dt}{\sqrt{1-t^2}},
        \qquad 0<x<1.
\ee
Finally, the function $h(x)$ is strictly increasing on $[0,1]$,
and hence it is $\geq\frac{4m}{2m-1}>0$ on $[-1,1]$.
\end{proposition}
\begin{proof}
Assuming  \eqref{eqhz} and either of the conditions \eqref{eqhzic}
the integral representation \eqref{eqsohz} can be found
using the integrating factor method.
The explicit formula \eqref{hz}
readily implies \eqref{eqhzic}
after elementary algebraic computations,
 and also that $h(x)$ is strictly increasing on $[0,1]$.
 
To prove \eqref{eqhz}, note that
from \eqref{hz}
$$
 \beta_{k}=\frac{4m}{2m-1}\frac{(-m+1)_k}{(-m+3/2)_k}
$$
where $(a)_k\equiv a(a+1)\cdots(a+k-1)$
 denotes the Pochhammer symbol.
As $(1)_k=k!$,
\be
\la{hzz}
\ba
   h (x) &= \frac{4m}{2m-1}
     \sum_{k=0}^{m-1} \frac{(1)_k(-m+1)_k}{(-m+3/2)_k} 
                \frac{(x^2)^{k}}{k!}\\ 
      &= \frac{4m}{2m-1}{\space_2F_{1}}(1,-m+1,-m+3/2;x^2)
\ea
\ee
(the series in ${\space_2F_{1}}$ terminates at $k=m-1$).
Thus $w(z)\equiv{}h (\sqrt{z})$ satisfies the hypergeometric equation
$$
   z(1-z)w^{\prime\prime} + (c-(a+b+1)z)w^\prime - abw =0
$$
with $a=1$, $b=-m+1$, $c=-m+3/2$ which becomes
\be\la{eqhge}
      z(1-z)w^{\prime\prime} + ((-m+3/2) + (m-3)z)w^\prime +(m-1)w =0.
\ee
In turn this equation can be rewritten as 
%Using $w^\prime(z)=h^\prime(\sqrt{z})/(2\sqrt{z})$, \eqref{eqhz}
\be\la{6.12}
     \frac{d}{dz}\Big( z(z-1)w^\prime + ((m-1/2)-(m-1)z)w\Big) = 0
\ee
and so
\be\la{6.12p}
     z(z-1)w^\prime + ((m-1/2)-(m-1)z)w = \const.
\ee
Letting $z\to1$ and using \eqref{eqhzic} we conclude that $\const=2m$.
But \eqref{6.12p} is just \eqref{eqhz} for $w(z)=h(\sqrt{z})$.
Of course \eqref{eqhz} can also be proved by substituting \eqref{hz}
directly into \eqref{eqhz}.
\end{proof}
\begin{remark} 
\la{rmpfprop1}
Differentiating \eqref{eqtheta} and using \eqref{eqhz},
we immediately obtain the proof of Lemma \ref{s5propde}.
\end{remark}
Now from \eqref{eqhz} it follows that
$$
            \frac{1}{1-x^2}\Big(\frac{4m}{h(x)}-1\Big)
              =2(m-1) - \frac{xh^{\prime}(x)}{h (x)}
$$
which together with \eqref{Itilde} and \eqref{ym} 
gives for $q=3,5,\cdots$
$$
\ba
   \tilde{I}(q) &=\frac{m}{4m} \frac2\pi\sin\frac{q\pi}2\int_{-1}^{1}
         \frac{\cos(q\arcsin x)}{1-x^2} \Big(\frac{4m}{h (x)}-1\Big)\,dx\\
      &=\frac1{2\pi}\sin\frac{q\pi}2\int_{-1}^{1}
         \cos(q\arcsin x)\,
             \Big(2(m-1) - \frac{xh^{\prime}(x)}{h (x)}\Big)\,dx.
\ea
$$
The first integral vanishes for odd $q\geq3$ by \eqref{eqcos},
and estimating the second by the absolute value of the integrand
(using $h^{\prime}\geq0$ on $[0,1]$ and the fact that $h$ is even) we obtain
the a priori estimate
$$
      | \tilde{I}(q)|
      \leq\frac1{\pi}\int_{0}^{1}
           \frac{xh^{\prime}(x)}{h (x)}\,dx
    =\frac1{\pi}\int_{0}^{1}
           x\Big(\log\frac{h(x)}{h(1)}\Big)^\prime\,dx
     =\frac1{\pi}\int_{0}^{1}
          \log\frac{4m}{h(x)}\,dx
$$
by \eqref{eqhzic}.
Now we use the estimate
$$
       \frac{4m}{h(x)}\leq (2m-1) - 2(m-1)x^2,\qquad 0\leq x\leq1,
$$
which follows from \eqref{eqhz} since $h,\,h^\prime\geq0$, 
to find 
\be
\la{eqcrude}  
           | \tilde{I}(q)| \leq L(m)
\ee
where
$$
        L(m)\equiv\frac{\log(2m-1) - 2
           + (t_m+1)\log(1+t_m^{-1}) - (t_m-1)\log(1-t_m^{-1})}
          {\pi}
$$
and $t_m\equiv\sqrt{\frac{2m-1}{2m-2}}$.
Using Maple, for example, 
it is straightforward to check that for $m=2,3,\cdots,51$
\be\la{6.13p}
      \big(1+L(m)\big)\,
         \Big(\frac12-\frac{(m!)^2}{m(2m)!}\cdot2^{2m-2}\Big)<1
\ee
and hence by Proposition \ref{s7prop1},
$T_{m-1}$ is invertible and therefore $\det{T_{m-1}}\neq0$
for $m\leq51$.
\begin{remark}
By Stirling's formula,
 $\frac{(m!)^2}{m(2m)!}\cdot2^{2m-2}\sim
\frac{\sqrt{\pi}}{4\sqrt{m}}$,
and so the LHS of \eqref{6.13p}
 grows logarithmically as $m\to\infty$.
\end{remark}
\subsection{%The case $m\leq99$. 
Riccati equation for $y_m(\theta)$}
\la{subsec73}
A consequence of the calculations in this Subsection
is that 
 $$
          \|\gamma_mX_{m-1}Y_{m-1}\|_{(\R^{m-1},\|\cdot\|)\to{}
                                 (\R^{m-1},\|\cdot\|)}<1
$$
for $m$ up to (the larger value) $99$.
In place of Proposition~\ref{s7prop1} we will use the following
result. The proof is immediate.
\begin{proposition}\la{s7prop3} 
Assume that for some matrix $\tilde{X}$
with nonnegative elements we have for all 
$j,k=1,\cdots,{m-1}$ that $|X_{j,k}|\leq \tilde{X}_{j,k}$.
Then 
\be\la{eqprop7.3} 
%      \big\| \gamma_m{}X_{m-1}Y_{m-1}\big\|
%           \leq \big\| \gamma_m{}\tilde{X}_{m-1}Y_{m-1}\big\|.
      \big\| \gamma_m{}X_{m-1}Y_{m-1}\big\|
           \leq  \gamma_m{}  \max_{i=1,\cdots,m-1}
           \sum_{j,k=1}^{m-1} \tilde{X}_{ik}Y_{kj}.
\ee
\end{proposition}
The main goal in this Subsection
is to prove the estimate \eqref{eqqestim}
below on $\tilde{I}(q)$.
This estimate is useful for $q$ in the range $O(\sqrt{m})\leq
q\leq O(m)$.
The matrix $\tilde{X}$ in Proposition \ref{s7prop3}
is then constructed by ``interpolating'' between
this estimate and the estimate \eqref{eqcrude} above, as indicated
in \eqref{eq56star} below.
Estimate \eqref{eqqestim} will be derived from the Riccati
equation \eqref{eqR} for $y_m(\theta)=\frac{m}{\cos\theta}
         \big(\frac{1}{h(\sin\theta)} 
        - \frac1{4m} 
         - \frac{\cos^2\theta}{2}\big)$
given in Proposition \ref{s7prop4} below.
Note from \eqref{hz} that $y_m(\theta)$ is even.
\begin{remark}
The specific form of $y_m(\theta)$ has the following consequences
for $\tilde{I}(q)=\frac{2}{\pi}\sin\frac{q\pi}{2}\int_{-\pi/2}^{\pi/2}
\cos(q\theta)\,y_m(\theta)\,d\theta$.
One can show that as $m\to\infty$,
$\frac1{h(x)}\to\frac{1-x^2}2=\frac{\cos^2\theta}{2}$
(formally at least, divide \eqref{eqhz}
by $m$ and let $m\to\infty$).
Hence for a fixed $q$, as $m\to\infty$, we expect
$\tilde{I}(q)\to-\frac12$ (recall \eqref{eqcos}).
On the other hand for odd $q=2l+1$,
$$
   \cos(q\theta)\,y_m(\theta) = \cos(2l\theta)\,(\cos\theta\,y_m(\theta) )
        -\sin(2l\theta)\,(\sin\theta\,y_m(\theta)) 
$$
and so $\tilde{I}(q)$ is a sum of Fourier coefficients
of the real analytic $\pi$-periodic functions
$(\cos\theta)\,y_m(\theta)$, $(\sin\theta)\,y_m(\theta)$,
and so for a fixed $m$, $\tilde{I}(q)$ decays exponentially
as $q\to\infty$ by the Paley--Wiener theorem.
As discussed in Remark \ref{rem2star} above,
 this indicates that there
should be a transition in the $(q,m)$-plane
between these two kinds of behavior.
And indeed, as we will see below, the transition region
is in the range $q\sim \sqrt{m}$.
\end{remark}
\begin{proposition}\la{s7prop4} 
For any $m\geq1$ the function $y_m(\theta)$
defined in \eqref{ym} 
satisfies a Riccati differential equation \eqref{eqR}
$$
       y_m^\prime = \frac4{\sin\theta}
             \bigg(y_m+\frac{2m+1}{4}\cos\theta\bigg)
            \bigg(y_m+\frac{1}{2\cos\theta}\bigg).
$$ 
Also $y_m(0)=-1/2$ and $y_m(\pi/2)=0$, see Fig.~\ref{figode1}.
\end{proposition}
\begin{proof} 
This is a straightforward computation using
the ODE \eqref{eqhz} for $h(x)$.
The interesting fact here is that the RHS in the Riccati equation
has real roots which, moreover, have simple expressions.
The proof that $y_m(0)=-1/2$ follows directly from \eqref{hvalues}
and the fact that $y_m(\pi/2)=0$ follows from \eqref{hvalues} together with 
L'H\^opital's rule.
\end{proof}
We will show next that $y_m$ is a unimodal nonpositive 
function on $[0,\pi/2]$,
and that its minimum satisfies a certain bound (see Fig.~\ref{figode2}).
\begin{figure}[tb]
\begin{center}
\epsfig{file=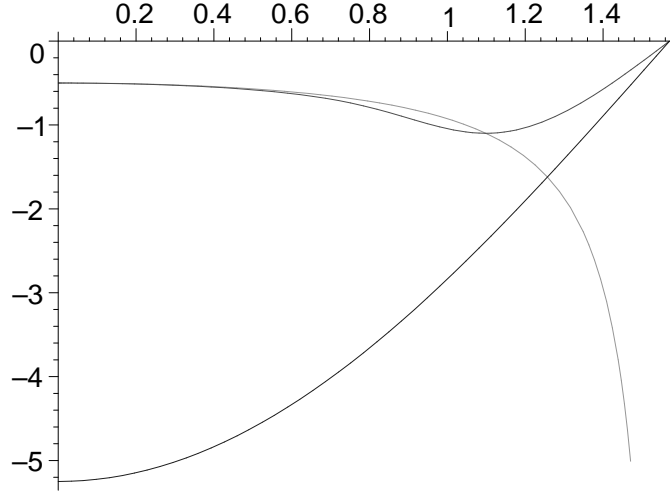,scale=1}
\caption{Graphs of $y_{m}(\theta)$, $-\frac{2m+1}4\cos\theta$ for $m=10$ and
$-1/(2\cos\theta)$ (the latter for $0\leq\theta\leq\frac\pi2-0.1$)}
\label{figode1}
\end{center}
\end{figure}
%
%\begin{comment}
\begin{figure}[tb]
\begin{center}
\epsfig{file=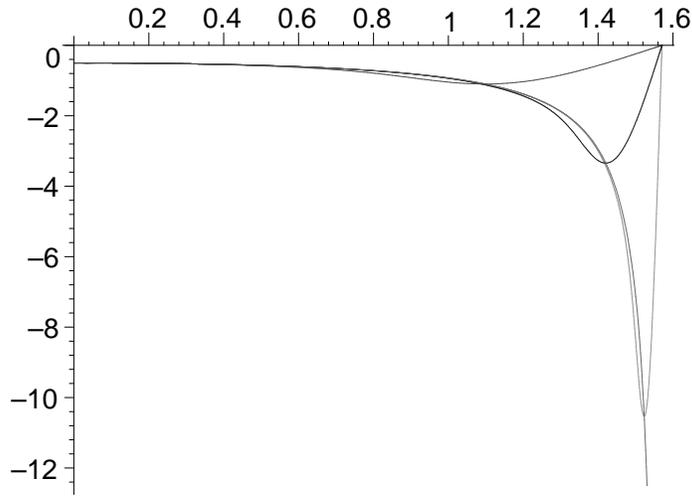,scale=1}
\caption{Graphs of $y_m(\theta)$ for $m=10$, $100$, $1000$
and $-\frac1{2\cos\theta}$ (the latter for $0\leq\theta\leq\frac\pi2-0.04$)}
\label{figode2}
\end{center}
\end{figure}
%\end{comment}
%
\begin{proposition}
\la{s7prop5} 
For any $m\geq2$ the function $y_m(\theta)$
on the interval $[0,\pi/2]$
strictly decreases from $y_m(0)=-1/2$ to the value $y_{m,\min}$
which is achieved at a unique point $\theta_{\min}\in(0,\pi/2)$,
and then strictly increases to $0$ at the point $\theta=\pi/2$.
The value $y_{m,\min}$ lies above the $y$-coordinate
of the point of intersection of the curves $-\frac1{2\cos\theta}$
and $-\frac{2m+1}{4}\cos\theta$, see Fig.~\ref{figode1}.
Finally, $y_m(\theta)\leq0$ and
\be\la{eqyminfo}
        y_{m,\min}\geq -\frac12\sqrt{m+1/2}.
\ee
\end{proposition}
\begin{proof} 
It is helpful to refer to Fig.~\ref{figode1} in the following argument.
Define $u(\theta)\equiv-\frac1{2\cos\theta}$
and $l_m(\theta)\equiv-\frac{2m+1}{4}\cos\theta$.
Note that by \eqref{hz} and \eqref{ym} it follows
that $y_m(\theta)$ is a smooth even function on $[-\pi/2,\pi/2]$.
This implies $y_m^\prime(0)=0$. Also it is clear that for all $m\geq2$,
$y_m(0)=u(0)>l_m(0)$. 
 Also $u^\prime(0)=0$ and 
$u^{\prime\prime}(0)=-1/2$.
Assume for a moment that 
\be\la{eqymppz}
       y_m^{\prime\prime}(0)=-\frac12\frac{2m-1}{2m-3}.
\ee
Then $y_m^{\prime\prime}(0)<-1/2$ for $m\geq2$.
Hence for small $\theta>0$, $y_m$ lies below $u$ and above $l_m$.
It follows in particular as $y_m(\pi/2)=0$
and $u(\theta)\to-\infty$ as $\theta\to\pi/2$,
 that $y_m$ must intersect $u$ at some point, $\theta_0$, say,
 in $(0,\pi/2)$.

It the argument that follows the signature
table for the vector field 
$$
   V(y_m,\theta)=\frac{4}{\sin\theta}(y_m - l_m(\theta))
               (y_m-u(\theta))
$$
plays a crucial role. We have
\begin{itemize}
\item $V(y_m,\theta)>0$ if $y_m$ lies above 
$l_m(\theta)$ and $u(\theta)$ or below $l_m(\theta)$ and $u(\theta)$
\item $V(y_m,\theta)<0$ if $y_m$ lies between
$l_m(\theta)$ and $u(\theta)$.
\end{itemize}
Let $\tilde{\theta}\in(0,\pi/2)$ be the (unique) intersection point
for $l_m(\theta)$ and $u(\theta)$, $l_m(\tilde{\theta})=u(\tilde{\theta})
\equiv\tilde{y}$.
From the signature table we see that $V(\tilde{y},\theta)<0$
for all $\theta\neq\tilde{\theta}$. 
Moreover if $y_m(\tilde{\theta})=
\tilde{y}$ then $y_m^\prime(\tilde{\theta})
=V(y_m(\tilde{\theta}),\tilde{\theta})=0$.
 But $u^\prime(\tilde{\theta})<0$
and $l_m^\prime(\tilde{\theta})>0$, and so $y_m(\theta)$ lies
between $l_m(\theta)$ and $u(\theta)$ for $\theta>\tilde{\theta}$,
$\theta$ close to $\tilde{\theta}$.
From the signature table it then follows that $y_m(\theta)$
decreases as $\theta$ crosses $\tilde{\theta}$.
In other words, we see that if $y_m(\theta)$ crosses the
level $\tilde{y}$, then it must decrease.
As $y_m(0)=-1/2>\tilde{y}$, and as $y_m(\pi/2)=0$,
 it follows, therefore, that
$y_m(\theta)$ cannot cross $\tilde{y}$.
In particular, $0<\theta_0<\tilde{\theta}$.
Now a similar argument shows that 
 for $\theta>\theta_0$, $\theta$ close to $\theta_0$,
$y_m(\theta)$ lies in the region above
 $l_m(\theta)$ and $u(\theta)$ 
 (and increases across $\theta_0$).
Suppose that $y_m(\theta)$
exits this region for the first time for some $\theta_1$,
$\theta_0<\theta_1<\pi/2$.
But then, again by the signature table, 
  $y_m^\prime(\theta)=V(y_m(\theta),\theta)>0$ for
  $\theta_0<\theta<\theta_1$, and hence 
  $y_m(\theta_1)>y_m(\theta_0)$.
In particular, as $u^\prime(\theta)<0$, it follows that
$y_m(\theta)$ does not exit the region through the curve $u(\theta)$.
Thus we must have $y_m(\theta_1)=l_m(\theta_1)$
and $\tilde{\theta}<\theta_1<\pi/2$.
But arguing as before we see that $y_m(\theta)$
must decrease as it crosses $\theta_1$: as $y_m(\pi/2)=0$
there must be some point $\theta_2$,
 $\theta_1<\theta_2<\pi/2$ for which 
 $y_m(\theta_2)=y_m(\theta_1)$
and $y_m^\prime(\theta_2)\geq0$.
But $V(y_m(\theta_2),\theta_2)<0$,
by the signature table, which is a contradiction.
The above arguments show that $y_m(\theta)$
crosses $u(\theta)$ at a unique point,
$\theta_{\min}=\theta_0$,
$0<\theta_{0}<\tilde{\theta}$: 
for $0<\theta<\theta_{0}$,
$y_m(\theta)$ lies between $l_m(\theta)$ and $u(\theta)$
and for $\theta_0<\theta<\pi/2$,
$y_m(\theta)$ lies above $l_m(\theta)$ and $u(\theta)$.
Thus $y_m(\theta)$ is unimodal, decreasing for $0<\theta<\theta_{0}$
and increasing for $\theta_0<\theta<\pi/2$.
 Clearly $y_{\min}=y_m(\theta_{\min})$
lies above $\tilde{y}=u(\tilde{\theta})=l_m(\tilde{\theta})=-\frac12\sqrt{m+1/2}$.
Note finally that for $0<\theta\leq\theta_{\min}$
we certainly have $y_m(\theta)\leq{}u(\theta)<0$.
On the other hand as $y_m^\prime(\theta)>0$ for
 $\theta_{\min}<\theta<\pi/2$, and as $y_m(\pi/2)=0$,
it follows that we must have $y_m(\theta)<0$
for all $\theta_{\min}<\theta<\pi/2$.
Thus $y_m(\theta)\leq0$
for all $0\leq\theta\leq\pi/2$.

It remains to prove \eqref{eqymppz}.
Divide the ODE \eqref{eqR} by $\sin\theta$ to obtain
\be\la{6.17p}
    \frac{y_m^\prime(\theta)}{\sin\theta}
      = 4\bigg(y_m+\frac{2m+1}{4}\cos\theta\bigg)
            \cdot\frac{y_m+\frac{1}{2\cos\theta}}{\sin^2\theta}.
\ee
Now as $y_m(0)=-1/2$,
$$
\ba
      \lim_{\theta\to0} \frac{y_m(\theta)+\frac{1}{2\cos\theta}}{\sin^2\theta}
   =   \lim_{\theta\to0} \frac{y_m^\prime(\theta)+\frac{\sin\theta}{2\cos^2\theta}}
          {2\sin\theta\cos\theta} %\\
  =\frac12 \lim_{\theta\to0} \frac{y_m^\prime(\theta)}{\sin\theta}
              +\frac14.
\ea
$$
Letting $\theta\to0$ on both sides of \eqref{6.17p}
we obtain $\lim_{\theta\to0} \frac{y_m^\prime(\theta)}{\sin\theta}
=-\frac12\frac{2m-1}{2m-3}$.
But $y_m^{\prime\prime}(0)=
\lim_{\theta\to0} \frac{y_m^\prime(\theta)}{\sin\theta}$,
and hence \eqref{eqymppz} follows.
This completes the proof of the Proposition.
\end{proof}
Now we integrate by parts to obtain
$$%\be
%\la{eqbyparts}
\ba
 |\tilde{I}(q)| &= \Big|\frac4{\pi}\int_{0}^{\pi/2} \cos(q\theta) y_m(\theta)\,d\theta\Big|
    =\frac4{q\pi}\Big|\int_0^{\pi/2}\sin(q\theta) y_m^\prime(\theta)\,d\theta\Big|\\
     &\leq \frac4{q\pi} \int_0^{\pi/2} \big|y_m^\prime(\theta)\big|\,d\theta 
     = \frac4{q\pi}\Big( \int_0^{\theta_{\min}} (-y_m^\prime(\theta))\,d\theta
                   +\int_{\theta_{\min}}^{\pi/2} y_m^\prime(\theta)\,d\theta\Big)
\ea
$$%\ee
where we have used $y_m(\pi/2)=0$ and also the unimodality of $y_m$
(Proposition~\ref{s7prop5}). 
Hence in view of \eqref{eqyminfo} for any $m\geq2$ and any odd $q\geq3$
we have
\be\la{eqqestim}
   |\tilde{I}(q)| \leq \frac8{q\pi} \big|y_{m,\min}\big| \leq \frac4{q\pi}\sqrt{m+1/2}.
\ee
In particular, $|\tilde{I}(q)|<1$
for odd $q$ such that
 $\big[\frac4{\pi}\sqrt{m+1/2}\big]+1\leq q\leq4m-5$.

Now we can combine the estimates \eqref{eqcrude}
and \eqref{eqqestim} at our discretion.
We define the estimating Toeplitz matrix 
$\tilde{X}_{m-1}\equiv(\tilde{X}_{j,k})_{j,k=1}^{m-1}$ in
Proposition \ref{s7prop3} as follows.
The diagonals are numbered by odd $q=3,5,\cdots,4m-5$ from the lower left corner.
Thus if $j,k=1,2,\cdots,m-1$ are the row and column indices starting from the usual
upper left corner, then $q=2(k-j)+(2m-1)$. 
Therefore, in view of  \eqref{eqcrude}, \eqref{eqqestim},
set
\be\la{eq56star}
      \tilde{X}_{j,k} \equiv 1 + \min\Big(L(m),
                   \frac4{(2(k-j)+(2m-1))\pi}\sqrt{m+1/2}\Big).
\ee
Also let
$
    Y_{m-1}\equiv(Y_{j,k})_{j,k=1}^{m-1}
$ where $   Y_{j,k}=  \binom{2m-1}{k-j} $ for $1\leq j\leq k\leq m-1$
and $0$ otherwise,
be the matrix of binomial coefficients as before.
Then one can check by computer (only sums and products are involved)
that the RHS in \eqref{eqprop7.3} is $<1$ for $m=2,3,\cdots,99$.
Thus by Proposition \ref{s7prop3},
$T_{m-1}\equiv I_{m-1}-\gamma_mX_{m-1}Y_{m-1}$
is invertible for $m\leq99$,
and hence $\det{}T_{m-1}\neq0$ for $m\leq99$.
\subsection{%The case $m\geq38$. 
Large $m$ asymptotics for $y_m(\theta)$ via an integral representation}
\la{subsec74}
This is the last and most technical computation. The main idea here
is to find for large enough $m$
an approximation to the solution $y_m(\theta)$
of the Riccati equation in 
$\int_0^{\pi/2}\cos(q\theta)y_m(\theta)d\theta$
such that the integral is well approximated 
in the range $q=3,5,\cdots,O(\sqrt{m})$. 
%The goal in this Subsection is the estimate \eqref{eq301} below
%for odd $q$ in the range $3\leq q\leq\big[\frac4\pi\sqrt{m+1/2}\big]$. 
We show that the estimate \eqref{eq301} below
that follows from this
approximation, together with \eqref{eqqestim} for
 $\big[\frac4{\pi}\sqrt{m+1/2}\big]+1\leq q\leq4m-5$,
gives the desired result 
for all $m\geq38$, so that there is a large overlap
with the region $m\leq99$ where the result of the preceding Subsection is valid.

%We will write below for brevity $y\equiv{}y_m$.
For $\rho\geq0$ set
\be\la{eqG}
  G(\rho)\equiv
  \frac
      {\int_0^1 (1+\rho v)^{-m-1/2} \frac{vdv}{\sqrt{1-v}} }
          {\int_0^1 (1+\rho v)^{-m-1/2} \frac{dv}{\sqrt{1-v}} }.
\ee
The first result we need is 
 the following 
integral representation for  $y_m(\theta)$.
Using $x=\sin\theta$,
\be\la{eq6.100}
\ba
   y_m(\theta)&\equiv \frac{m}{\sqrt{1-x^2}}\Big(\frac1{h (x)} -\frac1{4m}-\frac{1-x^2}2\Big)\\
        &=-\frac{\sqrt{1-x^2}}{2}  -\frac{x^2}{2\sqrt{1-x^2}}(m-1)
         \frac
      {\int_x^1 \big(\frac{x}t\big)^{2m}
        \big(\big(\frac{t}x\big)^{2}-1\big)\frac{dt}{\sqrt{1-t^2}} }
          {\int_x^1 \big(\frac{x}t\big)^{2m}\frac{dt}{\sqrt{1-t^2}} }\\
      &=-\frac{\sqrt{1-x^2}}{2}  -\frac{m-1}2{\sqrt{1-x^2}}
        \,G(\rho)
\ea
\ee
where 
\be\la{eq6.101}
    \rho \equiv x^{-2}-1 = \frac{\cos^2\theta}{\sin^2\theta},\qquad x=\sin\theta.
\ee
We will explicitly 
indicate the dependence of $h(x)$ on $m$
as $h_m(x)$ when needed.
To prove \eqref{eq6.100} we first note that for $m\geq2$
\be\la{eqhmhmo}
       h_m (x)
               =\frac{2m}{2m-1}(2+x^2h_{m-1}(x)),\qquad 0\leq x\leq 1,
\ee
which follows from the explicit formula \eqref{hz}
and in turn leads
(after adding and subtracting $\frac{2x^2}{2m-1}(m-1)h_m$)
to
$$
  (1+(2m-2)(1-x^2))h_m = 4m + 2x^2 (mh_{m-1}-(m-1)h_m)
$$
from which, dividing by $4mh_m$, we obtain
$$
      \frac1{h_m} - \frac1{4m} - \frac{1-x^2}2
       = -\frac{1-x^2}{2m} - \frac{x^2}{2m}\,\frac{mh_{m-1}-(m-1)h_m}{h_m}.
$$
Together with \eqref{eqsohz}, this
implies the second equality in \eqref{eq6.100}.
The third equality in \eqref{eq6.100} follows after changing the variable
%$t=xw$,  $w^2=1+s$, $s=\rho{}v$ 
$t=x\sqrt{1+\rho{}v}$
where $\rho$ is as in \eqref{eq6.101}.

Now from \eqref{eq6.100} for $q=3,5,\cdots$, and \eqref{eqcos},
$$
\ba
   \tilde{I}(q)&
 %\equiv\frac4\pi\sin\frac{q\pi}2
  %          \int_0^1 \cos(q\arcsin x) y_m(x)\frac{dx}{\sqrt{1-x^2}}\\&
    =-\frac{m-1}{\pi}2\sin\frac{q\pi}2
           \int_0^{\pi/2}\big(\cos(q\theta) \cos\theta\big) \,G(\rho(\theta))\,d\theta.
\ea
$$
Make now a change of variable $\phi=\frac\pi2-\theta$
giving
\be
\la{eq200}
 \rho=\tan^2\phi,\qquad
    d\rho=\frac{2\sin\phi}{\cos^3\phi}d\phi,\qquad
   \cos\phi=\frac1{\sqrt{1+\rho}},\qquad\sin\phi=\frac{\sqrt{\rho}}{\sqrt{1+\rho}}
\ee
to find
$$
 \tilde{I}(q)=-\frac{m-1}{\pi}\int_0^{\infty}\sin(q\phi(\rho))\,
       G(\rho)
       \,\frac{d\rho}{(1+\rho)^{3/2}}.
$$
We split the integral into two,
\be\la{eq190}
\tilde{I}(q) =-\frac{m-1}{\pi}\Big( \int_0^{m^{-1/2}}+\int_{m^{-1/2}}^{\infty}\Big)
     \sin(q\phi(\rho))\,
       G(\rho)
       \,\frac{d\rho}{(1+\rho)^{3/2}}.
\ee
Assume that $m\geq3$ (we only need the results below for $m\geq38$).
\begin{lemma}
\la{s7lem1}
We have
$$
-\frac{m-1}{\pi} \int_{m^{-1/2}}^{\infty} \frac{\sin(q\phi(\rho))\,
       G(\rho)}{(1+\rho)^{3/2}}
       \,d\rho
   = -\frac{m-1}{m-\frac32}\frac1\pi \int_{m^{-1/2}}^{\infty}
       \frac{\sin(q\phi(\rho))}{\rho(1+\rho)^{3/2}}d\rho + E_1(m,q)
$$
where the error term $E_1(m,q)$ satisfies
$$
         |E_1(m,q)|\leq\frac{A(m)+B(m)+C(m)}{Denom(m)}
$$
 uniformly for all $q=3,5,\cdots$,
where
\be\la{eqABCD}
\ba
 A(m)&\equiv \frac7{\pi\sqrt{2}}\frac{(m-1)\sqrt{m}}{(m-\frac12)(m-\frac52)}\\
 B(m)&\equiv \frac3{\pi\sqrt{2}}
     \bigg(\frac12(\log{m})
              \frac1{\big(1+\frac12\frac1{\sqrt{m}}\big)^{m-3/2}
                        }
            + \Big(\frac23\Big)^{m-\frac12} \bigg)\\
 C(m)&\equiv \frac{3}{\sqrt{2}-1}
        \frac1\pi \frac1{\big(1+\frac{\sqrt{2}-1}{\sqrt{2}}\frac1{\sqrt{m}}\big)^m
                                }\\
 Denom(m)&\equiv \frac{m-\frac32}{m-\frac12}
       \Big(1-e^{-\sqrt{m}\frac{1-\frac1{2m}}
                                              {1+\frac1{\sqrt{m}}}
                       }\Big).
\ea
\ee
\end{lemma}
\begin{proof}
%We perform integration by parts carefully keeping track of the 
%nonintegral terms.
We consider $\rho\geq{}m^{-1/2}$.
We need the following elementary inequality:
\be\la{eqE.4.0}
       e^{ay/(1+y)} \leq (1+y)^a \leq e^{ay},\qquad a,y\geq0. 
\ee
Set
\be\la{eqk}
             k\equiv m+1/2.
\ee
Splitting the interval $[0,1]$ into two and integrating by parts we find
\be\la{eqE.4.1}
\ba
         \int_0^1 (1+\rho v)^{-k}\,\frac{dv}{\sqrt{1-v}}
       = &\frac1{(k-1)\rho} 
             - \frac{\sqrt{2}}{(k-1)\rho}\,\frac1{(1+\rho/2)^{k-1}}\\
      &+\frac1{(k-1)\rho}
            \int_0^{1/2} (1+\rho v)^{-k+1}
                  \bigg(\frac1{\sqrt{1-v}}\bigg)^\prime\,dv\\
      &+\int_{1/2}^1 (1+\rho v)^{-k}\,\frac{dv}{\sqrt{1-v}}
\ea
\ee
and
\be\la{eqE.5.1}
\ba
         \int_0^1 (1+\rho v)^{-k}\,\frac{v\,dv}{\sqrt{1-v}}
       = &\frac1{(k-1)(k-2)\rho^2} 
             - \frac{1}{(k-1)\rho}\,\frac1{\sqrt{2}}\,\frac1{(1+\rho/2)^{k-1}}\\
       &- \frac{1}{(k-1)(k-2)\rho^2}\,\frac3{\sqrt{2}}\,\frac1{(1+\rho/2)^{k-2}}\\
      &+\frac1{(k-1)(k-2)\rho^2} 
             \int_0^{1/2} (1+\rho v)^{-k+2}
                  \bigg(\frac{v}{\sqrt{1-v}}\bigg)^{\prime\prime}\,dv\\
      &+\int_{1/2}^1 (1+\rho v)^{-k}\,\frac{v\,dv}{\sqrt{1-v}}.
\ea
\ee
This gives the leading behavior of the ratio 
in $G(\rho)$ for $m$ (and $k$) large
and motivates the definition
$$
       \Delta(\rho)\equiv G(\rho) - \frac1{(k-2)\rho}
                =\frac{(k-2)\rho\cdot(\textrm{RHS in \eqref{eqE.5.1})}
                      -(\textrm{RHS in \eqref{eqE.4.1}})}
                   {(k-2)\rho\int_0^1(1+\rho{}v)^{-k}\,\frac{dv}{\sqrt{1-v}}}
$$
which yields
\be\la{eq6.102}
\ba
   \Big( (k-2)\rho\int_0^1(1+\rho v)^{-k}\,&\frac{dv}{\sqrt{1-v}}\Big)
    \,\Delta(\rho) \\
 =&\frac1{(k-1)\rho} 
             - \frac{k-2}{k-1}\,\frac1{\sqrt{2}}\,\frac1{(1+\rho/2)^{k-1}}\\
       &- \frac{1}{(k-1)\rho}\,\frac3{\sqrt{2}}\,\frac1{(1+\rho/2)^{k-2}}\\
      &+\frac1{(k-1)\rho} 
             \int_0^{1/2} (1+\rho v)^{-k+2}
                  \bigg(\frac{v}{\sqrt{1-v}}\bigg)^{\prime\prime}\,dv\\
      &+(k-2)\rho\,\int_{1/2}^1 (1+\rho v)^{-k}\,\frac{v\,dv}{\sqrt{1-v}}\\
      &-\frac1{(k-1)\rho} 
             + \frac{\sqrt{2}}{(k-1)\rho}\,\frac1{(1+\rho/2)^{k-1}}\\
      &-\frac1{(k-1)\rho}
            \int_0^{1/2} (1+\rho v)^{-k+1}
                  \bigg(\frac1{\sqrt{1-v}}\bigg)^\prime\,dv\\
      &-\int_{1/2}^1 (1+\rho v)^{-k}\,\frac{dv}{\sqrt{1-v}}.
\ea
\ee
From this we derive an upper and a lower estimate on $\Delta(\rho)$.
Discarding the negative terms, we obtain 
\be\la{eq6.103}
\ba
   \Big( (k-2)\rho\int_0^1(1+\rho v)^{-k}\,&\frac{dv}{\sqrt{1-v}}\Big)
    \,\Delta(\rho) \\
 \leq&      \frac1{(k-1)\rho}  \int_0^{1/2} (1+\rho v)^{-k+2}
                  \bigg(\frac{v}{\sqrt{1-v}}\bigg)^{\prime\prime}\,dv\\
      &+(k-2)\rho\,\int_{1/2}^1 (1+\rho v)^{-k}\,\frac{v\,dv}{\sqrt{1-v}}\\
      &    + \frac{\sqrt{2}}{(k-1)\rho}\,\frac1{(1+\rho/2)^{k-1}}.
\ea
\ee
Note that the second derivative above is
 indeed nonnegative and that its maximum
over $[0,1/2]$ equals $7/\sqrt{2}$.
% Also $\int_0^1\frac{dv}{(1+\rho v)^{k-2}}\leq
%\frac{1}{(k-3)\rho}$.
 Next changing variables $\sqrt{1-v}=t$
and integrating by parts we obtain
$$
\ba
  \int_{1/2}^1 (1+\rho v)^{-k}\,&\frac{v\,dv}{\sqrt{1-v}}
      =2\int_{0}^{1/\sqrt{2}} \frac{1-t^2}{(1+\rho(1-t^2))^{k}}\,dt
      \leq   2\int_{0}^{1/\sqrt{2}} \frac{1-t^2}{(1+\rho(1-t))^{k}}\,dt\\
   &\leq 2\frac{(1+\rho(1-1/\sqrt{2}))^{-k+1}}{(k-1)\rho}\frac12
             -\frac2{(k-1)\rho}
           \int_{0}^{1/\sqrt{2}} \frac{-2t}{(1+\rho(1-t))^{k-1}}\,dt\\
   &\leq \frac1{(k-1)\rho}\frac2{(1+\rho(1-1/\sqrt{2}))^{k-1}}
                    \bigg(\frac12 + \int_{0}^{1/\sqrt{2}} 2t\,dt\bigg).
\ea
$$
Thus from \eqref{eq6.103}, \eqref{eqk}
\be\la{eq6.104}
\ba
   \Big( (m-3/2)\rho &\int_0^1(1+\rho v)^{-m-1/2}\,\frac{dv}{\sqrt{1-v}}\Big)
    \,\Delta(\rho) \leq   \frac7{\sqrt{2}}   \frac1{(m-1/2)(m-5/2)\rho^2}
              \\
      &+\frac2{(1+(1-1/\sqrt{2})\rho)^{m-1/2}}
           + \frac{\sqrt{2}}{(m-1/2)\rho}\,\frac1{(1+\rho/2)^{m-1/2}}.
\ea
\ee

Next discarding the positive terms in \eqref{eq6.102}, we find
\be\la{eq6.105}
\ba
   - \Big( &(k-2)\rho\int_0^1(1+\rho v)^{-k}\,\frac{dv}{\sqrt{1-v}}\Big)
    \,\Delta(\rho) \\
 &\leq   \frac{k-2}{k-1}\,\frac1{\sqrt{2}}\,\frac1{(1+\rho/2)^{k-1}}
      + \frac{1}{(k-1)\rho}\,\frac3{\sqrt{2}}\,\frac1{(1+\rho/2)^{k-2}}\\
       &      +\frac1{(k-1)\rho}
            \int_0^{1/2} (1+\rho v)^{-k+1}
                  \bigg(\frac1{\sqrt{1-v}}\bigg)^\prime\,dv
            +\int_{1/2}^1 (1+\rho v)^{-k}\,\frac{dv}{\sqrt{1-v}}.
\ea
\ee
Note that the first derivative above is indeed
 nonegative and that its maximum
over $[0,1]$ equals $\sqrt{2}$.
%also $\int_0^1\frac{dv}{(1+\rho v)^{k-1}}\leq
%\frac{1}{(k-2)\rho}$. 
Note finally that
$\int_{1/2}^1 (1+\rho v)^{-k}\,\frac{dv}{\sqrt{1-v}}\leq
   \frac1{(1+\rho/2)^{k}}\,\int_{1/2}^1\frac{dv}{\sqrt{1-v}}
   =\frac{\sqrt{2}}{(1+\rho/2)^{k}}$. 
Thus \eqref{eq6.105}, \eqref{eqk} give
\be\la{eq6.107}
\ba
   - \Big( &(m-3/2)\rho\int_0^1(1+\rho v)^{-m-1/2}\,\frac{dv}{\sqrt{1-v}}\Big)
    \,\Delta(\rho) \\
&\leq   \frac1{\sqrt{2}}\,\frac{m-3/2}{m-1/2}\,\frac1{(1+\rho/2)^{m-1/2}}
    + \frac3{\sqrt{2}}\,\frac{1}{(m-1/2)\rho}\,\frac1{(1+\rho/2)^{m-3/2}}\\
       & +\frac{\sqrt{2}}{(m-1/2)(m-3/2)\rho^2}
    +\frac{\sqrt{2}}{(1+\rho/2)^{m+1/2}}.
\ea
\ee
Now using \eqref{eqE.4.0},
we see that the factor in front of $\Delta(\rho)$ in \eqref{eq6.104}
satisfies
\be\la{eq6.108}
\ba
    \int_0^1 (1+\rho v)^{-k}\,\frac{dv}{\sqrt{1-v}}
   &\geq \int_0^1 (1+\rho v)^{-k}\,dv
            =\frac1{(k-1)\rho}
                    \bigg(1-\frac1{(1+\rho)^{k-1}}\bigg)\\
       &\geq \frac1{(k-1)\rho}
                    \bigg(1- e^{-(k-1)\frac{\rho}{1+\rho}}\bigg)
       \geq \frac1{(k-1)\rho}
                    \bigg(1- e^{-(k-1)\frac{m^{-1/2}}{1+m^{-1/2}}}\bigg)\\
       &\geq\frac1{(m-1/2)\rho}
                    \bigg(1- e^{-(1-\frac1{2m})
                                         \frac{\sqrt{m}}{1+\frac1{\sqrt{m}}}
                                      }\bigg),
\ea
\ee
by \eqref{eqk} since $-\frac{\rho}{1+\rho}$ decreases and $m^{-1/2}\leq\rho$.

We now consider the various terms in
 \eqref{eq6.104}, \eqref{eq6.107}
%(and slightly modify them if necessary)
to obtain an estimate on $|\Delta(\rho)|$, $\rho\geq m^{1/2}$.
Note that:
\begin{itemize}
\item 
   the 1st term in  \eqref{eq6.104} dominates
   the 3rd term in \eqref{eq6.107};
\item
   the 2nd term in  \eqref{eq6.107} dominates
   the 3rd term in \eqref{eq6.104};
\item
   both the 2nd term in \eqref{eq6.104}
   and the sum of the 1st and 4th terms in \eqref{eq6.107}
   are dominated by $\frac{3/\sqrt{2}}{(1+(1-1/\sqrt{2})\rho)^{m-1/2}}$.
\end{itemize}
Thus taking into account \eqref{eq6.108} and 
recalling the notation \eqref{eqABCD}
we conclude 
%find (maybe the only point that needs explanation
%is that the sum of the 1st and 4th terms in \eqref{eq6.107} is 
%dominated by the 2nd term in \eqref{eq6.104})
\be\la{eq6.110}
\ba
    |\Delta(\rho)| \leq &\frac1{Denom(m)} 
 \bigg( \frac7{\sqrt{2}}   \frac1{(m-1/2)(m-5/2)\rho^2}
              \\
      &   + \frac3{\sqrt{2}}\,\frac1{(m-1/2)\rho}\,
                   \frac1{(1+\rho/2)^{m-3/2}}
            + \frac3{\sqrt{2}}\,\frac1{(1+(1-1/\sqrt{2})\rho)^{m-1/2}}
         \bigg)
\ea
\ee
for all $m^{-1/2}\leq\rho<\infty$.

Now substituting $G(\rho)=\frac{1}{(m-3/2)\rho}+\Delta(\rho)$
into the integral $\int_{m^{-1/2}}^\infty$ in \eqref{eq190}, we obtain
$$
   -\frac{m-1}\pi \int_{m^{-1/2}}^{\infty}
     \frac{ \sin(q\phi(\rho))\,
       G(\rho)}{(1+\rho)^{3/2}}
       \,d\rho 
    = -\frac{m-1}{m-3/2}\frac1\pi \int_{m^{-1/2}}^{\infty}
       \frac{\sin(q\phi(\rho))}{\rho(1+\rho)^{3/2}}d\rho + E_1(m,q)
$$
where 
\be\la{eq6.115}
\ba
  |E_1(m;q)|
    \leq            &\frac1{Denom(m)} \frac{m-1}\pi
 \bigg( \frac7{\sqrt{2}}   \frac1{(m-1/2)(m-5/2)}
               \int_{m^{-1/2}}^{\infty}\frac{d\rho}{\rho^2(1+\rho)^{3/2}}
              \\
      &+ \frac3{\sqrt{2}}\,\frac1{(m-1/2)}
              \int_{m^{-1/2}}^{\infty}
               \frac{d\rho}{\rho(1+\rho/2)^{m-3/2}(1+\rho)^{3/2}}\\
       &+\frac3{\sqrt{2}}\,  \int_{m^{-1/2}}^{\infty}
               \frac{d\rho} {(1+(1-1/\sqrt{2})\rho)^{m-1/2}(1+\rho)^{3/2}}
 \bigg)
\ea
\ee
After elementary manipulations (splitting the integral as below)
the first term
gives $A(m)$ in \eqref{eqABCD}.
To estimate the second term, note that
$$
\ba
       \Big(\int_{m^{-1/2}}^1 + &\int_1^{\infty}\Big)
               \frac{d\rho}{\rho(1+\rho/2)^{m-3/2}(1+\rho)^{3/2}}\\
    &\leq  \frac1{(1+1/(2\sqrt{m}))^{m-3/2}}
          \int_{m^{-1/2}}^{1}  \frac{d\rho}\rho
      + \frac1{(3/2)^{m-3/2}} \int_{1}^{\infty}
               \frac{d\rho}{\rho^{5/2}}
\ea
$$
which gives $B(m)$ in \eqref{eqABCD}.
Finally in the third term in \eqref{eq6.115} we 
use 
$$
\int_{m^{-1/2}}^{\infty}
               \frac{d\rho} {(1+(1-1/\sqrt{2})\rho)^{m-1/2}(1+\rho)^{3/2}}
 \leq \int_{m^{-1/2}}^{\infty}
               \frac{d\rho} {(1+(1-1/\sqrt{2})\rho)^{m+1}}
$$
and arrive at $C(m)$ in \eqref{eqABCD}. 
This completes the proof of Lemma \ref{s7lem1}.
\end{proof}
In the integral $\int_0^{m^{-1/2}}$ in \eqref{eq190} we
approximate the function $G(\rho)$ by the function
\be\la{eqGa}
  G_a(\rho)\equiv
  \frac
      {\int_0^1 e^{-\rho(m+1/2)v} \frac{vdv}{\sqrt{1-v}} }
          {\int_0^1 e^{-\rho(m+1/2)v}  \frac{dv}{\sqrt{1-v}} }.
\ee
and use the following result.
\begin{lemma}\la{s7lem2} 
For $m\geq38$, we have
\be\la{eqs7lem2}
-\frac{m-1}{\pi} \int_{0}^{m^{-1/2}} \frac{\sin(q\phi(\rho))\,
       G(\rho)}{(1+\rho)^{3/2}}
       \,d\rho
   = -\frac{m-1}\pi \int_0^{m^{-1/2}}
       \frac{\sin(q\phi(\rho))}{(1+\rho)^{3/2}}\,G_a(\rho)\,d\rho + E_2(m,q)
\ee
where the error term $E_2(m,q)$ satisfies, uniformly for all $q=3,5,\cdots$,
$$
         |E_2(m,q)|\leq \frac1\pi\frac1{\sqrt{m}}\cdot6,
$$
where the number $6$ provides the following
 (crude) bound (proved in the Appendix) 
\be\la{eqApp1}
   \max_{m\geq38}
       \max_{s\geq0}s^2\frac{\int_0^1e^{-svR(m)}\,\frac{v^2\,dv}{\sqrt{1-v}}}
             {\int_0^1e^{-sv}\frac{dv}{\sqrt{1-v}}} \leq 6,
\ee
where 
\be\la{eqRm}
   R(m)\equiv\frac1{1+\frac1{\sqrt{m}}}.
\ee
As $R(m)$ increases with $m$, the
maximum on the LHS of \eqref{eqApp1}
is achieved at $m=38$.
\end{lemma}
\begin{proof}
We consider $0\leq\rho\leq{}m^{-1/2}$.
Recall the notation \eqref{eqk}
and denote
$$
   \Delta_k(s)\equiv
       s^2\,\frac{\int_0^1e^{-s vR(m)}\,\frac{v^2\,dv}{\sqrt{1-v}}}
             {\int_0^1e^{-s v}\frac{dv}{\sqrt{1-v}}}.
$$
Assume for a moment the estimates
\be\la{eqE.18.0}
 0\leq G(\rho) - G_a(\rho) \leq 
         \frac{\Delta_k(k\rho)}
             {k},\qquad0\leq\rho\leq m^{-1/2}.
\ee
Then \eqref{eqs7lem2} holds
where $E_2(m,q)$ satisfies
$$
\ba
       |E_2(m,q)|&\leq \frac{m-1}\pi
       \frac1k \int_0^{m^{-1/2}} \Delta_k(k\rho)\,\frac{d\rho}{(1+\rho)^{3/2}}\\
    &\leq\Big(\max_{s\geq0}\Delta_k(s)\Big) \frac1\pi
           \int_0^{m^{-1/2}} \frac{d\rho}{(1+\rho)^{3/2}}
   \leq6\cdot \frac1\pi\cdot\frac1{\sqrt{m}}
\ea
$$
uniformly for all $m\geq38$ and 
all $q=3,5,\cdots$.
This is the desired result.

It remains to prove \eqref{eqE.18.0}. First consider the upper bound.
In view of \eqref{eqG}, \eqref{eqGa}, \eqref{eqE.4.0}
$$
 G(\rho) \leq \frac
      {\int_0^1 e^{-k\rho v/(1+\rho)} \frac{vdv}{\sqrt{1-v}} }
          {\int_0^1 e^{-k\rho v}  \frac{dv}{\sqrt{1-v}} }
  = G_a(\rho) + 
       \frac
        {\int_0^1 \big(e^{-k\rho v/(1+\rho)} - e^{-k\rho v}\big) 
                            \,\frac{vdv}{\sqrt{1-v}} }
          {\int_0^1 e^{-k\rho v}  \frac{dv}{\sqrt{1-v}} }.
$$
Next, for all $0\leq v\leq1$
$$
\ba
 e^{-k\rho v/(1+\rho)} - e^{-k\rho v}
     &= \int_0^1 \frac{d}{dt}\Big(e^{-k\rho v(1-t + t/(1+\rho))}
                                               \Big)\,dt
     = \frac{k\rho^2v}{1+\rho}
                 \int_0^1  e^{-k\rho v(1-t + t/(1+\rho))} \,dt\\
    &\leq \frac{k\rho^2v}{1+\rho}
               e^{-k\rho v/(1+\rho)}  \int_0^1   dt
     \leq k\rho^2v\,  e^{-k\rho vR(m)}
\ea
$$
where the exponent is estimated by its maximal value at $t=1$
and we use the fact that $-\frac1{1+\rho}\leq-R(m)$ 
for $0\leq\rho\leq\frac1{\sqrt{m}}$ by the definition \eqref{eqRm}.
This proves the upper bound in \eqref{eqE.18.0}.

For the lower bound in \eqref{eqE.18.0} we note 
$$
\ba
 \big(G(\rho) &- G_a(\rho)\big)
   \cdot\Big(\int_0^1  (1+\rho v)^{-k}  \,\frac{dv}{\sqrt{1-v}}\Big)  
   \cdot\Big(\int_0^1  e^{-k\rho v}  \,\frac{dv}{\sqrt{1-v}}\Big)  \\
  &= \int_0^1\int_0^1  (1+\rho v)^{-k} e^{-k\rho w} 
                \,(v-w)\,\frac{dv}{\sqrt{1-v}}\,\frac{dw}{\sqrt{1-w}} \\
  &= \frac12 \int_0^1\int_0^1 \Big( 
           (1+\rho v)^{-k} e^{-k\rho w} 
           -(1+\rho w)^{-k} e^{-k\rho v} 
                \Big)\,(v-w)\,\frac{dv}{\sqrt{1-v}}\,\frac{dw}{\sqrt{1-w}} \\
  &= \frac12 \int_0^1\int_0^1 (v-w)\,(1+\rho v)^{-k}\,e^{-k\rho v} 
        \bigg[       e^{k\rho(v- w)} 
                                  -\bigg(\frac{1+\rho v}{1+\rho w}\bigg)^{k} 
          \bigg] \,\frac{dv}{\sqrt{1-v}}\,\frac{dw}{\sqrt{1-w}} .
\ea
$$
Consider $f(x)\equiv x^k$, $k\geq1$.
Note that for $a,b>0$,
 $f(a)-f(b)=f^\prime(\theta(a,b))(a-b)$ for some $\theta(a,b)$
between $a$ and $b$. As $\theta(a,b)>0$, we must
have $f^\prime(\theta(a,b))>0$.
Thus
\be\la{eq6.125}
\ba
 \big(G(\rho) &- G_a(\rho)\big)
   \cdot\Big(\int_0^1  (1+\rho v)^{-k}  \,\frac{dv}{\sqrt{1-v}}\Big)  
   \cdot\Big(\int_0^1  e^{-k\rho v}  \,\frac{dv}{\sqrt{1-v}}\Big)  \\
  &= \frac12 \int_0^1\int_0^1 (v-w)
        \bigg(       e^{\rho(v- w)} 
                                  -\frac{1+\rho v}{1+\rho w}
          \bigg) \,F_{k,\rho}(v,w)\,\frac{dv}{\sqrt{1-v}}\,\frac{dw}{\sqrt{1-w}} 
\ea
\ee
where $F_{k,\rho}(v,w)\geq0$ everywhere.
Consider now $f_w(v)\equiv e^{\rho(v- w)} 
                                  -\frac{1+\rho v}{1+\rho w}$.
Then $f_w(w)=0$ and $f_w^\prime(v)=\rho\big( e^{\rho(v- w)} 
                                  -\frac{1}{1+\rho w}\big)>0$ for $v>w\geq0$
(note $\rho>0$). Thus $f_w(v)>0$ for $v>w$.
Also  $f_w(v)= e^{\rho(v- w)}\frac{1+\rho v}{1+\rho w}
 \big( \frac{1+\rho w}{1+\rho v}-e^{\rho(w-v)}\big) 
$
which is $<0$ by the same argument for $0\leq v<w$.
We conclude that the indegrand in \eqref{eq6.125} is $\geq0$
for all $0\leq v,\,w\leq1$
which proves the lower bound in \eqref{eqE.18.0}.
The proof of Lemma \ref{s7lem2} is now complete.
\end{proof}
The integral that appears on the RHS in \eqref{eqs7lem2},
may be written in the form
\be
\la{eq201}
\ba
  -\frac{m-1}\pi &\int_0^{m^{-1/2}}
         \frac{\sin(q\phi(\rho))}{(1+\rho)^{3/2}}\,G_a(\rho)\,d\rho\\
   =&-\frac{m-1}{m+\frac12}\frac1\pi \int_0^{m^{-1/2}}
         \frac{\sin(q\phi(\rho))}{\rho(1+\rho)^{3/2}}\,
            \Big((m+1/2)\rho G_a(\rho)-1\Big)\,d\rho\\
     &-\frac{m-1}{m+\frac12}\frac1\pi \int_0^{m^{-1/2}}
         \frac{\sin(q\phi(\rho))}{\rho(1+\rho)^{3/2}}\,d\rho
\ea
\ee
The second integral on the right in \eqref{eq201} equals
$$
\ba
  -\frac{m-1}{m+\frac12}\frac1\pi \int_0^{m^{-1/2}}
         \frac{\sin(q\phi(\rho))}{\rho(1+\rho)^{3/2}}\,d\rho
  =-\frac{m-1}{m-\frac32}\frac1\pi \int_0^{m^{-1/2}}
         \frac{\sin(q\phi(\rho))}{\rho(1+\rho)^{3/2}}\,d\rho
   +E_3(m;q)
\ea
$$
where
\be\la{eq204}
\ba
  |E_3(m;q)|\equiv
    &\frac2\pi\frac{m-1}{(m+\frac12)(m-\frac32)} \int_0^{m^{-1/2}}
         \frac{|\sin(q\phi(\rho))|}{\rho(1+\rho)^{3/2}}\,d\rho\\
     &\leq\frac2\pi\frac{q}{m+\frac12} \frac{m-1}{m-\frac32}\int_0^{m^{-1/2}}
         \frac{|\phi(\rho)|}{\rho(1+\rho)^{3/2}}\,d\rho\\
    &\leq\frac2\pi\frac{q}{m+\frac12}\frac{m-1}{m-\frac32} \int_0^{m^{-1/2}}
         \frac{\sqrt{\rho}}{\rho(1+\rho)^{3/2}}\,d\rho\\
    &\leq\frac2\pi\frac{q}{m+\frac12}\frac{m-1}{m-\frac32} \int_0^{m^{-1/2}}
         \frac{d\rho}{\sqrt{\rho}}\\
    &\leq \frac4\pi\,\frac{q}{\sqrt{m+\frac12}}\,\frac{m-1}{m-\frac32}\,\frac1{m^{3/4}}
\ea
\ee
and we have used
\be\la{eq202}
   \phi(\rho)\leq\tan(\phi(\rho)) = \sqrt{\rho}
\ee
by \eqref{eq200}.
Denote the first integral on the right in \eqref{eq201}
 by $E_4(m;q)$
and set
$$
        F(t)\equiv t\,\frac{\int_0^1e^{-tv}\frac{v\,dv}{\sqrt{1-v}}}
                   {\int_0^1e^{-tv}\,\frac{dv}{\sqrt{1-v}}} - 1.
$$
Then $(m+1/2)\rho G_a(\rho)-1=F((m+1/2)\rho)$ and 
$$
\ba
 |E_4(m;q)| = 
   \bigg| &-\frac{m-1}{m+\frac12} \frac1\pi \int_0^{m^{-1/2}}
         \frac{\sin(q\phi(\rho))}{\rho(1+\rho)^{3/2}}\,F((m+1/2)\rho)\,d\rho \bigg|\\
  &\leq \frac{m-1}{m+\frac12} \frac1\pi \int_0^{m^{-1/2}}
         \frac{q\phi(\rho)}{\rho(1+\rho)^{3/2}}\,|F((m+1/2)\rho)|\,d\rho\\
  &\leq \frac1\pi q \int_0^{m^{-1/2}} |F((m+1/2)\rho)|\,
         \frac{d\rho}{\sqrt{\rho}}\\
  &\leq  \frac1\pi \frac{q}{\sqrt{m+\frac12}} 
       \int_0^{+\infty} |F(w)|\,
         \frac{dw}{\sqrt{w}}\\ 
  &\leq  \frac1\pi \frac{q}{\sqrt{m+\frac12}} 
       \int_0^{+\infty} 2|F(s^2)|\,ds
\ea
$$
where we have again used \eqref{eq202}.
Now we denote
$$
 H(s)\equiv 2 F(s^2) = 2\bigg(s^2\,\frac{\int_0^1e^{-s^2v}\frac{v\,dv}{\sqrt{1-v}}}
                   {\int_0^1e^{-s^2v}\,\frac{dv}{\sqrt{1-v}}} - 1\bigg)
$$
and refer to the (crude) bound proved in the Appendix
\be\la{eqApp2}
      \int_0^{+\infty}|H(s)|\,ds \leq 2.8
\ee
to conclude that 
\be\la{eq205} 
   |E_4(m;q)| 
    \leq  \frac1\pi \frac{q}{\sqrt{m+\frac12}} 
       \cdot 2.8.
\ee

Combining Lemma~\ref{s7lem1} and \ref{s7lem2} we find 
\be\la{eq210}
   \tilde{I}(q) = -\frac{m-1}{m-\frac32}\frac1\pi
       \int_0^\infty \frac{\sin(q\phi(\rho))}{\rho(1+\rho)^{3/2}}d\rho
 +\sum_{j=1}^4 E_j(m;q).
\ee
Returning to the variable $\theta=\frac\pi2-\phi(\rho)$
and using \eqref{eq200} we see that for any $q=3,5,\cdots$
$$
  \frac1\pi \int_0^\infty \frac{\sin(q\phi(\rho))}{\rho(1+\rho)^{3/2}}d\rho
 = \frac2\pi \int_0^{\pi/2} \frac{\sin(q\theta)}{\sin\theta}\,d\theta = 1.
$$
Hence \eqref{eq210} becomes
\be\la{eq210prime}
   \tilde{I}(q) = -1
 +\sum_{j=1}^5 E_j(m;q)
\ee
where 
$$
 E_5(m;q)\equiv-\frac12\frac1{m-\frac32},
       \qquad |E_5(m;q)|\leq\frac12\frac1{m-\frac32}.
$$
Recall that we need a bound on $1+\tilde{I}(q)$.
Assembling the above estimates
for $m\geq38$, $q=3,5,\cdots,4m-5$,
\be\la{eq300}
\ba
 |1+\tilde{I}(q)| &= \Big|\sum_{j=1}^5 E_j(m;q)\bigg| 
     \leq \sum_{j=1}^5 |E_j(m;q)|\\
    &\leq \frac{A(m)+B(m)+C(m)}{Denom(m)}
               +\frac1{\sqrt{m}}\,\frac{6}{\pi}+\frac12\frac1{m-\frac32}\\
      &\qquad+\frac{q}{\sqrt{m+1/2}}\,\bigg( \frac{2.8}{\pi}
             + \frac4\pi\,\frac{m-1}{m-\frac32}\,\frac1{m^{3/4}}\bigg).
\ea
\ee

We can now complete the proof of Theorem~\ref{thm7} for the 
case of large $m$.
We wish to apply Proposition \ref{s7prop1}.
We estimate the elements of the matrix $X_{m-1}$ as follows.
Enumerate the diagonals of $X_{m-1}$ by $q=3,5,\cdots,4m-5$
starting from the lower left corner as before. 
For a given {\em fixed\ }$m$,
and {\em large\ }odd $q$, that is odd $q$ which satisfy 
the condition $q\geq\big[\frac4\pi\sqrt{m+1/2}\big]+1$,
we estimate the elements on the corresponding diagonal
of $X_{m-1}$ as follows in view of \eqref{eqqestim}
\be\la{eqqq}
   |1+\tilde{I}(q)|\leq 1+|\tilde{I}(q)| 
     \leq 1 + \frac{\frac4\pi\sqrt{m+1/2}}
                  {\big[\frac4\pi\sqrt{m+1/2}\big]+1}
          \equiv C_1(m)<2.
\ee
%Hence for any fixed $m$ we will have
%$$
%    C_1(m)\,\Big(\frac12-\frac{(m!)^2}{m(2m)!}\cdot2^{2m-2}\Big)
%         <\frac{C_1(m)}2<1.
%$$
But for the elements in $X_{m-1}$ on the diagonals
corresponding to {\em small\ }odd $q$,
that is for $3\leq q\leq\big[\frac4\pi\sqrt{m+1/2}\big]$,
we have $\frac{q}{\sqrt{m+1/2}}\leq\frac4\pi$,
and for such $q$ we find from \eqref{eq300}
%substituting this in \eqref{eq300} (that we use for $q$ in this range)
%and find
\be\la{eq301}
\ba
 |1+\tilde{I}(q)| 
    &\leq \frac{A(m)+B(m)+C(m)}{Denom(m)}
     +\frac1{\sqrt{m}}\,\frac{6}{\pi}+\frac12\frac1{m-\frac32}\\
      &\qquad+\frac{4}{\pi}\,\bigg( \frac{2.8}{\pi}
             + \frac4\pi\,\frac{m-1}{m-\frac32}\,\frac1{m^{3/4}}\bigg)
        \equiv C_2(m).
\ea
\ee
Thus for $m\geq38$ and for $q=3,5,\cdots,4m-5$ we have
$$
   |1+\tilde{I}(q)| \leq C(m)
$$
where $C(m)=\max(C_1(m),C_2(m))$.
Now it is elementary to show that for $m\geq58$,
all the terms on the RHS of  \eqref{eq300}
decrease monotonically with $m$
and direct computation shows that $C_2(58)<1.997<2$.
Together with \eqref{eqqq} this implies $C(m)<2$
for all $m\geq58$ and hence $\det{}T_m\neq0$
by Proposition \ref{s7prop1} for $m$ in this range.

But direct evaluation shows that for $38\leq m\leq57$
$$
  C_2(m)\,\Big(\frac12-\frac{(m!)^2}{m(2m)!}
    \cdot2^{2m-2}\Big)<
      0.996<1
$$
and so, again using \eqref{eqqq}, we have
$$
  C(m)\,\Big(\frac12-\frac{(m!)^2}{m(2m)!}
    \cdot2^{2m-2}\Big)<1,\qquad 38\leq m\leq 57.
$$
Hence $\det{}T_{m-1}\neq0$ also for $38\leq m\leq 57$.
This completes the proof of Theorem~\ref{thm7}.
\section*{Appendix}
\setcounter{equation}{0}
\setcounter{theorem}{0}
\setcounter{section}{1}
\setcounter{subsection}{0}
\renewcommand{\thetheorem}{\Alph{section}.\arabic{theorem}}
\renewcommand{\theremark}{\Alph{section}.\arabic{remark}}
\renewcommand{\theequation}{\Alph{section}.\arabic{equation}}
\renewcommand{\thesubsection}{\Alph{section}.\arabic{subsection}}
Our goal here is to prove the two (crude) estimates,
\eqref{eqApp1} and \eqref{eqApp2} needed in the text.
For $s\in[0,+\infty)$, set
$$
     L(s) \equiv s^2\frac{\int_0^1 e^{-R_0sv}\frac{v^2\,dv}{\sqrt{1-v}}}
              {\int_0^1 e^{-sv}\,\frac{dv}{\sqrt{1-v}}}
$$
where $R_0\equiv0.855$. Note that $R(m)>R_0$ for all $m\geq38$.
Set also
\be\la{eqs8G}
  L_1(s)\equiv s\frac{\int_0^1 e^{-s^2v}\frac{v\,dv}{\sqrt{1-v}}}
              {\int_0^1 e^{-s^2v}\,\frac{dv}{\sqrt{1-v}}}
\ee
and 
$$
 H(s)\equiv 2\big(sL_1(s)-1\big).
$$
\begin{proposition}\la{s8prop1}
For $s\in[0,+\infty)$
$$
    0\leq L(s)\leq 6.
$$
\end{proposition}
\begin{proposition}\la{s8prop2}
We have
$$
    \int_0^{+\infty}|H(s)|\,ds\leq 2.8.
$$
The function $H(s)$ changes sign.
\end{proposition}
The functions $L(s)$ and $H(s)$
are plotted in Figs.~\ref{figapp1} and \ref{figapp2}
respectively, using Maple.
\begin{figure}[tb]
\begin{center}
\epsfig{file=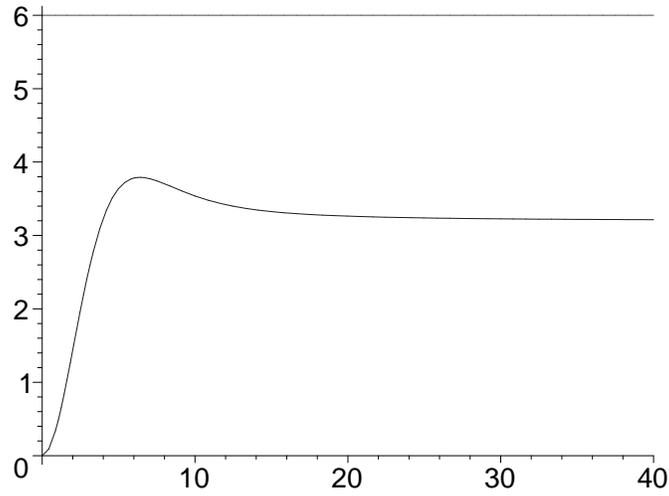,scale=1}
\caption{Graph of $L(s)$ and the line at height $6$}
\label{figapp1}
\end{center}
\end{figure}
\begin{figure}[tb]
\begin{center}
\epsfig{file=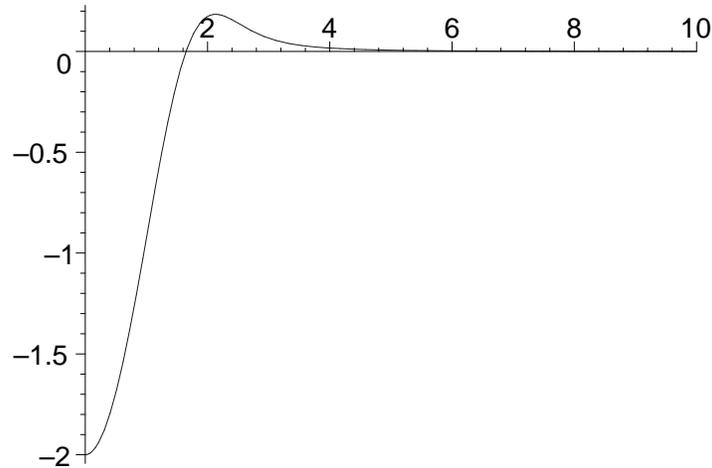,scale=1}
\caption{Graph of $H(s)$}
\label{figapp2}
\end{center}
\end{figure}
Fig.~\ref{figapp1} 
``proves'' Proposition \ref{s8prop1}.
Integrating by parts
we will prove below that
for all $s>0$
\be\la{s8eq2}
\begin{aligned}
%    G_1(s)&\leq \frac1s + \frac1{2s}\Big(\frac{1}{s^2}+\frac{30\sqrt{2}}{s^4}
 %           +2\sqrt{2}s^4e^{-s^2/2}\Big)\\
    H(s)&\leq\frac{1}{s^2}+\frac{30\sqrt{2}}{s^4}
            +2\sqrt{2}s^4e^{-s^2/2}
\ea
\ee
which together with the fact (``evident'' from Fig.~\ref{figapp2} and proved below)
that $H$ is positive on $[6,+\infty)$ implies 
$\int_6^\infty|H(s)|ds\leq0.233$. 
Again, using Maple, direct numerical integration
shows that $\int_0^6|H(s)|ds\leq2.407$.
This ``proves'' Proposition \ref{s8prop2} with $2.64$ on the right.
Given the importance of the numerical values
of the bounds \eqref{eqApp1} and \eqref{eqApp2}
in our Proof of Theorem~\ref{thm7},
we now give a rigorous
error analysis (in which we use the computer to the extent of computing
the four arithmetic operations and the exponential function only)
for these bounds.

In the error analysis both for $L$ and $H$, the integral 
$$
           INT(x)\equiv\int_0^1 e^{-xv}(1-v)^{-1/2}\,dv
$$
plays a basic role. Note that for any $0<a<1$, 
\be\la{eqs8STAR0}
   \int_a^1 e^{-xv}(1-v)^{-1/2}\,dv\leq e^{-ax}2\sqrt{1-a}.
\ee
Integrating by parts we find
\be\la{eqs8STAR}
\ba
       \int_0^a e^{-xv}(1-v)^{-1/2}\,dv 
        =&\frac1x\big[ 1-e^{-ax}(1-a)^{-1/2}\big]\\ 
&+\frac{1}{2x^2}\big[ 1-e^{-ax}(1-a)^{-3/2}\big]\\ 
&+\frac3{4x^3}\big[ 1-e^{-ax}(1-a)^{-5/2}\big]\\ 
&+\frac{15}{8x^4}\big[ 1-e^{-ax}(1-a)^{-7/2}\big]\\ 
&+\frac{105}{16x^4}\int_0^a e^{-xv}(1-v)^{-9/2}\,dv\\
\leq& \frac1x + \frac{1}{2x^2} + \frac3{4x^3} 
             +\frac{15}{8x^4}\,(1-a)^{-7/2}.
\ea
\ee
\subsection{Proof of Proposition \ref{s8prop1}}
Clearly $L(s)\geq0$ for $s\geq0$.
Note that 
\be\la{eqs8STARSTAR}
   \int_{0}^1 e^{-xv}(1-v)^{-1/2}\,dv \geq \frac{1-e^{-x}}{x}.
\ee
Expanding the square of $v=1-(1-v)$ and integrating by parts we find 
\be\la{eqs8STAR5}
\ba
       \int_0^1 e^{-xv}v^2 (1-v)^{-1/2}\,dv 
        =&-\frac1x - \frac{3}{2x^2}
    + \bigg(1+\frac1x + \frac{3}{4x^2}\bigg)\,INT(x).
\ea
\ee
Substututing here the estimates \eqref{eqs8STAR0}, \eqref{eqs8STAR} for
$a=1/2$ and using \eqref{eqs8STARSTAR}
we find (after simplifications)
an upper estimate on (nonnegative) $L(s)$
which implies that $L(s)\leq6$ for $s\geq25$.

Next, making the change of variable $t=\sqrt{1-v}$ in both integrals in $L(s)$
$$
     L(s) \leq s^2\frac{\int_0^1 e^{R_0s(t^2-1)}\,dt}
              {\int_0^1 e^{s(t^2-1)}\,dt}\leq s^2\,e^{s(1-R_0)}
      \leq 6
$$
for $0\leq s\leq2$.

Finally consider $s\in[2,25]$. Using \eqref{eqs8STARSTAR}
$$
      L(s)\leq\frac1{1-e^{-s}}\,L_2(s),\qquad
    L_2(s)\equiv s^3\int_0^1 e^{-R_0sv}\,\frac{v^2\,dv}{\sqrt{1-v}}.
$$
Note that the first factor is $\leq1.157$. Hence it suffices to prove
$$
          L_2(s)\leq 5.185,\qquad 2\leq s\leq 25.
$$
We do this as follows. Assume for a moment
that the following (crude)
a priori bound holds
\be\la{eqs8STAR4}
       |L_2^\prime(s)|\leq 75,\qquad 2 \leq s\leq 25.
\ee
Consider for some $N_e$ 
the mesh $s_j=2+jh$, $j=0,1,\cdots,N_e$ 
where $h=(25-2)/N_e$. By \eqref{eqs8STAR4}
\be\la{eqs8S9}
       |L_2(s)|\leq |L_2(s_j)| + 75h,
                \qquad s\in[s_{j-1},s_j],\qquad j=1,2,\cdots,N_e.
\ee
The value of $L_2$ at a point is estimated as follows.
$L_2$ is $s^3$ times \eqref{eqs8STAR5} with $x=R_0s$.
Making the change of variable $t=\sqrt{1-v}$ in $INT(x)$
 we obtain for any $N_i$
\be\la{eqs8S10}
 INT(R_0s) = 2\int_0^1e^{R_0s(t^2-1)}\,dt
            \leq \frac2{N_i}\sum_{k=1}^{N_i}
                 e^{R_0s((k/N_i)^2-1)}
\ee
where the integral is estimated by the right rectangle sum
using the monotonic growth of $e^{R_0s(t^2-1)}$ for $t\in[0,1]$.
Now we can just increase $N_e$, $N_i$ until 
$\max_{j=1,2,\cdots,N_e}$ of the RHS in \eqref{eqs8S9}
(where \eqref{eqs8S10}
has been substituted) becomes $\leq5.185$. It turns out that
$N_e=8,000$ and $N_i=12,000$ suffice and for these values
the maximum is $\leq5.162$. 
We must also check that the error does not
accumulate too much when we sum $12,000$ numbers,
the precision of each being $10$ digits. 
In the worst scenario,
the mantissa after each addition accumulates an error of  0.5
in the 10th digit.
After 12,000 additions the error is 6 in the 7th digit.
This means that our result has at least 5 correct digits.
And in the above explanation
we have kept only 4 digits in arriving at the number $5.162$.

It remains to prove the bound \eqref{eqs8STAR4}.
Differentiating $G_2(s)$ we find 
$$
       |G_2^\prime(s)|\leq 
           \max\Big\{
                  3s^2\int_0^1e^{-R_0sv}v^2\,\frac{dv}{\sqrt{1-v}},\,
                 R_0s^3\int_0^1e^{-R_0sv}v^3\,\frac{dv}{\sqrt{1-v}}
           \Big\}.
$$
The first integral is estimated  using \eqref{eqs8STAR5},
\eqref{eqs8STAR}. For the second integral: start by rewriting the expression
in terms of $INT(x)$ (as we have done in \eqref{eqs8STAR5})
and then use \eqref{eqs8STAR}. These estimates imply \eqref{eqs8STAR4}
for $s\geq2$.
\subsection{Proof of Proposition \ref{s8prop2}}
Writing $v=1-(1-v)$ and integrating by parts as above
we can rewrite $H(s)$
in terms of $INT(s^2)$
\be\la{eqs8S1}
       H(s) = 2s^2 - 1 - \frac{2}{INT(s^2)} 
         = 2s^2-1-\frac1{\int_0^1 e^{s^2(u^2-1)}\,du}.
\ee
Combining \eqref{eqs8S1}
with \eqref{eqs8STAR}, the estimate \eqref{s8eq2}
follows. 
We need also a {\em lower\ }estimate on $H(s)$.
As in \eqref{eqs8STAR}, for any $a\in(0,1)$
\be\la{eqs8S}
\ba
       \int_0^1 &e^{-xv}\,\frac{dv}{\sqrt{1-v}} \geq
  \int_0^a e^{-xv}\,\frac{dv}{\sqrt{1-v}}
        =\frac1x+\frac{1}{2x^2}+\frac3{4x^3}
                    +\frac{15}{8x^4}\\
&-e^{-ax}\bigg[ \frac1x(1-a)^{-1/2}
     +\frac{1}{2x^2} (1-a)^{-3/2}
     +\frac3{4x^3} (1-a)^{-5/2}
     +\frac{15}{8x^4} (1-a)^{-7/2} \bigg].
\ea
\ee
Together with \eqref{eqs8S1}, this implies
$$
    H(\sqrt{x}) \geq \frac{NUMER_a(x)}{DENOM_a(x)}
$$
where $DENOM_a(x)$ is the RHS in \eqref{eqs8S}
and $NUMER_a(x)$ is a certain, explicit function.
It is convenient to take $a=4/5$. Then it is elementary
to check $DENOM_{a}(x)>0$ for $x\geq9$.
We find
\be\la{eqs8ee}
\ba
  x^2NUMER(x)\geq 1+&\frac3x-\frac{15}{8x^2}
            -2\sqrt{5}x^2e^{-4x/5}\bigg(1+\frac{5}{2x}+\frac{75}{4x^2}
                    +\frac{1875}{8x^3}\bigg)\\
     &\geq\frac{211}{216} - \frac{27031}{17496}\sqrt{5}x^2e^{-4x/5}
\ea
\ee
where we have discarded $\frac3x$ and 
evaluated the monotonically decreasing 
terms $-\frac{15}{8x^2}$ and the term in the paranthesis at $x=9$.
It is now elementary to check that the RHS in \eqref{eqs8ee}
is $>0$ for $x\geq9$. 

We have proved that $H(s)>0$ for $s\geq3$. 
From this and \eqref{s8eq2} it follows that 
\be\la{eqs8one}
        \int_6^\infty|H(s)|ds\leq0.233.
\ee

On the interval $[3,6]$ where $H$ does not change sign and so $|H|$ is smooth
we use the trapezoidal integration scheme with a rigorous error
estimation.
However on $[0,3]$ (where $H$ may and indeed does change sign)
we use the following argument.
Fix $N_e$ and $N_i$ and consider 
the mesh $s_j=jh$, $j=0,1,\cdots,N_e$ 
where $h=(3-0)/N_e$. 
Using the left and right rectangles as in \eqref{eqs8S10}
we find that
$$
  H_j^{-} \leq H(s) \leq H_j^+,\qquad s\in[s_{j},s_{j+1}],
         \quad j=0,1,\cdots N_e
$$
where for any $N_i$
$$
\ba
    H_j^-&\equiv 2s_j^2 -\frac1{ \frac1{N_i}\sum_{k=0}^{N_i-1}
                 e^{s_{j+1}^2((k/N_i)^2-1)}}\\
    H_j^+&\equiv 2s_{j+1}^2 -\frac1{ \frac1{N_i}\sum_{k=0}^{N_i-1}
                 e^{s_{j}^2((k/N_i)^2-1)}}.
\ea
$$
This gives
$$
    |H(s)| \leq \max\big\{|H_j^-|,|H_j^+|\big\},\qquad s\in[s_{j},s_{j+1}],
         \quad j=0,1,\cdots N_e-1
$$
and
$$
         \int_0^3 |H(s)|\,ds\leq\frac3{N_e}\sum_{j=0}^{N_e-1}
                     \max\big\{|H_j^-|,|H_j^+|\big\}
$$
for any choice of $N_{i,e}$. Increasing these numbers we improve
the estimate. It turns out that for $N_{e}=N_i=3,000$ we get
$
         \int_0^3 |H(s)|\,ds\leq 2.242
$
(Maple numerical integration suggests the estimate $2.200$).
Again we only use the computer to evaluate
arithmetic operations and exponents. Also
the precision is 10 digits so that after $9\cdot10^6$ summations
a worst possible error of $0.5$ in the 10th digit at each step
becomes an error of $4.5$ in the 4th digit. 
Thus we write
\be\la{eqs8two}
         \int_0^3 |H(s)|\,ds\leq 2.247.
\ee

Finally we evaluate the integral over $[3,6]$ (where $H$ is positive)
via the trapezoidal scheme to obtain
$$
\ba
         \int_3^6 H(s)\,ds &= h \sum_{j=1}^{N_e-1}H(s_j) + \frac{H(3)+H(6)}2\,h
                 + ERR_e + ERR_i\\
    |ERR_e| &\leq \frac {(6-3)\,h^2}{12}\,
               \max_{s\in[3,6]} |H^{\prime\prime}(s)|\\
    |ERR_i|&\leq\textrm{(maximal error in computation of $H(s_j)$ over all
                       steps)$\times(6-3)$}
\ea
$$
where $s_j=3+jh$, $j=0,1,\cdots,N_e$, $h=(6-3)/N_e$.
At each $s_j$ we will compute $H(s_j)$ also by the trapezoidal scheme
with an explicit error estimate. We demand 
$|ERR_e|\leq0.05$ and $|ERR_{i}|\leq0.05$. Let us first find the required $N_e$.
Assume for a moment a (crude) estimate
\be\la{eqHpp}
      |H^{\prime\prime}(s)|\leq 24s^4 + 20 s^2 + 4,\qquad s\geq0.
\ee
Then 
$$
    \frac3{12}\bigg(\frac3{N_e}\bigg)^2\,(24\cdot6^4+2\cdot6^2+4)
          \leq0.05\qquad
         \textrm{implies}\qquad N_e=1,197.
$$
Now consider $N_i$. For any $s\in[3,6]$ we compute $\int_0^1 F_s(u)du$
where  $F_s(u)\equiv e^{s^2(u^2-1)}$ (see \eqref{eqs8S1})
by the trapezoidal scheme. Note
$$
        \max_{u\in[0,1]}\bigg|\frac{\partial^2}{\partial u^2} F_s(u)\bigg| 
               = 4s^4+2s^2.
$$
Now if $\int_0^1F_s(u)du=\sum(s)+err_i(s)$, then 
the error in the computation of $1/\int_0^1F_s(u)du$ (which enters $H(s)$) is 
$$
      \bigg|\frac1{\int_0^1} - \frac1{\sum(s)}\bigg| \leq 
                 \frac{|err_i(s)|}{|\int_0^1|\cdot|\sum(s)|}.
$$
From $\int_0^1e^{s^2(u^2-1)}du
=\frac12\int_0^1e^{-s^2t}\,\frac{dv}{\sqrt{1-v}}
\geq\frac{1-e^{-s^2}}{2s^2}$ 
we get a lower bound 
$$
   \min_{u\in[3,6]}\int_0^1e^{s^2(u^2-1)}du\geq 0.01388.
$$
When computing the sums $\sum(s)$ below
we also store the minimal
value of $\sum(s_j)$ over $j=0,1,\cdots,N_e$
 and verify at the end that it is also $\geq0.01388$.
Thus we demand
$$
\ba
        |ERR_i| &\leq (6-3)\,\max_{j=0,1,\cdots, N_e}
            \frac{|err_i(s_j)|}{(0.01388)^2} \\
           &\leq \frac1{12}\max_{j=0,1,\cdots, N_e}
                      \bigg(\frac1{N_i(s_j)}\bigg)^2
           \,4(s_j^4+s_j^2/2)\,\frac3{(0.01388)^2} \leq 0.05
\ea
$$
which holds if we choose
$$
        N_i(s_j)\geq 323\sqrt{s_j^4+s_j^2/2}, \qquad j=0,1,\cdots, N_e
$$
where $s_j=3+j\frac{6-3}{N_e}$.
A computer evaluation (again using only elementary operations)
then shows that $\int_3^6H(s)ds\leq0.208$ with our 
guaranteed precision $\pm0.1$. Also we keep 10 digits
and the total number of summands is 
$\leq1,197\cdot (323\sqrt{6^4+6^2/2}+1)\leq1.5\cdot10^8$.
Hence if the worst error of 0.5 in the 10th digit is made at each step,
the accumulated error does not exceed $7.5$ in the 4th digit
(in the mantissa of the number $2.08\cdot10^{-1}$).
Thus 
\be\la{eqs8three}
   \int_3^6 H(s)\,ds \leq 0.208+0.1+0.00075 = 0.30875\leq0.309.
\ee
(Maple numerical integration gives
$\int_3^6 H(s)\,ds \leq 0.208$.)

Collecting \eqref{eqs8one}, \eqref{eqs8two}, \eqref{eqs8three}
we conclude
$$
   \int_{0}^{\infty}|H(s)|\,ds \leq 0.233+2.247+0.309 = 2.789<2.8
$$ 
which completes the proof up to \eqref{eqHpp}.
To prove \eqref{eqHpp} denote
$$
      K(x)\equiv -2 + 2x\cdot\frac{\int v}{\int 1},
        \qquad \int v^l\equiv \int_0^1 e^{-xv}\frac{v^l\,dv}{\sqrt{1-v}},
         \quad l=0,1,2,\cdots.
$$
Then 
\be\la{eqs8extra}
     H(s)=K(s^2),\qquad H^\prime(s)=2s\cdot K^\prime(s^2).
\ee
We find
$$
         K^\prime(x) = 2\cdot\frac{\int v}{\int 1}
           +2x\cdot\bigg[-\frac{\int v^2}{\int 1} 
           + \bigg(\frac{\int v}{\int 1}\bigg)^2\bigg].
$$
If we single out the positive and the negative terms
and use the fact that each fraction is bounded from above by $1$,
we obtain
$$
         -2x \leq K^\prime(x) \leq 2 + 2x,\qquad x\geq0.
$$
and from \eqref{eqs8extra}
$$
     |H^\prime(s)| \leq 4s^3 + 4s,\qquad s\geq0.
$$
After computing $H^{\prime\prime}(s)$
in terms of  $K^\prime(x)$, $K^{\prime\prime}(x)$ 
in the same way,
we arrive at \eqref{eqHpp}. 
\bibliographystyle{plain}

\end{document}